\documentclass[a4paper,12pt]{article}
\pdfoutput=1
\usepackage{epsfig,graphicx,xcolor,amsbsy,amssymb,latexsym,amsfonts,amsmath,setspace}
\usepackage{pstricks}
\usepackage{color}
\usepackage{soul}
\usepackage[numbers,square,comma, compress]{natbib}
\usepackage{placeins}
\usepackage{wrapfig}
  \usepackage[bookmarksopen, bookmarksnumbered, bookmarksopenlevel=2]{hyperref}
\usepackage{eurosym}
\usepackage{nccmath}
\usepackage[a4paper]{geometry}
\usepackage{graphicx}
\usepackage{tikz}
\usetikzlibrary{decorations.pathreplacing}
\usepackage{xcolor}
\usepackage{amsmath,amssymb,amsfonts,pstricks,setspace}
\usepackage[enableskew]{youngtab}
\numberwithin{equation}{section}

\newcommand{\bea}{\begin{eqnarray}\displaystyle}
\newcommand{\eea}{\end{eqnarray}}

\newcommand\scalemath[2]{\scalebox{#1}{\mbox{\ensuremath{\displaystyle #2}}}}

\DeclareMathOperator{\img}{img}

\begin{document}
\pagestyle{empty}
\baselineskip=14pt
\parskip 5pt plus 1pt 

\vspace*{2cm}
\begin{center}
{\Large \bfseries
Beyond Large Complex Structure: Quantized Periods \\[.2cm] 
and Boundary Data for One-Modulus Singularities
}~\\[.3cm]

\vspace{1cm}
{\bf Brice Bastian}, {\bf Damian van de Heisteeg},$^{2}$ {\bf Lorenz Schlechter}$^1$

\vskip 11 mm
\small ${}^{1}$ 
{\small
Institute for Theoretical Physics, Utrecht University\\ Princetonplein 5, 3584 CC Utrecht, The Netherlands\\[3mm]
}

\small ${}^{2}$ 
{\small
Center of Mathematical Sciences and Applications, Harvard University \\
20 Garden Street, Cambridge, MA 02138, USA\\[3mm]
}

\vspace*{1.5em}

\end{center}

\begin{abstract} \noindent
We study periods in an integral basis near all possible singularities in one-dimensional complex structure moduli spaces of Calabi-Yau threefolds. Near large complex structure points these asymptotic periods are well understood in terms of the topological data of the mirror Calabi-Yau manifold. The aim of this work is to characterize the period data near other boundaries in moduli space such as conifold and K-points. Using results from Hodge theory, we provide the general form of these periods in a quantized three-cycle basis. Based on these periods we compute the prepotential and related physical couplings of the underlying supergravity theory. Moreover, we elucidate the meaning of the model-dependent coefficients that appear in these expressions: these can be identified with certain topological and arithmetic numbers associated to the singular geometry at the moduli space boundary. We illustrate our findings by studying a wide set of examples.

\end{abstract}

\begin{center}

\vspace*{16em}

{\small \verb "brice_bastian@hotmail.com, dvandeheisteeg@fas.harvard.edu," } \\
{\small \verb "l.k.schlechter@uu.nl" }

\end{center}

\newpage
\setcounter{footnote}{0}
\setcounter{page}{1}
\pagestyle{plain}
\vspace{-1.5cm}
\tableofcontents
\newpage
\section{Introduction}
Effective field theories arising from compactifications of string theory vary over scalar field spaces that often extend across a number of distinct phases. In a given phase one may typically compute physical couplings as expansion series around a singularity. However, these expansions break down as one passes from one phase to another, signaling the emergence of a dual description taking over. In order to get a complete picture of the resulting landscape of effective field theories, one is thus required to carefully examine each of the phases in the field space. 

In this work we consider 4d $\mathcal{N}=2$ (and $
\mathcal{N}=1$) supergravity theories obtained from Type IIB (orientifold) compactifications on Calabi-Yau threefolds. We focus on field spaces spanned by the complex structure moduli: these moduli spaces exhibit a rich pattern of possible phases \cite{Witten:1993yc, Aspinwall:1994ay}, the most well-known type being the so-called large complex structure (LCS) regime. Our understanding of the LCS phase has benefited tremendously from mirror symmetry \cite{Candelas:1990rm}, which identifies them with large volume regimes in the K\"ahler moduli space of a mirror-dual Calabi-Yau threefold. The aim of this paper is to illuminate the underlying structure in other phases, away from this LCS lamppost, to a similar degree. For simplicity we restrict our attention to the one-modulus setting, i.e.~$h^{2,1}=1$, in which case the other possible singularities are conifold points and K-points (following the nomenclature of \cite{vanstraten2017calabiyau}).

A prerequisite for our goal is to characterize the model-dependent parameters appearing in the physical couplings --- which we will refer to as boundary data --- of the effective field theory. In the LCS phase we know from mirror symmetry that these numbers can be interpreted as certain topological data associated to the mirror manifold. In fact, this angle opens up a window to generate large numbers of explicit examples from the Kreuzer-Skarke database \cite{Kreuzer:2000xy} using toric geometry methods, cf.~\cite{Demirtas:2022hqf} for the current state-of-the-art technology. In order to put other phases in complex structure moduli space on an equal footing, we must first identify the appropriate asymptotic structures associated to these limiting regimes. To this end we turn to asymptotic Hodge theory \cite{Schmid, CattaniKaplan, CKS}, which specifies a so-called limiting mixed Hodge structure (LMHS) for each of these phases. Following the work of \cite{KerrLMHS}, we show how the required boundary data is characterized by the LMHS. This approach produces a set of data analogous to the topological numbers one encounters in the LCS phase.

In general, the boundary data associated to a given phase is closely connected to the degenerate geometry that arises at its singularity. In order to make this statement more precise, it is convenient to sort the boundary data into three classes: monodromy data, rigid periods and extension data. The monodromy matrix contains certain intersection information: for the LCS phase it specifies the triple intersection number of the mirror manifold, while for the other phases it encodes similar data about the degenerate geometry. Rigid periods are related to periods of a resolution of the degenerate geometry (keeping $h^{2,1}=0$) at the singularity. Finally, extension data describes mixing among the periods: in the LCS phase this is parameterized by the second Chern class, and for other phases similar mixings have to be taken into account.

With the boundary data at hand, we next lift this information into full-fledged period vectors expanded in a given phase in moduli space. To this end we use the work of \cite{Bastian:2021eom} where a general method, based on the mathematical work of \cite{CattaniFernandez2000, CattaniFernandez2008}, was developed to describe these periods explicitly. In \cite{Bastian:2021eom} these techniques were already applied to describe the periods for all possible one- and two-moduli phases. The upshot of our paper is that we use the extension part of the boundary data to rotate to an integral basis; previously just a real basis was considered. Having this quantization is the crucial piece needed to match the data in the periods precisely with the geometrical data associated to the phase. For physical applications it is also essential e.g.~in the quantization of the fluxes for moduli stabilization scenarios.

In addition to the quantization of the basis, another take-away from asymptotic Hodge theory is the choice of coordinate for a given phase. A starting point is to take a so-called covering coordinate: the singularity then lies at imaginary infinity, while winding around it induces integer shifts. Such a coordinate is however not uniquely defined, as one can still perform divisor-preserving transformations, i.e.~shifts of the covering coordinate by exponentially small terms. In the LCS phase this ambiguity is resolved by the mirror map, which fixes the complex structure coordinate in terms of the K\"ahler coordinate on the dual side. In the other phases we find that similar natural choices can be made based on general Hodge-theoretic considerations. Remarkably, when applied to the LCS phase this method reproduces the mirror map, with a similar approach as \cite{morrison1992mirror,localbehavior}.

From our periods we compute the physical couplings in the 4d $\mathcal{N}=2$ and $\mathcal{N}=1$ supergravity theories. We provide an exhaustive summary of all relevant quantities such as prepotentials, K\"ahler metrics and flux potentials. The resulting couplings elucidate the boundary data that underlies the phase from the physical side. For instance, the gauge-kinetic functions are given directly by rigid periods and intersection data associated to the degenerate variety at the singularity. Expansions of the Yukawa coupling in examples also indicate that our Hodge-theoretic coordinate has a deeper interpretation: in some cases it even yields a series of rational coefficients (up to a common factor), reminiscent of the Gopakumar-Vafa invariants in the LCS phase \cite{Gopakumar:1998ii,Gopakumar:1998jq}. 

We demonstrate our models on an assortment of examples taken from the database \cite{van2004monodromy,almkvist2005tables} of one-parameter Calabi-Yau threefolds. Our starting point is given by the class of manifolds for which the periods are hypergeometric functions, from which we take the three conifold points and three K-points at infinity that have a semisimple monodromy factor. In these models the boundary data is algebraic, consisting of mostly rational coefficients. In order to portray a more genuine reflection of the landscape, we supplement this dataset by considering manifolds where increasingly more transcendental numbers show up.

We explain this pattern in the boundary data from the presence of this semisimple factor, i.e.~of finite order, in the monodromy. Namely, such a factor must be a symmetry of the boundary data (more precisely: an automorphism of the LMHS), thereby imposing certain polynomial constraints on the model-dependent coefficients. The number of constraints is given by the number of distinct eigenvalues of this finite order factor, which for the hypergeometric examples this is indeed maximal. We provide a complete classification of these semisimple monodromy factors, giving the allowed boundary data at each order. As a byproduct we obtain a stronger upper bound on the order $n$ of these semisimple factors, finding $n\leq 6$ for conifold and K-points rather than $n \leq 12$ for one-modulus singularities.

The boundary data appearing in these examples --- to be precise, the rigid periods and extension data --- has a precise arithmetic interpretation. Studying such connections between string theory and modularity was first explored in \cite{Moore:1998pn}, where evidence was found for rank-two attractor points to be located at special arithmetic loci in the moduli space such as complex multiplication points. Recently there has been much progress, both from the direction of these attractor points away from symmetric loci in moduli space \cite{Candelas:2019llw,Candelas:2021mwz}, as well as for supersymmetric flux vacua \cite{Kachru:2020sio, Schimmrigk:2020dfl, Kachru:2020abh, Candelas:2023yrg}. For our purposes the main point is that the geometrical data can be specified as certain L-function values for modular forms associated to the manifold. For conifold points this aspect has recently been studied in detail in \cite{Bonisch:2022mgw}. In this work, guided by asymptotic Hodge theory, we add to these results both by performing a similar study of K-points, as well as by considering conifold points with monodromies with a semi-simple part. In comparison to the LCS phase, it is amusing to remark that $\zeta(3)$ --- which appears with the Euler characteristic correction to the volume --- may be viewed as a simple example of a such an L-value.

The structure of this paper is as follows. In section \ref{sec:review} we review the machinery underlying period vectors of Calabi-Yau manifolds: in subsections \ref{ssec:analytic} and \ref{ssec:Hodge1} we discuss the necessary background involving Picard-Fuchs equations and asymptotic Hodge theory; in subsection \ref{ssec:Hodge2} we describe our approach for defining a natural coordinate and integral basis.
 
In section \ref{sec:summary} we give a physical summary of our results, where we describe how the geometrical data associated to the phases can be understood through the couplings in the associated supergravity theories. Section \ref{sec:construction} contains the bulk of our Hodge-theoretic analysis, where we derive the most general form of the periods for the three singularity types in an integral basis. In section \ref{sec:geometryinput} we discuss the geometrical meaning of the period data, identifying the analogue counterparts for intersection numbers and transcendental numbers like $\zeta(3)$ at conifold and K-points. This discussion is partly based on evidence from section \ref{sec:examples}, where we provide a variety of examples for which we computed this data explicitly. This data has been summarized in tables \ref{table:C} and \ref{table:K} for conifold and K-points respectively. We conclude in section \ref{sec:conclusions}. To the submission of this file we have attached two notebooks, detailing the computations for the three conifold and K-points at infinity in the hypergeometric models.

\section{Review on Calabi-Yau Periods}\label{sec:review}
In this section we introduce the necessary background on period vectors of Calabi-Yau threefolds. It is divided into two components: an analytical part reviewing how these periods are computed in geometrical examples, and a Hodge-theoretic part describing the mathematical structure underlying the asymptotic behavior. In subsection \ref{ssec:analytic} we discuss the computation of these periods by using the methods of Picard-Fuchs differential equations. In particular, we cover how to determine the transition matrices that rotate their solutions into an integral symplectic basis, a crucial aspect for many applications in physics or mathematics. 
In subsection \ref{ssec:Hodge1} we discuss these periods from the perspective of asymptotic Hodge theory: we review the near-boundary form of these periods given by the nilpotent orbit theorem, and introduce the limiting mixed Hodge structure associated to these singularities. The reader familiar with (either of) these two perspectives may skip to subsection \ref{ssec:Hodge2}, where we discuss the two key Hodge-theoretic concepts to this work: (1) the construction of an integral basis by using extension data following \cite{KerrLMHS}, and (2) the choice of a natural coordinate for the boundary.

\subsection{Analytic perspective: Picard-Fuchs systems and transition matrices}\label{ssec:analytic}
Here we give a short review of the periods of Calabi-Yau threefolds and their computations in examples. We focus mostly on the case of a single modulus, i.e.~$h^{2,1}=1$, since this will be the main setting throughout this work. We also introduce the class of 14 models with hypergeometric periods of \cite{Doran:2005gu}, which serve as a guideline later for understanding the geometric meaning of model-dependent coefficients in the periods near conifold points and K-points.
Periods arise from a geometrical perspective as integrals for differential forms over certain cycles of the underlying manifold. For a Calabi-Yau threefold we focus on its unique holomorphic (3,0)-form $\Omega_3\in H^{3,0}$. Choosing any integral basis $\gamma_I \in H^3(Y_3, \mathbb{Z})$ of 3-cycles (with $I=1, \ldots,2(h^{2,1}+1) $), components of the period vector are given by 
\begin{equation}
   \tilde{\Pi}^I=\int_{\gamma_I} \Omega_3 \;. \label{eq:IntPer}
\end{equation}
The period integrals are rather hard to evaluate directly. It turns out to be much more convenient to use the fact that the integrals fulfill a system of differential equations, the Picard-Fuchs (PF) equations, which are a special case of the Gel’fand-Kapranov-Zelevinsky (GKZ) system. In this paper we will be mainly interested in the special case of 1-parameter hypergeometric systems, i.e. we consider mostly geometries with $h^{2,1}=1$ and whose PF system consists out of a single operator of the form 
\begin{equation}\label{eq:PF}
    L = \theta^4-\mu z(\theta+a_1)(\theta+a_2)(\theta+a_3)(\theta+a_4)\;,
\end{equation}
where $z$ is a local coordinate on the complex structure moduli space and we have denoted $\theta=z \frac{d}{dz}$. The coefficients $0\leq a_i\leq 1$ are rational numbers determined by the geometry under consideration, which we take to be in ascending order $a_1 \leq a_2 \leq a_3 \leq a_4$. Note that this form fixes the complex structure coordinate $z$. We choose it to be centered around the LCS point and fix the scaling by demanding that the conifold singularity is positioned at $z=1/\mu$. We note that the position of the conifold or any other point has no special meaning on the complex structure side. For example, in the hypergeometric cases the moduli space is a 3-punctured sphere, and coordinate changes allow one to freely choose their positions. We note, however, that on the mirror side we associate fixed values to these singularities for the K\"ahler modulus under the mirror map. These latter values have a physical meaning as volumes associated to the Calabi-Yau manifold, and mathematically these numbers show up as well as data specifying the boundary Hodge structures. Nevertheless, there exists a natural choice for the coordinate scale $\mu$ on the complex structure side, which is given by 
\begin{equation}
    \mu^{-1}=e^{4\gamma_E+\psi(a_1)+\psi(a_2)+\psi(a_3)+\psi(a_4)},
\end{equation}
where $\gamma_E$ is the Euler–Mascheroni constant and $\psi(x)$ the digamma function. Remarkably this results in an integer number for $\mu^{-1}$ in the hypergeometric cases. 
We can similarly determine parts of the topological data associated to the mirror Calabi-Yau threefold --- the integrated second Chern class and Euler characteristic --- from the indices $a_i$ as follows
\begin{align}
    c_2\cdot D&=\kappa\left(-2+\frac{3}{\pi^2}\sum_{i=1}^4\psi_1( a_i)\right)=\kappa\left(-2+3\sum_{i=1}^2\csc(\pi a_i)^2\right)\;,\\
    \chi&=\frac{\kappa}{6}\left(8  +\frac{1}{\zeta(3)}\sum_{i=1}^4 \psi_2(a_i)\right)\, ,
\end{align}
where $\psi_i(x)$ denotes the $i$-th polygamma function. These formulas are obtained by explicitly evaluating the formulas found in equation (4.21) of \cite{Hosono:1994ax} for general complete intersection CYs and specializing to the simpler hypergeometric case. This leaves only one topological number, the triple intersection number, undetermined. But the condition that one period vanishes at the conifold in the integer symplectic basis gives one more constraint, which allows the determination of all topological numbers purely from the knowledge of the indices.
We listed the 14 hypergeometric examples based on their values for the $a_i$ in table \ref{table:hypergeom}. As the values of the indices carry a lot of information, it is conventional to gather them in a so-called Riemann symbol, which lists the position of all special points in moduli space together with the associated local exponents for the periods. In the hypergeometric cases the symbol takes the form\footnote{In the general case the singularities are located at infinity and zero, as well as at the zeroes of the coefficient polynomial of the highest order derivative in the Picard-Fuchs operator, i.e.~$1-\mu z=0$ for \eqref{eq:PF}.}
\begin{equation}\label{eq:P}
{\cal  P} \left\{\begin{array}{ccc}
0& 1/\mu& \infty\\ \hline
0& 0 & a_1\\
0& 1 & a_2\\
0& 1 & a_3\\
0& 2 & a_4 
\end{array}\right\}\ 
\end{equation}
In order to derive these local exponents it is useful to introduce a local coordinate around the other boundaries. A hypergeometric PF system has by definition $3$ regular singular points, which with our convention are positioned at $z=\{0,1/\mu,\infty\}$. Thus we introduce the additional coordinates
\begin{equation}\label{eq:coors}
    z_c=z-\frac{1}{\mu}\, ,\qquad\qquad z_\infty=1/z,
\end{equation}
which are centered around the conifold and the point at infinity respectively. The indices in \eqref{eq:P} are then obtained by rewriting the PF equation in terms of the local coordinate, subsequently making a solution ansatz with arbitrary exponents, and solving the resulting leading-order equation for these exponents. For the hypergeometric case \eqref{eq:PF} the singularity at $z=0$ will always be a large complex structure point, and at $z=1/\mu$ a conifold point. The singularity at $z=\infty$ can be of any monodromy type, such as a finite order monodromy, a conifold, a K-point or a second LCS point. These types can be read off from the values of the coefficients $a_i$
\begin{itemize}
    \item LCS point: all are equal $a_1 = a_2=a_3=a_4$.
    \item Conifold point: the middle two are equal $a_2=a_3$, but $a_1 \neq a_2$ and $a_3\neq a_4$.
    \item K-point: the first and last two are equal $a_1=a_2$ and $a_3=a_4$ (but $a_2 \neq a_3$).
    \item F-point: all four are distinct $a_1 \neq a_2 \neq a_3 \neq a_4$ and rational.\footnote{A closely related class are apparent singularities, where the $a_i$ are integers but with a gap, e.g.~$a_i=0,1,3,4$.}
\end{itemize} 
The last singularity type corresponds to a finite order monodromy. The case $a_1=a_2=a_3\neq a_4$ does not arise, as will be elaborated upon below \eqref{eq:singtypes} in the light of asymptotic Hodge theory. Let us also mention that, while we referenced the indices $a_i$ particular to the hypergeometric cases here, this classification based on the local exponents applies singularities in any one-dimensional complex structure moduli space. In the following we will therefore use the indices $a_i$ as the local exponents for the singularity under consideration, and not just the point at $z=\infty$.

\begin{table}[h!]
{{ 
\begin{center}
	\begin{tabular}{|c|c|c|c|c|c|c|c|c|}
		\hline
	      Type&  $N_1$	&$N_\infty$ & $a_1,a_2,a_3,a_4$						        & $\mu$	& Mirror $M$			& $\kappa$	& $c_2 \cdot D$	& $\chi(M)$\\\hline
M&8  	& & $\frac{1}{2},\frac{1}{2},\frac{1}{2},\frac{1}{2}$			& $2^8$		& $X_{2,2,2,2}(1^8)$		& $16$		& $64$		& $-128$\\[2mm]
F&9  	& & $\frac{1}{4},\frac{1}{3},\frac{2}{3},\frac{3}{4}$			& $2^63^3$	& $X_{4,3}(1^5 2^1)$		& $6$		& $48$		& $-156$\\[2mm]
C&16 	&32 & $\frac{1}{4},\frac{1}{2},\frac{1}{2},\frac{3}{4}$			& $2^{10}$	& $X_{4,2}(1^6)$		& $8$		& $56$		& $-176$\\[2mm]
F&25 	&& $\frac{1}{5},\frac{2}{5},\frac{3}{5},\frac{4}{5}$			& $5^5$		& $X_5(1^5)$			& $5$		& $50$		& $-200$\\[2mm]
K&27& 27	& $\frac{1}{3},\frac{1}{3},\frac{2}{3},\frac{2}{3}$			& $3^6$		& $X_{3,3}(1^6)$		& $9$		& $54$		& $-144$\\[2mm]
K&32& 16	& $\frac{1}{4},\frac{1}{4},\frac{3}{4},\frac{3}{4}$			& $2^{12}$	& $X_{4,4}(1^4 2^2)$		& $4$		& $40$		& $-144$\\[2mm]
C&36& 9	& $\frac{1}{3},\frac{1}{2},\frac{1}{2},\frac{2}{3}$			& $2^43^3$	& $X_{3,2,2}(1^7)$		& $12$		& $60$		& $-144$\\[2mm]
C&72& 108	& $\frac{1}{6},\frac{1}{2},\frac{1}{2},\frac{5}{6}$			& $2^83^3$	& $X_{6,2}(1^5 3^1)$		& $4$		& $52$		& $-256$\\[2mm]
F&108&	& $\frac{1}{6},\frac{1}{3},\frac{2}{3},\frac{5}{6}$			& $2^43^6$	& $X_{6}(1^4 2^1)$			& $3$		& $42$		& $-204$\\[2mm]
F&128&	& $\frac{1}{8},\frac{3}{8},\frac{5}{8},\frac{7}{8}$			& $2^{16}$	& $X_{8}(1^4 4^1)$		& $2$		& $44$		& $-296$\\[2mm]
F&144&	& $\frac{1}{6},\frac{1}{4},\frac{3}{4},\frac{5}{6}$			& $2^{10}3^3$	& $X_{6,4}(1^3 2^2 3^1)$		& $2$		& $32$		& $-156$\\[2mm]
F&200&	& $\frac{1}{10},\frac{3}{10},\frac{7}{10},\frac{9}{10}$			& $2^85^5$	& $X_{10}(1^3 2^1 5^1)$		& $1$		& $34$		& $-288$\\[2mm]
K&216& 12	& $\frac{1}{6},\frac{1}{6},\frac{5}{6},\frac{5}{6}$			& $2^83^6$	& $X_{6,6}(1^2 2^2 3^2)$		& $1$		& $22$		& $-120$\\[2mm]
F&864&	& $\frac{1}{12},\frac{5}{12},\frac{7}{12},\frac{11}{12}$		& $2^{12}3^6$	& $X_{12,2}(1^4 4^1 6^1)$	& $1$		& $46$		& $-484$\\ [ 2 mm]
		\hline
	 \end{tabular}	
\end{center}}}
\caption{\label{table:hypergeom}Data of the 14 hypergeometric examples, adapted from \cite{Bonisch:2022mgw}. The type in the first column indicates the singularity type at $z=\infty$ as a maximally unipotent monodromy point (M) (also known as an LCS point), a conifold point (C) or a finite order monodromy point (F). The numbers $N_1$ indicate the level of the weight four cusp forms associated to the conifold at $z=1$ as found in \cite{Bonisch:2022mgw}; the numbers $N_\infty$ are the levels we find for the modular forms associated to the conifold points and K-points at $z=\infty$.}
\end{table}

Let us now elaborate on the construction of this local solution to the PF system, known as the Frobenius method. One takes the ansatz for a period $\omega_i$ to be
\begin{equation}
    \omega_i= \sum_{n=0}^\infty c_{i,n} x^{a_i+n}\;,
\end{equation}
where the $c_{i,n}$ are rational numbers and the $a_i$ are the indices. For repeated indices the solutions become more complicated and one has to introduce logarithms. For $a_i=a_j$ one has as the second solution
\begin{equation}
    \omega_i= \frac{1}{2\pi i}\omega_j\log(x)+\frac{1}{2\pi i}\sum_{n=0}^\infty c_{i,n} x^{a_i+n}\;.
\end{equation}
At LCS points all four $a_i$ are equal, thus there are three solutions containing a logarithm. We use the so-called complex Frobenius basis, which is normalized as
\begin{equation}
\begin{aligned}
    \omega_0&=f_0\, ,\\
    2\pi i\; \omega_1&=f_0\log(x)+f_1\, ,\\
    (2\pi i)^2\; \omega_2&=\frac{1}{2}f_0\log(x)^2+f_1\log(x)+f_2\, ,\\
    (2\pi i)^3\; \omega_3&=\frac{1}{6}f_0\log(x)^3+\frac{1}{2}f_1\log(x)^2+f_2\log(x)+f_3\,,
\end{aligned}
\end{equation}
where the $f_j$ are holomorphic power series. The PF operators do not fix the coefficients $c_{i,n}$ of these power series completely. We normalize the solutions by choosing $c_{i,0}=1$ and setting as many leading $c_{i,n}=0$ as possible. Other choices of the $c_{i,n}$ correspond to a different choice of basis. 
Around a K-point the Frobenius solution takes the form
\begin{equation}\label{Kfrobansatz}
\begin{aligned}
    \omega_0(w)&=w^a f_0(w) \, ,\\
    \omega_1(w)&=w^a f_0(w) \frac{\log(w)}{2 \pi i} + w^{a+1}f_1(w)\, , \\
    \omega_2(w)&=w^{1-a} f_2(w)  \, ,\\
    \omega_3(w)&=w^{1-a} f_2(w) \frac{\log(w)}{2 \pi i} + w^{2-a} f_3(w)\, ,
\end{aligned}
\end{equation}
while for a conifold point it reads
\begin{equation}\label{Cfrobansatz}
\begin{aligned}
    \omega_0(w)&=w^a f_0(w)\, , \\
    \omega_1(w)&=w^{b} f_1(w) \, , \\
    \omega_2(w)&=w^{b} f_1(w) \frac{\log(w)}{2 \pi i} + w^{b+1}f_2(w)\, , \\
    \omega_3(w)&=w^{1-a} f_3(w)\, .
\end{aligned}
\end{equation}
We denote the period vector as obtained from the Frobenius method by $\omega_s=(\omega_0,\omega_1,\omega_2,\omega_3)^T$, where the subscript makes reference to the singularity around which we expand. 

The main difference between the boundaries from this solution basis lies in the logarithmic structure, which encodes the infinite order monodromy around the singularity. In general, the monodromy $T$ induced by encircling a singular point in moduli space will be quasi-unipotent \cite{10.2307/1996622}, meaning that there exist positive integers $n$ and $d$ such that
\begin{align}
    (T^n- \mathbb{I})^{d+1}=0 \,.
\end{align}
As such, the monodromy operator may be decomposed into two commuting factors
\begin{equation}\label{Tdecomp}
T = T_{ss} T_u\, 
:\qquad (T_{ss})^n=\mathbb{I}\, , \quad (T_u)^n = e^{N}\, ,
\end{equation}
where $T_{ss}$ denotes a semisimple part of finite order $n$; this order is bounded through the Euler totient function by the dimension of the vector space, $\phi(n) \leq 2h^{2,1}+2$, which for $h^{2,1}=1$ restricts to $n=2,3,4,5,6,8,10,12$.\footnote{Later, in section \ref{ssec:semisimple}, we will obtain stricter bounds by considering the underlying limiting mixed Hodge structure, limiting us to $n=2,3,4,6$ for conifold and K-points.} The unipotent part $T_u$ may be expressed in terms of the log-monodromy matrix $N$, which is nilpotent of degree $d$, i.e.~$N^{d+1}=0$. For Calabi-Yau threefolds the maximal degree of nilpotency is $d=4$ and occurs for the LCS points. It is important to note that, while $T \in \text{Sp}(4,\mathbb{Z})$, the factors in \eqref{Tdecomp} are in general only rational $T_u, T_{ss} \in \text{Sp}(4,\mathbb{Q})$. For this reason we have taken the $n$-th power in \eqref{Tdecomp} in the definition of the log-monodromy matrix $N$, so that $(T_u)^n$ does correspond to an integral monodromy matrix. We will come back to the log-monodromy operator $N$ in the next subsection \ref{ssec:Hodge1} in more detail when we review the Hodge-theoretic perspective on periods. 

For the moment, let us have a closer look at the finite order part $T_{ss}$. This finite order part will play an important role in our example study in section \ref{sec:examples}, as it turns out to strongly constrain the transition matrices to the integral basis. In fact, this will allow us to determine more easily the numerical coefficients by forcing them to be rational numbers. We discuss these features in more detail in section \ref{sec:geometryinput}. Given a special point in moduli space with Riemann indices $a_1,a_2,a_3,a_4$, the finite order part has a simple form in the Frobenius basis
\begin{align}
   T_{ss}= \begin{pmatrix}
        e^{2 \pi i a_1} & 0 & 0 & 0 \\
        0 & e^{2 \pi i a_2} & 0 & 0 \\
        0 & 0 & e^{2 \pi i a_3} & 0 \\
         0 & 0 & 0 & e^{2 \pi i a_4} 
    \end{pmatrix}\, . \label{eq:FiniteMon}
\end{align}
It is clear from the expression that the finite monodromy is only non-trivial for Riemann indices that are fractional. Furthermore, depending on their exact values, it has more or less degenerate eigenvalues. For computational purposes, it is often more convenient to go to a finite cover of moduli space by a coordinate transformation of the form 
\begin{align}
    z \mapsto z^l \,,
\end{align}
where $l$ refers to the order of $T_{ss}$ as discussed above. After this transformation the resulting monodromy is unipotent, meaning that we are left with only the infinite order part $T_u$ when we circle the boundary. By no means is the information about $T_{ss}$ lost, as the periods still enjoy a $\mathbb{Z}_l$ symmetry under $z \to e^{2\pi i/l}z$. More importantly to our work, such coordinate redefinitions do not affect the structure that the finite order part of the monodromy imposes on the transition matrices.

Another important aspect of the story is the choice of basis, which we want to be integral. For physics this is imposed upon us by charge quantization, while for mathematics it is important in order to properly reveal the geometric information about the manifold carried by its periods.  Unfortunately, the Frobenius solution $\omega$ considered so far will not directly be in an integral symplectic basis. The different basis choices are related by a linear transformation of the basis vectors, which is encoded in a transition matrix acting on the periods in the complex Frobenius basis. It is important to note that the integral frame is not unique as any further $Sp(4,\mathbb{Z})$ transformation preserves it. Among the set of symplectic frames, there is a privileged subset that is, as is made more precise with \eqref{eq:mLCS}, adapted to the LCS points 
\begin{equation}
    \Pi_{LCS} = f_{LCS}(z) \ m_{LCS} \,\omega_{LCS}\,, \label{eq:LCSPeriod}
\end{equation}
where $\Pi_{LCS}$ denotes the period vector in the integral LCS frame, and $m_{LCS}$ the appropriate transition matrix. The rescaling by the holomorphic function $f_{LCS}(z)$ corresponds to a K\"ahler transformation which reflects the physical gauge freedom. We will come back to the exact form of $f_s(z)$ at a later point. The LCS frame is also natural from the point of view of mirror symmetry, where it allows for a direct identification of the entries of the period vector with the $D(2p)$-branes on the mirror IIA side. 

These solutions for the periods in the LCS phase do not cover the whole moduli space, but typically only converge up to the radius set by the closest singularity. Describing all phases thus requires analytically continuing the periods to the phases around these other singularities. In practice this is achieved in two steps. First one computes the local Frobenius solution for the singularity, given for instance by \eqref{Kfrobansatz} and \eqref{Cfrobansatz}. This solution for the periods is in a complex basis and not yet an integral basis. As a second step, one therefore goes to an intermediate point in the overlap of the two regions of convergence: here the integral periods of one phase can be matched numerically to the complex Frobenius periods of another by computing a transition matrix. By these means one can then chain together transitions from the LCS point to subsequent singularities and regular points in moduli space. For more details we refer to the references \cite{Candelas:1990rm, Alim:2012ss, CaboBizet:2016uzv,  Blumenhagen:2016bfp, Blumenhagen:2018nts, Joshi:2019nzi} and the appendix \ref{appendix:transition}.

However, depending on the singularity type and its location in moduli space, this LCS frame is not always the most appropriate. While it certainly is an integral basis of cycles all over moduli space, it might obscure some characteristic features of other singularity types and make them difficult to compare. As will be argued later on, Hodge theory provides guidance on how to choose an appropriate subset of integral frames for each singularity type that reveals their universal structure, making it clear that there are only three different forms of period vectors one has to study in order to cover all infinite order singularities that can occur in one-parameter families. We call this adapted subset of integral frames the \textit{monodromy weight frame}. Analogously to \eqref{eq:LCSPeriod} above, we write 
\begin{align}
    \hat{\Pi}_s= f_s(z) \ \hat{m}_s \,\omega_s \, ,\label{eq:HodgePeriod}
\end{align}
where $s$ now labels the singularity type under consideration, the period vector $\hat{\Pi}_s$ denotes the period vector in the integral monodromy weight frame expanded around this singularity and $\hat{m}_s$ the transition matrix from the local Frobenius basis to this monodromy weight basis. Again we allow by $f_s(z)$ for a possible K\"ahler transformation, which may be different from the choice at LCS. It might appear tautological, but we nevertheless write it out, $\Pi_{LCS}=\hat{\Pi}_{LCS}$, i.e. the LCS frame is the monodromy weight frame for LCS points. 

Having specified the different choices of integral frame, we must emphasize that it is generally a highly non-trivial task to determine the transition matrices analytically. Depending on the case, it contains more or less potentially transcendental numbers that are difficult to identify. Nevertheless, the transition matrices from the examples computed in section \ref{sec:examples}, give a pretty good idea of the universal structure, also in accordance with the recent results of \cite{Bonisch:2022mgw} for conifold points. Currently, the only completely understood case is for the LCS points, where one can immediately write down the transition matrix in terms of the topological numbers of the manifold, i.e.~the Euler characteristic $\chi$, the triple intersection numbers $\kappa$ and the second Chern class $c_2$:
\begin{equation}\label{eq:mLCS}
m_{\rm LCS}= \left(
\begin{array}{cccc}
 1 & 0 & 0 & 0 \\
 0 & 1 & 0 & 0 \\
- \frac{c_2}{24} & -\sigma  & \kappa  & 0 \\
 -\frac{i \chi  \zeta (3)}{8 \pi ^3} & -\frac{c_2}{24} & 0 & -\kappa  \\
\end{array}
\right)  \;,
\end{equation}
where $\sigma=\frac{\kappa}{2} \text{ mod } 1$. This transition matrix corresponds to the ordering of the symplectic periods as \footnote{As the coordinates $s$ and $t$ are identical around the LCS point we use here $t$ to comply with the usual conventions. In our conventions the prepotential coordinate would be denoted $s$.}
\begin{equation}\label{eq:piprepot}
    \Pi_{LCS}=\left(
\begin{array}{cccc}
 1  \\
 t  \\
 \partial_t F \\
  2F-t \partial_t F \\
\end{array}
\right) \, ,
\end{equation}
where $t$ is the Kähler coordinate obtained via the mirror map
\begin{equation}
    t(z)=\frac{\Pi_{LCS,2}}{\Pi_{LCS,1}}(z)=\frac{\log(z)}{2\pi i}+\mathcal{O}(z)\;.
\end{equation}
As various different coordinates are introduced throughout the paper we have summarized them in table \ref{table:conventions}.
\begin{table}[h!]
\centering
\renewcommand*{\arraystretch}{2.0}
\begin{tabular}{| c || c | c | c |}
\hline 
& coordinate & boundary location & monodromy \\ \hline \hline
local normal coordinate & $z$ & $z=0$ & $z\to e^{2\pi i}z$ \\ \hline
natural coordinate & $q$ & $q=0$ & $q \to e^{2\pi i}q$ \\ \hline
natural covering coordinate & $t= x+iy$ & $t=i\infty$ & $t\to t+1$ \\ \hline
prepotential coordinate & $s$ & $\begin{aligned}[t]
\text{LCS}&: s=i\infty \\
\text{K}&: s=i\infty \\
\text{C}&: s=0 \\ \vspace*{-0.02cm}
    \end{aligned}$ & $\begin{aligned}[t]
\text{LCS}&: s\to s+1\\
\text{K}&: s\to s+1 \\
\text{C}&: s\to e^{2\pi i}s \\ \vspace*{-0.02cm}
    \end{aligned}$ \\ \hline
\end{tabular}
\caption{\label{table:conventions} Summary of coordinate conventions. We use $z$ as generic normal coordinate, which may be redefined by divisor-preserving transformations, i.e.~keeping the point $z=0$ fixed. Hodge theory singles out a natural normal coordinate $q$ that fixes this ambiguity. Its accompanying covering coordinate is $t= \log[q]/(2\pi i)$, with $x$ and $y$ its axionic and saxionic part respectively. We use $s$ as coordinate to compute physical couplings from the prepotential. For LCS-points the latter two agree as $t=s$ and correspond to the so-called mirror map; for conifold and K-points the relation between $t$ and $s$ is more complicated, cf.~\eqref{eq:Cz} and \eqref{eq:zs}. We stress that $t$ is also the covering coordinate for conifold and K-points, and is not reserved exclusively for the mirror map at LCS.}
\end{table}
The prepotential $F(t)$ around the LCS point takes the form
\begin{equation}\label{eq:LCSprepotential}
F= \frac{\kappa}{6}t^3-\frac{\sigma}{2}t^2-\frac{c_2}{24} t-\frac{\zeta(3) \chi}{2(2\pi i )^3}+\mathcal{O}(e^{2\pi i t}) \, , 
\end{equation}
where we dropped worldsheet instanton corrections to the prepotential for convenience. The symplectic pairing matrix associated to the above periods is given by
\begin{equation}\label{eq:pairing}
 \Sigma= \left(
\begin{array}{cccc}
 0 & 0 & 0 & 1 \\
 0 & 0 & 1 & 0 \\
 0 & -1  & 0  & 0 \\
 -1 & 0 & 0 & 0  \\
\end{array}
\right)  \;.
\end{equation}
Note that for K-points there is a different natural choice of symplectic pairing, $Q$ cf.~\eqref{eq:pairings}.
\begin{equation}\label{eq:pairing2}
Q= \left(
\begin{array}{cccc}
 0 & 0 & 1 & 0 \\
 0 & 0 & 0 & 1 \\
 -1 & 0  & 0  & 0 \\
 0 & -1 & 0 & 0  \\
\end{array}
\right)  \;.
\end{equation}
Moreover, note that the form of the transition matrices is dependent on the coordinate choice. As the periods may contain logarithms, the constant term gets shifted by a rescaling of the coordinate, i.e.
\begin{equation*}
    x\rightarrow \mu x\,:\qquad c_0+\log(x)\rightarrow c_0+\log(\mu)+\log(x)\;.
\end{equation*}
As we have chosen to normalize $c_0=1$, a shift in the constant term induces a change of the local basis. As the integral symplectic basis is fixed, this leads to a modification of the transition matrix. The transition matrix corresponding to this change around the LCS point is
\begin{equation}
    \left(
\begin{array}{cccc}
 1 & 0 & 0 & 0 \\
 \frac{i \log (\mu )}{2 \pi } & 1 & 0 & 0 \\
 -\frac{\log ^2(\mu )}{8 \pi ^2} & \frac{i \log (\mu )}{2 \pi } & 1 & 0 \\
 -\frac{i \log ^3(\mu )}{48 \pi ^3} & -\frac{\log ^2(\mu )}{8 \pi ^2} & \frac{i \log (\mu )}{2 \pi } & 1 \\
\end{array}
\right)\;,
\end{equation}
and similar matrices for the other boundaries. Throughout this paper we keep the coordinate normalization such that the mirror map starts with $(2\pi i)^{-1}\log(x)$, which corresponds to the conifold being positioned at $1/\mu$ in the hypergeometric cases.

\subsection{Hodge-theoretic perspective (1): nilpotent orbits and mixed Hodge structures}\label{ssec:Hodge1}
We now proceed and describe the Hodge-theoretic structure that underlies these period vectors. In this first subsection we introduce the necessary background such as nilpotent orbit approximations and limiting mixed Hodge structures. The reader familiar with this material can skip to subsection \ref{ssec:Hodge2}, where we describe how asymptotic Hodge theory provides us with a natural choice of integral basis and local coordinate near different moduli space boundaries.

In the physics literature we are used to describe the dependence of physical couplings on the complex structure moduli through the periods of the $(3,0)$-form or, being even more pragmatic, through prepotentials such as \eqref{eq:LCSprepotential} by using special geometry. On the other hand, Hodge theory takes a much more democratic approach by considering the periods of all $(p,q)$-forms. To this end, let us therefore begin by writing down the Hodge decomposition of the middle cohomology as
\begin{align}\label{eq:hodgedecomp}
H^{3}(Y, \mathbb{C})= H^{3}(Y, \mathbb{Z}) \otimes \mathbb{C} =H^{3,0} \oplus H^{2,1} \oplus H^{1,2} \oplus H^{0,3} \,,
\end{align}
where $\overline{H^{p,q}}=H^{q,p}$. For the dimensions of these vector spaces we write $h^{p,q}=\dim_{\mathbb{C}} H^{p,q}$. This decomposition is also known as a pure Hodge structure, where pure refers to the fact that the all the spaces $H^{p,q}$ have weight three, i.e. their superscripts add up to three. This is in contrast to the concept of mixed Hodge structure which will be introduced later. 

The next step is to consider a natural framework for variations in the complex structure moduli. While the $(3,0)$-form periods are holomorphic, for the other $(p,q)$-forms this is not the case. It is therefore convenient to switch to the so-called Hodge filtration of vector spaces $F^p$ (with $0 \leq p \leq 3$), which are related to those in the Hodge decomposition \eqref{eq:hodgedecomp} by
\begin{align}
    F^p = \bigoplus_{k=p}^3 H^{k,3-k} \,, \qquad H^{p,q} = F^p \cap \overline{F^q} \,.
\end{align}
The vectors spanning these spaces $F^p$ do vary holomorphically in the complex structure moduli, as can be illustrated by the identities
\begin{align}
    \partial F^p \subset F^{p-1} \, , \qquad \bar{\partial} F^p \subset F^p \,,
\end{align}
which is most easily checked in practice by considering moduli-dependent basis vectors $\omega_p \in F^p$ and taking their derivatives. We now want to describe how these vector spaces degenerate as we move towards a boundary in complex structure moduli space. This degeneration will give rise to the concept of a limiting mixed Hodge structure (LMHS). For the sake of clarity, we specialize to the case of a single modulus, while the generalization to higher dimensional moduli spaces is straightforward and has been discussed for example in \cite{Grimm:2018cpv} in the physics literature. One typically starts by considering a local normal coordinate $z$ on moduli space in which the singularity is located at the origin $z=0$, cf.~\eqref{eq:coors}. Encircling the boundary then corresponds to the transformation  
\begin{equation}
   z \mapsto e^{2\pi i} z\, , \label{eq:BdCircling}
\end{equation} 
It follows from the nilpotent orbit theorem \cite{Schmid} that the Hodge filtration takes the asymptotic form
\begin{equation}\label{eq:Fpnil}
    F^p_{\rm nil}(t) = e^{\frac{\log(z)}{2 \pi i} N}F^p_0\, ,
\end{equation}
where we dropped corrections terms in $z$ compared to the full fledged Hodge filtration $F^p(z)$ in the bulk of the moduli space. As there was only a single modulus to begin with, the limiting filtration $F^p_0$ associated to the boundary is independent of the complex structure modulus $z$. This allows us to read off the monodromy transformation $F^p_{\rm nil} \to e^N F^p_{\rm nil}$ under \eqref{eq:BdCircling}, where $N$ is the log-monodromy matrix of the unipotent part of the monodromy.\footnote{For the purposes of this discussion we have assumed that any semi-simple part of the monodromy has been removed by an appropriate coordinate redefinition $z \to z^l$ (with $l$ the order of the semi-simple part of the monodromy). }

For our purposes it will be useful to specialize the above nilpotent orbit approximation \eqref{eq:Fpnil} to the period vector of the $(3,0)$-form.   It allows us to expand the period vector into a leading part and subleading corrections as 
\begin{equation}\label{eq:periodexpansion}
    \hat{\Pi}_{s}(t) = e^{\frac{\log(z)}{2 \pi i} N} (\textbf{a}_0 + \sum_{n>0} z^n \textbf{a}_n )\, ,
\end{equation}
where the vectors $\textbf{a}_0$ and $\textbf{a}_n$ are the coefficients in the expansion of the period vector. The nilpotency of the log-monodromy matrix $N$ ensures that there are only finitely many terms in the expansion in $\log(z)$, up to at most $\log(z)^3$, as occurs for singularities of large complex structure type. 

The vector spaces $F^p_0$ of the limiting filtration, combined with the log-monodromy matrix $N$, encode a more refined splitting -- the limiting mixed Hodge structure -- associated to the singularity under consideration. In order to define this splitting we first introduce another set of vector spaces, known as the monodromy weight filtration $W_\ell(N)$ ($\ell=0,\ldots,6$), associated to the log-monodromy matrix $N$. These can be computed from kernels and images of $N$ as
\begin{equation}\label{Wfiltr}
W_{\ell}(N)= \!\!\!\! \sum_{j \geq \max(-1,\ell-3)} \ker N^{j+1} \cap \img N^{j-\ell+3}\, .
\end{equation}
From a physical perspective this monodromy weight filtration allows one to sort charges in $H^3(Y_3,\mathbb{Z})$ based on the asymptotic behavior of their physical couplings, resulting in power-law behaviors labelled by the indices $\ell$ of the $W_\ell$. For example, for large complex structure points the vector spaces $W_{2p}$ corresponds precisely to the decomposition into D($2p$)-branes known from mirror symmetry, with the power-law behavior of the couplings indeed captured by the dimensions of the cycles wrapped by the D-branes. One of the upshots of asymptotic Hodge theory is that it provides such a decomposition for any boundary, which has served as a powerful tool in the study of the swampland program recently \cite{Grimm:2018ohb,Grimm:2018cpv,Corvilain:2018lgw,Font:2019cxq,Grimm:2019wtx,Grimm:2019ixq,Bastian:2020egp,Grimm:2020ouv,Palti:2021ubp,Grana:2022dfw}.

From a more abstract perspective, one can also define the monodromy weight filtration without making a reference to the explicit decomposition \eqref{Wfiltr}. The vector spaces $W_\ell(N)$ are then defined as the unique increasing filtration satisfying
\begin{enumerate}
    \item Action by $N$ lowers the index by two: $N W_\ell \subseteq W_{\ell-2}$.
    \item The map $N^\ell: Gr_{3+\ell} \to Gr_{3-\ell}\,$ is an isomorphism where $Gr_\ell = W_\ell/W_{\ell -1}$. \label{isomorphism}
\end{enumerate}
Condition (1) characterizes how the log-monodromy matrix $N$ lowers charges in the weight filtration; going back to the large complex structure example from before, it means it lowers a $2p$-cycle into a $2(p-1)$-cycle on the mirror dual side. Condition (2) gives us a sense of electro-magnetic duality, allowing to identify charges in the higher-end of the weight filtration with those in the lower-end; from the perspective of large complex structure this means we associate the 6-cycle to the 0-cycle and 4-cycles to 2-cycles. The quotient $Gr_\ell=W_\ell/W_{\ell-1}$ is taken because $W_\ell$ includes all elements up to weight $\ell$; for LCS it means we consider e.g.~a 4-cycle modulo 2- and 0-cycles, since these do not affect the asymptotic behavior of the couplings. 

The triple $(H^{3}(Y,\mathbb{Z}),F_0^p,W_{\ell}(N))$ defines what is called a limiting mixed Hodge structure, which can be associated to a given boundary where the underlying Calabi-Yau manifold degenerates. All of this structure can be conveniently encoded in the so called Deligne splitting
\begin{equation}\label{Ipq}
I^{p,q} = F_{0}^{p} \cap W_{p+q} \cap \bigg( \bar{F}_{0}^{q}\cap W_{p+q}+\sum_{j\geq 1} \bar{F}_{0}^{q-j}\cap W_{p+q-j-1} \bigg)\, .
\end{equation}
This cumbersome expression reduces in the special case where the limiting mixed Hodge structure is $\mathbb{R}$-split, i.e.~$\bar{I}^{p,q} = I^{q,p}$, to the much simpler form $I_{\mathbb{R}}^{p,q} = F_{0}^{p} \cap \bar{F}^q_0 \cap W_{p+q} $. We denote the dimensions of the vector spaces in the Deligne splitting by $i^{p,q} = \dim_{\mathbb{C}} I^{p,q}$, which satisfy some symmetries 
\begin{equation}\label{eq:ipqsyms}
    i^{p,q}=i^{q,p}\, , \qquad i^{p,q}=i^{q-3,p-3} \quad (\text{for }p+q>3)\, , \qquad h^{p,q} = \sum_{r} i^{p,r}\, ,
\end{equation} 
where we included the relation with the Hodge numbers in the usual pure Hodge structure. We should also point out that these vector spaces satisfy certain positivity conditions with respect to the symplectic pairing $\langle \, \cdot \, , \, \cdot \, \rangle$, given by
\begin{equation}\label{eq:pol}
    \omega \in I^{p,q} \cap \ker N^{p+q-2}: \quad i^{p-q} \langle \omega , N^{p+q-3} \bar{\omega} \rangle  > 0\, .
\end{equation}
The restriction to $\ker N^{p+q-2}$ ensures us that we take only primitive elements in $I^{p,q}$, i.e.~they do not descend by action of $N$ from a higher-positioned $I^{r,s}$ ($r>p, s>q$). 

While the whole concept of a limiting mixed Hodge structure can seem rather abstract, there exists a convenient diagrammatic representation known as Hodge-Deligne diagrams. In the one-modulus case, there exist only three different types of boundaries having the following diamonds
\begin{equation}\label{eq:singtypes}
 \mathrm{I}_1:\begin{tikzpicture}[baseline={([yshift=-.5ex]current bounding box.center)},scale=0.65,cm={cos(45),sin(45),-sin(45),cos(45),(15,0)}]
  \draw[step = 1, gray, ultra thin] (0, 0) grid (3, 3);

  \draw[fill] (0, 3) circle[radius=0.05];
  \draw[fill] (1, 1) circle[radius=0.05];
  \draw[fill] (2, 2) circle[radius=0.05];
  \draw[fill] (3, 0) circle[radius=0.05];
\end{tikzpicture} \, , \quad
\mathrm{II}_0:\begin{tikzpicture}[baseline={([yshift=-.5ex]current bounding box.center)},scale=0.65,cm={cos(45),sin(45),-sin(45),cos(45),(15,0)}]
  \draw[step = 1, gray, ultra thin] (0, 0) grid (3, 3);

  \draw[fill] (0, 2) circle[radius=0.05];
  \draw[fill] (1, 3) circle[radius=0.05];
  \draw[fill] (2, 0) circle[radius=0.05];
  \draw[fill] (3, 1) circle[radius=0.05];
\end{tikzpicture} \, , \quad
\mathrm{IV}_1:\begin{tikzpicture}[baseline={([yshift=-.5ex]current bounding box.center)},scale=0.65,cm={cos(45),sin(45),-sin(45),cos(45),(15,0)}]
  \draw[step = 1, gray, ultra thin] (0, 0) grid (3, 3);

  \draw[fill] (0, 0) circle[radius=0.05];
  \draw[fill] (1, 1) circle[radius=0.05] node[above]{};
  \draw[fill] (2, 2) circle[radius=0.05] node[above]{};
  \draw[fill] (3, 3) circle[radius=0.05];
\end{tikzpicture}\, ,
\end{equation}
where each dot at position $(p,q)$ on the roster denotes a  vector space $I^{p,q}$ of dimension one. We also included the labels $\mathrm{I}_1$, $\mathrm{II}_0$ and $\mathrm{IV}_1$ for the singularity types, which were introduced in the mathematics literature in \cite{Kerr2017}. From this depiction we can understand the role of the quotient vector spaces $Gr_\ell=W_\ell/W_{\ell-1}$ better. To begin with, we can represent a given $Gr_\ell$ by the union of spaces $I^{p,q}$  with $p+q=\ell$, i.e.~those positioned on the $\ell$-th horizontal line in the diamond. Here we count these rows such that the middle line corresponds to $Gr_3$. It is worth pointing out that on the individual spaces $Gr_\ell$, one again has pure Hodge structures, similar to \eqref{eq:hodgedecomp}. The difference is that the weight of the pure Hodge structure --- the row $\ell$ for a given $Gr_\ell$ --- can be lower or larger than three now, as can be seen from the diagrams.

This classification by mixed Hodge structures can be straightforwardly related to the exponents $a_i$ we encountered previously in \eqref{eq:PF} in the study of the Picard-Fuchs system: for each set of $k$ equal indices $a_{i_1}=\ldots = a_{i_k}$, there is a column of length $k$ in the Hodge-Deligne diamond. More explicitly, one observes for the above types that:
\begin{itemize}
    \item conifold point ($\mathrm{I}_1$): $a_1<a_2=a_3<a_4$, and therefore a column made up of two dots at $(2,2)$ and $(1,1)$, accompanied by two single dots at $(3,0)$ and $(0,3)$ for $a_1$ and $a_4$.
    \item K-point ($\mathrm{II}_0$): $a_1=a_2<a_3=a_4$, resulting in two columns with one made up of $(3,1)$ and $(2,0)$ and the other by $(1,3)$ and $(0,2)$.
    \item LCS point ($\mathrm{IV}_1$): four equal indices $a_1=\ldots = a_4$, and therefore a single column made up of $(3,3)$ up to $(0,0)$. 
\end{itemize}
From this point of view we can also heuristically explain the absence of singularities with three equal indices $a_1=a_2=a_3 \neq a_4$: this would result in a Hodge-Deligne diamond with a column consisting of three points, i.e.~a singularity of some type $\mathrm{III}_c$. However, in order to make the diamond symmetric according to \eqref{eq:ipqsyms} it would therefore require six points in total, while the vector space is only of dimension $2(h^{2,1}+1)=4$ for $h^{2,1}=1$.

\newpage
\subsection{Hodge-theoretic perspective (2): integral basis and coordinates}\label{ssec:Hodge2}
Having reviewed the rudiments of asymptotic Hodge theory, we next want to discuss how insights from Hodge theory can be used to fix the freedoms that one encounters when computing periods: the choice of integral basis, the K\"ahler frame and the local coordinate on complex structure moduli space. The concepts laid out in this part will be applied to the possible one-modulus boundaries in section \ref{sec:construction}. It can thus be helpful to read this section with some examples at hand. First, we will discuss how the weight filtration singles out a set of privileged integral frames: this provides a universal framework for each singularity type which brings all geometric examples on the same footing, thereby allowing us to classify and compare their boundary data more easily. Secondly, we will lay out our conventions for fixing the K\"ahler gauge, i.e.~the freedom of holomorphic rescalings of the periods. This allows us to reformulate the period expansion \eqref{eq:periodexpansion} in terms of the so-called instanton map. In this reformulation we subsequently explain how one can read off a natural choice of local coordinate for any given boundary. In addition, we comment on the formulation of the periods in terms of a prepotential. 
\subsubsection{Integral basis and boundary data}
We begin with the quantization of the basis for our periods. From a physical perspective we need this information in order to appropriately quantize the charges or fluxes in our effective theories. While the Picard-Fuchs equations presented in section \ref{ssec:analytic} are easily solved to an arbitrary order, the resulting Frobenius periods $\omega_s$ are not automatically in an integral basis. In the present literature, there is only a complete understanding of how to rotate the Frobenius solution into an integral symplectic basis for LCS boundaries. For conventional conifold points, e.g.~those located at $z=1/\mu$ in the hypergeometric examples, significant progress has been made recently in \cite{Bonisch:2022mgw}. However, a general understanding for one-modulus boundaries still appears to be missing. To this end, let us therefore try to understand what the limiting mixed Hodge structure looks like in an appropriate integral basis. This serves as a first step towards a uniform way of writing down the transition matrix from the local Frobenius basis to the integral monodromy weight basis.
\paragraph{Preferred integral frame.} We will construct our integral basis by making use of the monodromy weight filtration $W_\ell(N)$ of the LMHS. This approach benefits from the quantization of the monodromy matrices in finding a set of integral basis vectors. Moreover, it allows us to fix (part of) the freedom in basis choice under Sp$(4,\mathbb{Z})$ transformations. In the following, we refer to the set of integral frames that respect the weight filtration as the \textit{monodromy weight frame}. Using the latter makes the singularities inside a given class easily comparable, and will highlight the relevant pieces of geometrical data one needs to understand. It is shown in \cite{KerrLMHS} that there exists an integral basis for the periods which is adapted to the monodromy weight filtration $W_j(N)$. To be explicit, one can simply take the natural choice of basis vectors 
\begin{align}
    e_0= \begin{pmatrix}
        0 \\ 0 \\ 0 \\ 1
    \end{pmatrix} \,, \quad 
    e_1= \begin{pmatrix}
        0 \\ 0 \\ 1 \\ 0
    \end{pmatrix} \,, \quad 
    e_2= \begin{pmatrix}
        0 \\ 1 \\ 0 \\ 0
    \end{pmatrix} \,, \quad 
    e_3= \begin{pmatrix}
        1 \\ 0 \\ 0 \\ 0
    \end{pmatrix} \,, \label{eq:BasisWF}
\end{align}
which span the relevant weight filtrations for the one-modulus cases at hand as
\begin{itemize}
    \item Conifold point: $e_0 \in W_2$, $e_1, e_2 \in W_3$ and $e_3 \in W_4$,
    \item K-point: $e_0, e_1 \in W_2$ and $e_2, e_3 \in W_4$,
    \item LCS point: $e_0 \in W_0$, $e_1 \in W_2$, $e_2 \in W_4$ and $e_3 \in W_6$.
\end{itemize}
In terms of this basis for the weight filtration, the appropriate bilinear forms are chosen to be
\begin{align}\label{eq:pairings}
   \text{Conifold/LCS:} \ \ \  \Sigma=\begin{pmatrix}
        0 & 0 & 0 & 1 \\
        0 & 0 & 1 & 0 \\
        0 & -1 & 0 & 0 \\
        -1 & 0 & 0 & 0
    \end{pmatrix} \, , \qquad    \text{K:}\ \ \ Q=\begin{pmatrix}
        0 & 0 & 1 & 0 \\
        0 & 0 & 0 & 1 \\
        -1 & 0 & 0 & 0 \\
        0 & -1 & 0 & 0
    \end{pmatrix} \, .
\end{align}
One can also understand our choice of integral basis naturally from the logarithmic structure in the periods. Take for example the conifold point: in a basis that respects the weight filtration the period vector should only contain a single logarithm, located in the entry along $e_3$. While there exist integral symplectic transformations that would mix the logarithm into all periods, this would be a rather unnatural frame. 
Equivalently, the monodromy weight frame brings the transition matrices $\hat{m}_s$ for a given singularity type into a similar form: this can be seen from how it will map logarithmic solutions in the Frobenius basis into the same entries in the monodromy weight basis. For example, for conifold singularities one finds a log-monodromy matrix $N$ with just a single non-zero integer entry in the monodromy weight basis. However, in another integral basis this simplicity could have been obscured by many non-vanishing entries, even though the matrix would still have rank one. 

After fixing a basis adapted to the weight filtration, there is still a non-trivial subgroup of basis transformations $G \subseteq \text{Sp}(4,\mathbb{Z})$ that is compatible with this choice
\begin{equation}\label{Gdef}
    G \, W_{\ell}\subseteq W_{\ell}\, ,
\end{equation}
In the natural basis \eqref{eq:BasisWF} the group $G$ corresponds to all lower block diagonal matrices that have integral entries and respect the symplectic structure. Furthermore, such a basis choice together with the general properties given below \eqref{Wfiltr} poses two constraints for the log-monodromy matrix $N$. The integral structure entails that the entries of $e^N$ have to be integers, while the compatibility with the weight filtration, $N W_{\ell} \subseteq W_{\ell-2} $, forces $N$ to be strictly lower triangular. In this form the entries of the monodromy matrix display more directly the geometric information about the degenerating variety, as we will verify explicitly in examples in section \ref{sec:examples}. The most well known example of this would be for the LCS point where the topological numbers appear, see \eqref{eq:LCSNtop} for a reminder.  

\paragraph{Distinction of boundary data.} 
Next, we want to characterize how  the boundary data appears as coefficients in the limiting mixed Hodge structure. We can make this description more precise by using the monodromy weight basis we just introduced. For this, one writes the elements of the complex vector spaces $I^{p,q}$ (or equivalently $F^p_0$) in terms of the natural basis for the weight filtration. Among the entries appearing in the vectors $ \omega_{p,q} \in I^{p,q}$, we distinguish between between two different kinds of boundary data, depending on whether we project to the same row $Gr_{p+q}$ or a lower row $Gr_{\ell<p+q}$:
\begin{enumerate}
    \item The first set of boundary data comes from the graded pieces $Gr_{\ell}$, corresponding to certain \textit{rigid periods} associated to the degenerated manifold. These rows host pure Hodge structures induced by the LMHS, cf.~the discussion below the diagrams \eqref{eq:singtypes}, to which we can associate this period data. Given a vector $\omega_{p,q} \in I^{p,q}$ in the Deligne splitting, the entries that lie in $I^{p,q} \cap Gr_{p+q}$ can be interpreted as periods of these pure Hodge structures. For the one-modulus cases the pure Hodge structures are either one- or two-dimensional, i.e.~there are one or two dots on any given row in the Hodge-Deligne diamonds. Furthermore, these pure Hodge structures are rigid, since our single modulus has been taken to a limit. As a result this period data is constant along a given boundary, encoding particular information about the degenerated manifold in this limit, in accordance with recent results of \cite{Bonisch:2022mgw}.  
    
    \item The second set of boundary data comes from the lower-lying graded pieces $Gr_{\ell<p+q}$, typically referred to as \textit{extension data}. In a generic situation the entries of $\omega_{p,q} \in I^{p,q}$ that lie in $I^{p,q} \cap Gr_{\ell < p+q}$ are non-zero. If they all vanish,\footnote{Possibly up to a basis transformation by an element in the group $G \subseteq \text{Sp}(4,\mathbb{Z})$ defined by \eqref{Gdef}.} the LMHS is said to split over $\mathbb{Z}$ and given by simply a direct sum of pure Hodge structures on the weight graded pieces $Gr_{\ell}$. The coefficients that encode the non-triviality of how these pure Hodge structures are stacked together to make up the LMHS are referred to as \textit{extension data}. For a more mathematically oriented discussion we refer to \cite{KerrLMHS}. We will simply follow the notation from the math literature, so given two pure Hodge structures of weight $w$ and $w+k$ respectively, we denote the extension as $\text{Ext}^1(H^{w+k},H^w)$. By the isomorphism property \ref{isomorphism} of the map $N$, some of these extension are dual to another and should carry the same information as a result.
\end{enumerate}
For the sake of concreteness, we illustrate the concepts that we just outlined for the three possible one-modulus cases. It is helpful to consider the relevant Hodge-Deligne diagrams \eqref{eq:singtypes} when reading what follows. Given the basis $e_i$ chosen above the entries of the vectors $\omega_{p,q} \in I^{p,q}$ are distinguished in the following way  
\begin{align}
 \text{Conifold point:} \quad \quad   \omega_{3,0} = \begin{pmatrix}
       0 \\ \hline * \\ * \\ \hline *
    \end{pmatrix} \begin{matrix}
     \\      \\ \text{Periods on } \, Gr_3 \\ \\ \text{Ext}^1 (Gr_3,Gr_2) 
    \end{matrix}\,, \quad \omega_{2,2}= \begin{pmatrix}
        * \\ \hline * \\ * \\ \hline *
    \end{pmatrix}
    \begin{matrix}
    \text{Period on } \, Gr_4 \\      \\ \text{Ext}^1 (Gr_4,Gr_3) \\ \\ \text{Ext}^1 (Gr_4,Gr_2) 
    \end{matrix}
\end{align}
where $Gr_3 \cong H^{3,0} \oplus H^{0,3}$ resembles the middle cohomology of a rigid CY3 and $Gr_4$ and $Gr_2$ are simply one-dimensional pure Hodge structures of weight four and two respectively. As these two spaces are isomorphic under the action of $N$, the extensions $\text{Ext}^1(Gr_4,Gr_3)$ and $\text{Ext}^1(Gr_3,Gr_2)$ are dual to another.
\begin{align}
\text{K-point:} \quad \quad   \omega_{3,1} = \begin{pmatrix}
       * \\  * \\\hline * \\  *
    \end{pmatrix} \begin{matrix}
          \text{Periods on } \, Gr_4 \\  \\ \text{Ext}^1 (Gr_4 ,Gr_2) 
    \end{matrix}\,, \quad \omega_{2,0}= \begin{pmatrix}
        0 \\  0 \\ \hline * \\  *
    \end{pmatrix}
    \begin{matrix}
           \\  \\ \text{Periods on } \, Gr_2 
    \end{matrix}
\end{align}
where $Gr_4 \cong H^{3,1} \oplus H^{1,3}$ and $Gr_2 \cong H^{2,0} \oplus H^{0,2}$ resemble the middle cohomolgy of a rigid K3\footnote{It could seem strange at first that the weights of $Gr_4\cong H^{3,1} \oplus H^{1,3}$ are shifted with respect to the usual expectation for a rigid K3. Without going into details, one should simply think of it as bookkeeping on where this pure Hodge structure sits in the Hodge-Deligne diamond.  }, that are isomorphic to each other through the property \ref{isomorphism} of the log-monodromy operator.
\begin{align}\label{eq:IV1extension}
 \text{LCS point:} \quad \quad   \omega_{3,3} = \begin{pmatrix}
       * \\ \hline * \\ \hline * \\ \hline *
    \end{pmatrix} \begin{matrix}
          \text{Period on }Gr_6  \\
          \text{Ext}^1 (Gr_6 ,Gr_4) 
          \\  \text{Ext}^1 (Gr_6 ,Gr_2)  \\  \text{Ext}^1 (Gr_6 ,Gr_0)  
    \end{matrix}\,, \quad \omega_{2,2}= \begin{pmatrix}
        0 \\ \hline * \\ \hline * \\ \hline *
    \end{pmatrix}
    \begin{matrix}
           \\
          \text{Period on }Gr_4 
          \\  \text{Ext}^1 (Gr_4 ,Gr_2)  \\  \text{Ext}^1 (Gr_4 ,Gr_0)  
    \end{matrix}
\end{align}
where all the $Gr_i$ are only one-dimensional. This makes the LCS point rather special as there is no non-trivial period structure appearing, i.e. we can always set a one-dimensional piece to unity. As is well known, all the geometric information for the LCS points sits in the monodromy matrix and in the extension data. We will briefly review this in section \ref{sec:construction}. 

Throughout the discussion, we kept the non-zero entries arbitrary for the moment as we are merely interested in the overall structure provided by the weight filtration in the integral basis. The individual entries of the vectors will be made more specific in section \ref{sec:construction}, when we review the general construction of the LMHS in an integral basis. The zeroes can simply be inferred by the compatibility with the action of the log-monodromy matrix, e.g.~$\omega_{2,2}$ in \eqref{eq:IV1extension} has no piece along $Gr_6$.

The above distinction of the boundary data entails a natural splitting of the group $G \subseteq \text{Sp}(4,\mathbb{Z})$ of basis transformations between different monodromy weight frames, recall \eqref{Gdef} for the precise definition of $G$. To be explicit, this splitting takes the form
\begin{equation}\label{Gsplit}
G = G_p \times G_{ext}\, .
\end{equation}
On one hand, we have a part $G_p$ rotating the rigid periods --- typically by some subgroup SL$(2,\mathbb{Z}) \subseteq \text{Sp}(4,\mathbb{Z})$ ---which acts on the pure Hodge structures on the graded pieces. On the other, we have a factor $G_{ext}$ acting on the extension data by shifts of these parameters. This splitting proves to be convenient in section \ref{sec:examples} in our example study, since it helps us in identifying the appropriate integral monodromy weight frame that simplifies the rigid periods and extension data. We refer to section \ref{sec:construction} for the explicit form of the generators of these subgroups $G_p$ and $G_{ext}$ for the three different one-modulus singularities.

\subsubsection{Fixing the K\"ahler gauge}
The period vector of the $(3,0)$-form is unique up to holomorphic rescalings, which physically correspond to K\"ahler transformations. Fixing this K\"ahler gauge is commonly done by rescaling the Frobenius periods $\omega_s$ by a particular holomorphic function, denoted by $f_s(z)$ in \eqref{eq:LCSPeriod} and \eqref{eq:HodgePeriod}. A natural candidate is the first Frobenius period $f^{-1}_s(z)=(\omega_s)_0$, since this period is holomorphic at all boundaries, i.e.~it does not contain any logarithms. For LCS points there is only one such holomorphic solution, but at other boundaries multiple Frobenius periods will be free of logarithms. For the LCS-point there is thus the only one natural choice to consider, given by
\begin{align}
    \text{LCS point: } \, f_{LCS}(z)=(\omega_0)^{-1}\, ,
\end{align}
which sets the first entry in the period vector to one. For the other two one-modulus boundaries, however, we can in principle choose any linear combination of the holomorphic periods at these singularities. Below we demand convenient orthogonality conditions on the derivative of the period vector to fix this ambiguity.

\paragraph{Conifold point.} For conifold points the holomorphic periods we can use for the rescaling are the three Frobenius periods $\omega_0$, $\omega_1$ and $\omega_3$ as
\begin{align}
 f_C(z) =(\omega_0+r_{1} \, \omega_1+r_{2}\, \omega_3)^{-1} \, . \label{eq:GFC}
\end{align}
where $r_{1}$ and $r_{2}$ are coefficients particular to the example under consideration. We fix $r_{1},r_{2}$ by demanding an orthogonality condition on $\partial_z \Pi$. From the perspective of the limiting mixed Hodge structure this choice will ensure that this derivative lies solely along $I^{2,2}$ and has no piece along $I^{3,0}$. Imposing this orthogonality amounts to the condition
\begin{equation}\label{eq:KTC}
    \langle \partial_z \Pi \,, \ \omega_{0,3} \rangle = 0\, ,
\end{equation}
where by taking the pairing with $\omega_{0,3}$ we project onto the piece along $I^{3,0}$, as can be seen from the polarization conditions \eqref{eq:pol}. We can solve this orthogonality condition for the rescaling function $f_C(z)$ as
\begin{equation}
    \partial_z \log[f_C] = -\partial_z \log\big[\langle \frac{1}{f_C}\Pi \, , \ \omega_{0,3} \rangle\big]\, ,
\end{equation}
where $\Pi/f_C$ denotes the unrescaled period vector. This is solved simply by\footnote{We stress that this is a direct expression for $f_C$. By lack of a better notation $\Pi/f_C$ is actually independent of $f_C$, since $\Pi$ denotes the period vector already rescaled by $f_C$.}
\begin{equation}\label{eq:fC}
f_C(z) = \frac{1}{ \langle \tfrac{1}{f_C} \Pi\, , \ \omega_{0,3} \rangle}\, ,
\end{equation}
where the pairing of the unrescaled periods $\Pi/f_C$ with $\omega_{0,3}$ thus simply picks out the coefficients in the relation \eqref{eq:KTC}. In section \ref{sec:examples} we find that $r_{1}=r_{2}=0$ for the three conifold points at infinity in the hypergeometric examples, while $r_{2} \neq 0$ for the other two examples we considered.

\paragraph{K-point.} For K-points we consider the two holomorphic Frobenius periods $\omega_0$ and $\omega_2$, giving us as rescaling
\begin{align}
f_K(z) =(\omega_0+r\,  \omega_2)^{-1} \, .\label{eq:GFK}
\end{align}
where the coefficient $r$ is fixed by the underlying example. Similar to the conifold point, we fix our rescaling by demanding an orthogonality condition on $\partial_z \Pi$. From the perspective of the limiting mixed Hodge structure this choice will ensure that the derivative lies solely along $I^{2,0}$ and has no piece along $I^{3,1}$. This is imposed by the orthogonality condition
\begin{equation}
    \langle \partial_z \Pi \, , \ \omega_{0,2}\rangle = 0 \, ,
\end{equation}
where by taking the pairing with $\omega_{0,2}$ we project onto the piece along $I^{3,1}$, as can be seen from the polarization condition \eqref{eq:pol}. Analogous to the conifold point in \eqref{eq:fC}, we can solve this orthogonality condition by the following choice of rescaling
\begin{equation}\label{eq:fK}
    f_K(z) = \frac{1}{\langle\frac{1}{f_K}\Pi\, , \ \omega_{0,2}\rangle }\, ,
\end{equation}
where $\Pi/f_K$ denotes the unrescaled period vector. Thus the rescaling is simply given by computing the pairing of the unrescaled Frobenius periods with the vector $\omega_{0,2}$ of the LMHS.

\subsubsection{Instanton map and coordinate choices}
Finally, we review the instanton map formulation of the asymptotic expansion of the period vector. The main point of our discussion here is to illustrate how this description gives a natural choice of canonical coordinate for every boundary type; for a more in-depth review of the instanton map itself we refer to appendix \ref{app:instantonmap}. For LCS points we find that this coordinate choice reproduces the familiar mirror map, as we will review in section \ref{ssec:LCS}. Our prescription extends this idea to the other two types of boundaries --- conifold and K-points --- and allows us to make a canonical coordinate choice there as well.

\paragraph{Instanton map formulation.} Let us begin with a brief recap of this reformulation of the asymptotic periods. Previously we encountered the expansion \eqref{eq:periodexpansion} of the periods, divided into a leading polynomial part and an infinite series of exponential corrections. This infinite series can be conveniently reorganized in terms of a single moduli-dependent matrix $\Gamma(z)$, dubbed the \textit{instanton map}, as
\begin{equation}\label{eq:instantonmap}
\hat{\Pi}_s = e^{\frac{\log(z)}{2 \pi i} N} e^{\Gamma(z)}\textbf{a}_0\, .
\end{equation}
This operator $\Gamma$ has been studied for abstract variations of Hodge structure in  \cite{CattaniFernandez2000,CattaniFernandez2008} in the math literature. In the physics literature it has been used to model one- and two-moduli periods in \cite{Bastian:2021eom} in order to describe couplings in 4d supergravities arising from string compactifications. The naming `instanton map' introduced in the latter reference stems from the physical origin of this data at large complex structure boundaries, where the correction terms coming from $\Gamma$ have a precise interpretation as instantons on the mirror side.

The upshot of this approach is that the instanton map $\Gamma(z)$ is not simply any matrix, but satisfies certain constraints imposed by the underlying boundary structure. These conditions can be summarized as:
\begin{itemize}
    \item It has to satisfy $\Gamma^T \, \Sigma + \Sigma \, \Gamma = 0$ with respect to the symplectic pairing \eqref{eq:pairing}, similar to the log-monodromy matrices. In other words, it has to be valued in its Lie algebra, which in our case is simply $\Gamma(z) \in \mathfrak{sp}(4,\mathbb{C})$.
    \item It has to act on the Deligne splitting \eqref{Ipq} by lowering the first index by at least one degree. To be more explicit, let us write out this action as
    \begin{equation}\label{eq:gamma-action}
\Gamma(z)\,  \omega_{p,q} \in \bigoplus_{r<p} \bigoplus_{s} I^{r,s}\, ,
\end{equation}
for all $\omega_{p,q}\in I^{p,q}$.
\end{itemize}
It is convenient to decompose the instanton map according to the action above as 
\begin{align}\label{eq:Gammadecomp}
    \Gamma=\Gamma_{-1}+\Gamma_{-2}+\Gamma_{-3}\,,
\end{align}
where the subscript refers to the lowering of the first index as  $\Gamma_{-p}I^{r,s} \subseteq \oplus_q I^{r-p,q}$. This sum truncates at minus three since we are dealing with Calabi-Yau threefolds. The first condition above imposes that $\Gamma$ is not just any matrix, but must be a symplectic matrix. The second condition enforces a particular action on the boundary Deligne splitting. In fact, this condition may be rewritten as a set of differential equations among the $\Gamma_{-p}$, cf.~\eqref{eq:recursiongamma}. These can be understood as recursion relations among the components, fixing $\Gamma_{-2},\Gamma_{-3}$ uniquely in terms of $\Gamma_{-1}$, as was shown in \cite{CattaniFernandez2000}.

In order to make the instanton map reformulation more intuitive, it proves to be useful to represent the components by  arrows in the Deligne diamond. As it suffices to focus only on $\Gamma_{-1}$, we have the following diagrams
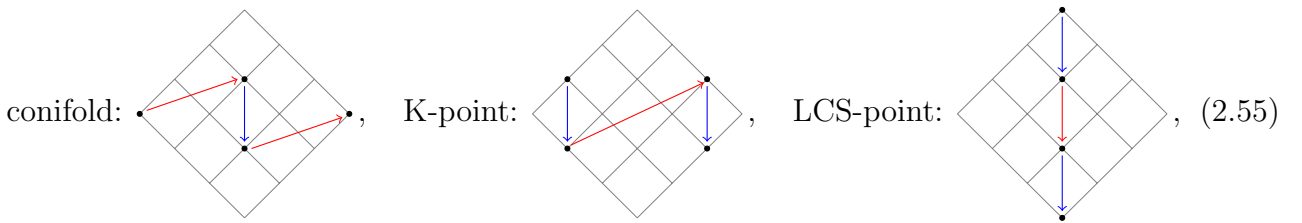
\begin{equation}\label{eq:instantonarrows}
\text{conifold: } \begin{tikzpicture}[baseline={([yshift=-.5ex]current bounding box.center)},scale=0.65,cm={cos(45),sin(45),-sin(45),cos(45),(15,0)}]
  \draw[step = 1, gray, ultra thin] (0, 0) grid (3, 3);

  \draw[fill] (0, 3) circle[radius=0.05];
  \draw[fill] (1, 1) circle[radius=0.05];
  \draw[fill] (2, 2) circle[radius=0.05];
  \draw[fill] (3, 0) circle[radius=0.05];

\draw[->, red] (0.15,2.95) -- (1.9,2.1);
\draw[->, red] (1.1,0.9) -- (2.85,0.05);

\draw[->, blue] (1.9,1.9) -- (1.1,1.1);
\end{tikzpicture}\, , \quad \text{K-point: }\begin{tikzpicture}[baseline={([yshift=-.5ex]current bounding box.center)},scale=0.65,cm={cos(45),sin(45),-sin(45),cos(45),(15,0)}]
  \draw[step = 1, gray, ultra thin] (0, 0) grid (3, 3);

  \draw[fill] (0, 2) circle[radius=0.05];
  \draw[fill] (1, 3) circle[radius=0.05];
  \draw[fill] (2, 0) circle[radius=0.05];
  \draw[fill] (3, 1) circle[radius=0.05];

\draw[->, red] (0.1,2) -- (2.9,1);

\draw[->, blue] (0.9,2.9) -- (0.1,2.1);
\draw[->, blue] (2.9,0.9) -- (2.1,0.1);
\end{tikzpicture}\, , \quad \text{LCS-point: } \begin{tikzpicture}[baseline={([yshift=-.5ex]current bounding box.center)},scale=0.65,cm={cos(45),sin(45),-sin(45),cos(45),(15,0)}]
  \draw[step = 1, gray, ultra thin] (0, 0) grid (3, 3);

  \draw[fill] (0, 0) circle[radius=0.05];
  \draw[fill] (1, 1) circle[radius=0.05];
  \draw[fill] (2, 2) circle[radius=0.05];
  \draw[fill] (3, 3) circle[radius=0.05];

\draw[->, blue] (0.9,0.9) -- (0.1,0.1);
\draw[->, blue] (2.9,2.9) -- (2.1,2.1);

\draw[->, red] (1.9,1.9) -- (1.1,1.1);
\end{tikzpicture}\, ,
\end{equation}
where same-colored components have to be linearly combined to ensure that $\Gamma$ is $\mathfrak{sp}(4)$-valued. We thus find that there are only two independent functions in the instanton map at all one-modulus boundaries. The blue arrow corresponds to log-monodromy operator $N$ and its coefficient function will be denoted by $n(z)$, while for the other map in red it will be called $A(z)$. For the large complex structure point $n(z)$ corresponds to the holomorphic part of the mirror map and $A(z)$ is related to the instanton corrections to the Yukawa coupling, as we will see in more detail in section \ref{ssec:LCS}. 

\paragraph{Canonical coordinate.} Finally, let us elaborate on how the instanton map furnishes a canonical coordinate for every boundary. A priori we are free to perform any coordinate redefinitions of the form
\begin{align}
    z = \lambda \, q \, \exp [2 \pi i \, n(q)] \,,  \label{eq:Canco}
\end{align}
where $\lambda \in \mathbb{C}$ and the holomorphic series satisfies $n(0)=0$. Coordinate changes as \eqref{eq:Canco} ensure that the boundary structure is preserved, i.e.~in the new coordinate the boundary is still given by $q=0$. The redefinition \eqref{eq:Canco} produces the following shifts to the instanton map
\begin{equation}
\Gamma_{-1}(z) \to \Gamma_{-1}(q) + n(q) N\, .
\end{equation}
We can thus choose the coordinate transformation such that it kills off any piece along $N$ in the instanton map $\Gamma$, i.e.~the red arrows in \eqref{eq:instantonarrows}, hence the naming of the function in the reparametrization \eqref{eq:Canco}. Note that while this procedure uniquely fixes the functional part of \eqref{eq:Canco}, it does not immediately tell us how to pick the constant $\lambda$. The real part of $\lambda$ can always be chosen to make the LMHS $\mathbb{R}$-split and we choose to have the imaginary part vanish, so that $\lambda \in \mathbb{R}$. \\ \\
This choice of canonical coordinate is natural from the perspective of the LCS point, as we review in more detail in section \ref{ssec:LCS}. In these examples the function $n(z)$ can be identified with the inverse mirror map. The remaining functional degree of freedom in the instanton map --- the function $A(q)$ corresponding to the red arrow in \eqref{eq:instantonarrows}, when expressed in the $q$ coordinate --- is directly related to the Yukawa coupling. We thus find that both functions $n(z),A(z)$ in the instanton map $\Gamma_{-1}$ have a precise physical meaning for LCS point. 

Moving to conifold or K-points instead, we find that this formalism gives us analogue structures specialized to these other boundaries as well. Again one can determine a canonical coordinate for this singularity from the instanton map $\Gamma$, which is now defined by the expansion \eqref{eq:instantonmap} around these conifold or K-points instead of the LCS point. This gives us an analogue to the mirror map coordinate for these other boundary points. Similarly, the function $A(q)$ can be identified with the Yukawa coupling when expanded around these singularities in this special coordinate $q$. In fact, these expansions provide us with compelling evidence in favor of our special coordinate, as --- up to a common transcendental factor --- the expansion coefficients take rational values in some examples, cf.~tables \ref{table:C} and \ref{table:K}. This could be a hint that there is an underlying physical interpretation for the data,
similar to the LCS point expansion in terms of the mirror coordinate, where these coefficients are then the genus-zero GV invariants. In fact, for K-points this can be matched explicitly in certain hypergeometric examples with recent GLSM computations performed by \cite{Erkinger:2022sqs}, as we elaborate upon below \ref{eq:KpointSeries}.

\newpage
\paragraph{Prepotential formulation.} In addition to the asymptotic Hodge-theoretic formulation given above, let us briefly compare to the $\mathcal{N}=2$ language for the vector multiplet sector more familiar in the physics literature. In this setting the information about the moduli space and the physical couplings is encoded in a single holomorphic function known as the prepotential. For instance, the period vector may be obtained straightforwardly by taking derivatives of the prepotential as described by \eqref{eq:piprepot}. Note, however, that this description requires crucially the use of the correct prepotential coordinate. For LCS points this coordinate agrees with the canonical coordinate $q$ introduced above given by the inverse mirror map, but for other boundaries this is not the case, as we will find in the construction in section \ref{sec:construction}. For that matter we introduce another coordinate $s$, to which we refer as the \textit{prepotential coordinate} for conifold and K-points. While this coordinate $s$ is convenient from a supergravity point of view in using prepotentials, let us stress that it is not the appropriate coordinate in which e.g.~the rational expansion of the Yukawa coupling becomes manifest. Were we to bypass the canonical coordinate $q$ and only use the prepotential formulation in terms of $s$, this arithmetic structure would have been lost on us. In the following section \ref{sec:summary} we will use both formulations, including both the prepotentials ready for direct physical applications as well as the rewriting in terms of the canonical coordinate.

\section{Physical summary: prepotentials and couplings}\label{sec:summary}
In this section we summarize the main findings of this paper for one-modulus singularities in complex structure moduli spaces. We focus on the physical side of the story, centered around 4d $\mathcal{N}=2$ supergravity couplings such as prepotentials and gauge kinetic functions. We refer to section \ref{sec:construction} for the construction of this data, where we explicitly derive the asymptotic periods in an integral basis for all one-modulus models. We also include some brief comments about the geometric origin of the model-dependent data, and refer to section \ref{sec:geometryinput} for a more detailed discussion.

\subsection{Large complex structure points}
To start off our survey of one-modulus singularities, we begin with large complex structure points. These boundaries have been considered extensively in the literature, and therefore form the ideal ground to set the stage for the other two lesser-studied boundaries. Moreover, it allows us to draw a direct comparison between the form of the couplings at large complex structure and these other conifold and K-point singularities.

\paragraph{Prepotential and periods.} We first recall the prepotential and periods for the large complex structure point. In the canonical coordinate the boundary is located at $q=0$, or equivalently at $t=\log[q]/2\pi i = i\infty$. The prepotential \eqref{eq:LCSprepotential} reads\footnote{The prepotential and covering coordinates $s$ and $t$ are identical around the LCS point, so we use $t$ here to comply with the usual conventions from the literature, deviating slightly from the conventions laid out in \ref{table:conventions}.} 
\begin{equation}
F= \frac{\kappa}{6}t^3-\frac{\sigma}{2}t^2-\frac{c_2}{24} t-\frac{\zeta(3) \chi}{2(2\pi i )^3}+\frac{1}{(2\pi i )^3 }C(e^{2\pi i t}) \, , 
\end{equation}
where we assume that the first two periods are fixed to $X^I = (1,t)$. By using mirror symmetry the parameters in this prepotential can be understood as: $\kappa$ is the triple intersection of the mirror Calabi-Yau threefold, $c_2$ the integrated second Chern class, $\chi$ the Euler characteristic, and $\sigma = \kappa/2\mod 1$. The function $C(q)$ encodes the non-perturbative expansion in worldsheet instanton corrections of the periods; the extracted factor of $(2\pi i )^3$ guarantees that the series expansion of $C(q)$ in $z$ has rational coefficients. From this data we compute the periods to be
\begin{equation}
\Pi = \begin{pmatrix}
1 \\
t \\
\partial_t F \\
2 F - t \partial_t F 
\end{pmatrix} = \begin{pmatrix}
 1 \\
 t \\
 \frac{1}{2}\kappa t^2-\sigma t -\frac{1}{24}c_2 + \frac{1}{(2\pi i)^2} e^{2\pi i t} C'(e^{2\pi i t}) \\
 -\frac{1}{6}\kappa t^3 -\frac{1}{24} c_2 t -\frac{\chi \zeta(3)}{2(2\pi i )^3}+\frac{2}{(2\pi i)^3} C(e^{2\pi i t})-\frac{1}{(2\pi i)^2}e^{2\pi i t} C'(e^{2\pi i t})  
\end{pmatrix}\, .
\end{equation}
\paragraph{Subleading polynomial corrections.} In these periods there is a clear distinction between geometrical parameters that are crucial for the leading asymptotic behavior (the intersection number $\kappa$), and subleading polynomial corrections by $c_2, \sigma, \chi$. From a physical perspective this latter set can be understood as $\alpha'$-corrections by using mirror symmetry. From a mathematical perspective they are regarded as so-called extension data, cf.~\cite{KerrLMHS}. More precisely, it captures how the boundary splitting associated to the singularity -- the limiting mixed Hodge structure -- is realized over $\mathbb{Q}$. In other words, it characterizes how the integral basis of the periods should be quantized. In the physics literature we are, in fact, already familiar with these quantization conditions: take for instance the inclusion of the second Chern class $c_2$ in the quantization condition of the fluxes \cite{Palti:2008mg}. Taking a more pragmatic approach, we can encode this data into an Sp$(4, \mathbb{C})$ basis transformation.  This approach was taken in \cite{Palti:2008mg, Escobar:2018rna, Marchesano:2021gyv} in the study of moduli stabilization, where this rotation allows one to absorb these $\alpha'$-corrections into the flux quanta. Here we recall the details for the large complex structure point, since we later extend this methodology to the other two one-modulus boundaries. In this case the rotation reads
\begin{equation}\label{eq:LCSlambda}
\Lambda = 
\left(
\begin{array}{cccc}
 1 & 0 & 0 & 0 \\
 0 & 1 & 0 & 0 \\
 \frac{c_2}{24} & -\sigma  & 1 & 0 \\
 -\frac{i \chi  \zeta (3)}{8 \pi ^3} & \frac{c_2}{24}
   & 0 & 1 \\
\end{array}
\right): \qquad \Pi \to \Lambda \, \Pi \, ,
\end{equation}
One can straightforwardly verify that this basis transformation effectively sets the extension data parameters to zero $c_2 = \sigma = \chi = 0$. The Euler characteristic $\chi$ is what makes this rotation over $\mathbb{C}$, and if one prefers to work over $\mathbb{R}$ it can be left out of $\Lambda$. Another way to see that these parameters $c_2,\sigma,\chi$ correct the periods at subleading order is from the lower-triangular form of the rotation matrix $\Lambda$; it means that periods at a given order $t^k$ get corrected by terms at order $t^{k-1}$ and lower.\footnote{From a Hodge-theoretic perspective, this can be understood as that $\Lambda$ acts in a downwards fashion on the limiting mixed Hodge structure \eqref{eq:LCSIpq}. To be precise, it maps $(\Lambda-1)I^{3,3}\subset I^{1,1} \oplus I^{0,0}$, $(\Lambda-1)I^{2,2} \subset I^{1,1} \oplus I^{0,0}$, while it annihilates $I^{1,1}$ and $I^{0,0}$.} This perspective will prove to be useful for the other two boundaries, where there is no mirror symmetry-based argument for the nature of these corrections, but it still allows us to distinguish parameters associated to leading order terms from subleading polynomial corrections.

\subsubsection*{Physical couplings}
Having characterized the large complex structure prepotential and its model-dependent parameters, we now turn to some of the physical couplings to illustrate how this data affects the underlying supergravity theories. We intend to characterize both the polynomial terms as well as some of the structure underlying the exponential corrections.

\paragraph{K\"ahler potential and metric.} Let us begin with the K\"ahler potential and computing the corresponding metric on the moduli space. We can ignore the real components in the basis transformation \eqref{eq:LCSlambda}, since this part of the rotation drops out of the K\"ahler potential. On the other hand, the Euler characteristic crucially corrects the volume of the mirror Calabi-Yau as a term at order $(\alpha')^3$. To be precise, the K\"ahler potential takes the form
\begin{equation}
K = -\log\big[ \frac{4}{3} \kappa y^3 + \frac{\chi \zeta(3)}{4\pi^3} + \mathcal{K}_{\rm inst} \big]\, ,
\end{equation}
where we expanded $t=x+i y$, and the instanton corrections to the K\"ahler potential are given by
\begin{equation}
\begin{aligned}
\mathcal{K}_{\rm inst} &= \frac{1}{4\pi^3} \text{Re}[C(e^{2\pi i t}] + \frac{ y}{2\pi^2} e^{-2\pi y} \text{Re}[ e^{2\pi i x} C'(e^{2\pi i t})] \\
&=  \frac{1}{2\pi^3} \sum_k A_k e^{-2\pi k y} \cos[2\pi k x](1+2k\pi y)\, ,
\end{aligned}
\end{equation}
In the second line we expanded in terms of the series coefficients $C(z) = \sum_k A_k z^k$, with $A_k$ rational numbers. The corresponding metric takes the form
\begin{equation}
K_{t \bar t} = \partial_t \partial_{\bar t} K = \frac{\kappa y \big(\frac{4}{3}\kappa y^3-\frac{\chi \zeta(3)}{2\pi^3}\big)}{\big(\frac{4}{3}\kappa y^3+ \frac{\chi \zeta(3)}{4\pi^3} \big)^2} + K^{\rm inst}_{t\bar t}
\end{equation}
Note that the metric asymptotes to $K_{t\bar{t}} \sim \tfrac{3}{4y^2}$ for large $y \to \infty$. The complete form of the instanton corrections to the metric is not particularly enlightening, but let us for completeness write down the first correction
\begin{equation}
K_{t \bar t}^{\rm inst} \supseteq -\frac{2A_1}{\pi}  e^{-2\pi y} \cos[2\pi x]\frac{(\frac{4}{3}\kappa y^3-\frac{\chi\zeta(3)}{8\pi^3})(\frac{4}{3}\kappa y^3 +\frac{2}{\pi}\kappa y^2 +\frac{1}{\pi^2}\kappa y^2 +\frac{\chi \zeta(3)}{4\pi^2})}{\big(\frac{4}{3}\kappa y^3+ \frac{\chi \zeta(3)}{4\pi^3} \big)^3}\, .
\end{equation} 

\paragraph{Yukawa coupling.} Another coupling of the 4d supergravity theory underlying these periods that we can compute is the Yukawa coupling. In this coupling the physical importance of the instanton corrections becomes quite manifest, as it takes the form
\begin{equation}
Y = \langle \Pi, \partial_t^3 \Pi \rangle = \kappa +(\frac{1}{2\pi i} \partial_t)^3 C(e^{2\pi i t}) = \kappa +\sum_k k^3 A_k e^{2\pi i k t}  \, .
\end{equation}

\paragraph{Gauge couplings.} The other set of couplings we study are the gauge kinetic functions that describe the couplings of the field strengths of the vectors in 4d $\mathcal{N}=2$ supergravity theories arising from Calabi-Yau compactifications. These gauge kinetic functions can be computed straightforwardly from the prepotential as \eqref{eq:NIJ}, yielding
\begin{align}
\text{Re}(\mathcal{N}) &= \begin{pmatrix}
\frac{1}{3}\kappa x^3  & -\frac{1}{2} \kappa x^2 -\frac{c_2}{24}  \\
-\frac{1}{2} \kappa x^2-\frac{c_2}{24} & \kappa x-\sigma 
\end{pmatrix} +\frac{\kappa y^2 \chi \zeta(3)}{8\pi^3(\frac{4}{3} \kappa y^3 - \frac{\chi \zeta(3)}{8\pi^3})} \begin{pmatrix} 
3x & 2 \\
2 & 0 \\
\end{pmatrix} \, ,\\
\text{Im}(\mathcal{N}) &= \begin{pmatrix}
\frac{\kappa}{6} (3x^2+y^2)y & -\frac{1}{2}\kappa xy \\
-\frac{1}{2}\kappa xy & \frac{1}{2}\kappa y
\end{pmatrix} + \frac{\kappa \chi \zeta(3)}{16\pi^3(\frac{4}{3} y^3 - \frac{\chi \zeta(3)}{8\pi^3})} \begin{pmatrix} 
-\frac{1}{3}y^3 +3 x^2 y - \frac{\chi \zeta(3)}{4\pi^3} & -3 x y \\
-3 x y & 3 y
\end{pmatrix}\, . \nonumber
\end{align}
As the form of the polynomial corrections -- in particular those involving the Euler characteristic $\chi$ -- is already rather unwieldy, we have suppressed any exponential corrections. In principle one could rotate the electric and magnetic charges of BPS states to simplify the polynomial terms to some extend, since the rotation \eqref{eq:LCSlambda} (with $\chi=0$) sets $c_2=0=\sigma$. We postpone this kind of analysis to the next coupling, since this rotation appears more naturally in the study of flux potentials.

\paragraph{Scalar flux potential.} Here, as our final sort of coupling, we look at the scalar potential induced by three-form fluxes near large complex structure points. For ease of notation we use the formalism of flux-axion polynomials, which allows us to incorporate the polynomial dependence on $x$ in the flux quanta. Moreover, we rotate the flux quanta by the basis transformation \eqref{eq:LCSlambda} to hide away some of the $\alpha'$-corrections. Altogether we find that the scalar potential reads
\begin{equation}
V = \rho^T \scalebox{0.9}{$\left(
\begin{array}{cccc}
 \frac{\kappa  y^3 \left(2 \kappa  y^3-3 \xi \right)}{4 \left(3 \xi +4 \kappa  y^3\right)} & 0 & \frac{2 \kappa  y^4-3 \xi  y}{6 \xi +8 \kappa  y^3} & 0 \\
 0 & \frac{\kappa  y \left(3 \xi +\kappa  y^3\right)^2}{-9 \xi ^2+8 \kappa ^2 y^6-6 \kappa  \xi  y^3} & 0 & 3 \kappa  y^2 \left(\frac{1}{3 \xi +4 \kappa  y^3}+\frac{1}{3 \xi -2
   \kappa  y^3}\right) \\
 \frac{2 \kappa  y^4-3 \xi  y}{6 \xi +8 \kappa  y^3} & 0 & \frac{2 \kappa  y^3-3 \xi }{4 \kappa ^2 y^4+3 \kappa  \xi  y} & 0 \\
 0 & 3 \kappa  y^2 \left(\frac{1}{3 \xi +4 \kappa  y^3}+\frac{1}{3 \xi -2 \kappa  y^3}\right) & 0 & \frac{36 \kappa  y^3}{-9 \xi ^2+8 \kappa ^2 y^6-6 \kappa  \xi  y^3} \\
\end{array}
\right)$}  \rho\, , 
\end{equation}
where the flux-axion polynomials are given by
\begin{equation}
\begin{aligned}
\rho &= \Lambda \, e^{x N} \, G_3 \\
&= \scalebox{0.9}{$\left(
\begin{array}{c}
 g_1 \\
 g_1 t+g_2 \\
 g_1 \left(\frac{1}{2} \left(\kappa  t^2-2 \sigma  t\right)-\frac{c_2}{24}\right)+g_2 (\kappa  t-\sigma )+g_3 \\
 g_1 \left(\frac{1}{12} \left(-c_2 t-2 \kappa  t^3\right)+\frac{c_2 t}{24}+\frac{i \chi  \zeta (3)}{8 \pi ^3}\right)+g_2 \left(-\frac{c_2}{24}+\frac{1}{2} \left(-\kappa  t^2-2
   \sigma  t\right)+\sigma  t\right)-g_3 t+g_4 \\
\end{array}
\right)$}\, ,
\end{aligned}
\end{equation}
where we wrote $G_3 = (g_1,g_2,g_3,g_4)$ for $G_3 = F_3 - S \, H_3  $ with the shorthands $g_a=f_a- S \, h_a$, with $S$ the axio-dilaton.

\subsection{Conifold points}
Here we study one-modulus models for conifold points. We begin by recalling the form of the prepotential from section \ref{ssec:conifoldconstruction} and characterize its model-dependent coefficients. We then set out and compute the couplings in the supergravity theory from this prepotential, which elucidates the physical meaning of some of this data. In comparison to other parts of this paper, we stress that only in this section on conifold points we differ by a basis transformation
\begin{equation}
    \begin{pmatrix}
        0 & 1 & 0 & 0 \\
        1 & 0 & 0 & 0 \\
        0 & 0 & 0 & 1 \\
        0 & 0 & 1 & 0
    \end{pmatrix} \in \mathrm{Sp}(4,\mathbb{Z})\, .
\end{equation}
The reason is that the other period basis is better aligned with the monodromy weight filtration, while the basis used here matches with the standard conventions of the prepotential formulation. 

\paragraph{Prepotential and periods.} We first recall the prepotential \eqref{eq:prepotentialconifold}
and compute the corresponding periods. We parametrize the region around the conifold point by a coordinate $s$ such that it is located at $s=0$, and circling the boundary corresponds to $s \to e^{2\pi i} s$. We can write down the prepotential as an expansion in $s$ as\footnote{\label{footnote:conifold}See \eqref{eq:prepotentialconifold} for a description of what higher order corrections to the prepotential look like. In particular, there do not appear any terms involving logarithms $\log(s)$.}
\begin{equation}\label{eq:Fc}
F = F_{(0)} + s F_{(1)} + s^2 F_{(2)} + s^3 F_{(3)} +\mathcal{O}(s^4)\, ,
\end{equation}
where the expansion terms read
\begin{equation}
\begin{aligned}
 F_{(0)}  &= \frac{\tau}{2} \, ,  \\
 F_{(1)}  &= \delta-\gamma \tau \, , \\
 F_{(2)} &= \frac{i k}{8 \pi} \left(3- 2 \log(s)+2 \log(A_1) \right)- \frac{1}{2}\gamma(\delta-\gamma \tau) \, , \\
 F_{(3)} &= -\frac{i k \left(3 A_1^2 \gamma -A_2\right)}{12 \pi  A_1^2}\, .
\end{aligned}
\end{equation}
We note that the logarithm $\log[s]$ only appears at second order $s^2$, and the coefficients all higher-order terms in the prepotential expansion are independent of $s$. In order to compute the periods we choose as projective coordinates  $X^I=(1,s)$. From this prepotential we can then compute the periods to be
\begin{align}\label{eq:summaryperiodsconifold}
    \Pi(s) &= \begin{pmatrix}
    1 \\ s \\  \partial_s F \\ 2 \, F - s \, \partial_s F 
    \end{pmatrix} \nonumber \\
    &= \begin{pmatrix}
    1 \\
    s \\
    (\delta-\gamma \tau)  + \left( \frac{i k}{8 \pi} \left(1- 2 \log(s)+2 \log(A_1) \right)- \frac{1}{2}\gamma(\delta-\gamma \tau) \right)s \\
    \tau + (\delta-\gamma \tau)s -\dfrac{i k}{4 \pi} s^2
    \end{pmatrix}\, .
\end{align}
In addition to the prepotential coordinate $s$ used above, for later applications to physical couplings it is useful to introduce a covering coordinate for the asymptotic regime. To leading order this coordinate is given by
\begin{equation}
    e^{2\pi it}=  s/A_1 +\mathcal{O}(s^2)\, ,
\end{equation}
where we refer to \eqref{eq:Cz} for the form of the corrections. The singularity corresponds to the large field limit $t\to i \infty$ and circling around to $t \to t+1$. Applying also a rescaling of the periods, we recover the periods given
in \eqref{eq:Cpiasymp}, which we repeat for completeness in the basis used here
\begin{equation}\label{eq:summaryperiodsconifoldres}
    \Pi = \begin{pmatrix}
        1 \\
        0 \\
        \delta-\gamma \tau \\
        \tau 
        \end{pmatrix}
        +A_1e^{2\pi i t}\begin{pmatrix}
            \gamma  \\
             1\\
             t -k/2\pi i\\
            \delta 
        \end{pmatrix}
        +e^{4\pi i t}\begin{pmatrix}
            \gamma A_2-\frac{k}{8\pi \tau_2}(A_1)^2 \\
            A_2 \\
            A_2 (t -k/4\pi i) +(\delta-\gamma \bar \tau)(A_1)^2\\
            \delta A_2-\frac{k \bar\tau}{8\pi \tau_2} (A_1)^2
        \end{pmatrix}+\ldots\, ,
\end{equation}
Before we use these periods to compute physical couplings, let us summarize some of the main features of the model-dependent parameters appearing in these periods:
\begin{itemize}
    \item Rigid period $\tau=\tau_1+i\tau_2$. As suggested by the choice of symbol, this parameter can be viewed as a period living in the upper half plane $\tau_2>0$. Moreover, a subgroup of the symplectic duality group $\text{SL}(2,\mathbb{Z}) \subseteq \text{Sp}(4,\mathbb{Z})$ acts on this as the modular group, see \eqref{Eq:ExtShift} for more details. From a geometric point of view, one can understand it from the rigid ($h^{2,1}=0$) conifold geometry that arises at the singularity as its period vector $(1,\tau)$. From the periods in the bulk of the moduli space this data can be read of as the constant part $\Pi \supseteq (1,0,0,\tau)$ (when the extension data vanishes $\gamma=\delta=0)$). And as we will see in more detail later, it corresponds precisely to the complexified gauge coupling of the graviphoton, cf.~\eqref{eq:Ccouplings}.
    \item Extension data $\gamma,\delta$. These parameters mix the rigid period $\tau$ with the periods along the conifold cycle and its dual (the period containing a linear term in $t$). This mixing propagates through most of the physical couplings, as is illustrated for instance by the off-diagonal components of the gauge kinetic functions \eqref{eq:Ccouplings}. Let us note, however, that they can be treated systematically in most cases. For example, in the scalar potential \eqref{eq:Cpotential} we can redefine the flux quanta according to \eqref{eq:Crho} in order to remove these mixed terms. The sort of numbers one encounters in examples for $\gamma,\delta$ can be rational or transcendental, depending on the example under consideration, cf.~table \ref{table:K}. This aspect is correlated with the finite order part of the monodromy around the conifold point: if there is a finite order part then $\gamma,\delta \in \mathbb{Q}$, while otherwise $\gamma,\delta \in \mathbb{R}$. For a more detailed discussion on how the finite order part of the monodromy affects the extension data $\gamma,\delta$ we refer to subsection \ref{ssec:geometryI1}. 
    \item Massless states. At conifold points there is always a BPS state becoming light \cite{Candelas:1989ug, Candelas:1989js}: for ordinary conifold points close to the LCS (in the sense that the regions of convergence of their local solutions overlap) it is the mirror D6-brane state, but when multiple conifold points are present another state becomes light at the other singularities. This state has been listed in table \ref{table:C} for the examples we considered, corresponding to D6-branes with additional D4- and D0-brane charges.\footnote{This linear combination of D-branes can be obtained by applying the inverse of the transition matrix \eqref{eq:conifoldmatrix} on the vector $(0,0,0,1)$, which is the conifold state in the basis considered there.} 
    
    In addition, for certain examples there is a candidate for a second light state at the conifold point, as has been noted already in \cite{Huang:2006hq} for the $X_{3,2,2}$. These states are present whenever the extension data $\gamma,\delta$ is rational: the rational vector $q=(\gamma,1,0,\delta)$ has a vanishing product with the leading term $(1,0,\delta-\gamma \tau,\tau)$ in the periods \eqref{eq:summaryperiodsconifold}. We may rescale this rational vector appropriately to obtain a candidate BPS state with integer charges; due to its vanishing product with the leading part of the periods its mass would be zero at the conifold point.
    \item Order $k$. Geometrically, the massless conifold state corresponds to a three-cycle that shrinks to zero size. This three-cycle can be represented by a three-sphere $S^3$ quotiented by some finite subgroup of order $k$ \cite{Gopakumar:1997dv}. In the periods this data enters as an overall factor for the monodromy, which is seen most clearly in the log-monodromy matrix
    \begin{equation}
    N = \begin{pmatrix}
    0 & 0 & 0 & 0 \\
    0 & 0 & 0 & 0 \\
    0 & k & 0 & 0 \\
    0 & 0 & 0 & 0 
    \end{pmatrix}\, .
    \end{equation}
    \item Instanton numbers $A_1,A_2,\ldots$. These coefficients are the series coefficients in the expansion in $s$ of the prepotential. In principle these could be arbitrary complex numbers to give a set of well-defined periods, however, in geometrical examples there is an underlying arithmetic structure. We find that we can extract a common factor from the coefficients $A_l$, which can be specified in terms of certain L-values associated to the geometry; the residual coefficients then take rational values. In table \ref{table:C} we have summarized these findings, giving both the common factor and the rational part of the first few instanton coefficients. 
\end{itemize}

\begin{table}[h!]
\centering
\renewcommand*{\arraystretch}{2.0}
\rotatebox{270}{
\begin{tabular}{| c || c | c | c | c | c | c |}
\hline  Example & \hyperref[parX42]{$X_{4,2}$} & \hyperref[parX322]{$X_{3,2,2}$} & \hyperref[parX62]{$X_{6,2}$} & \hyperref[par217]{2.17} & \hyperref[par262]{2.62} & \hyperref[parX5]{$X_5$} \\ \hline \hline 
$\tau$ & $\frac{1}{2}+\frac{i}{2}$  & $\frac{1}{2}+\frac{i}{2\sqrt{3}}$ & $\frac{1}{2}+\frac{i\sqrt{3}}{2}$ & $i\frac{L_f(2)}{8 \pi L(f,1)}  $ & $\frac{1}{2}+\frac{i}{2}$ & $\frac{1}{2}+i\frac{125 L(f_{25},2)}{8\pi L(f_{25},1)}$\\ \hline
$(\gamma,\delta)$ & $(0,\frac{1}{2})$ & $(0,0)$ & $(\tfrac{2}{3},\tfrac{1}{3})$ & $(0,\frac{1}{8})$ & $(0,\frac{1}{2})$ & $0.08641 \cdot (2,1)$ \\ \hline
$k$ & 2 & 6 & 1 & 1 & 42 & 1\\ \hline
Instanton coeff. & 
$\begin{aligned}[t]
    \hat{A}_1&: 1\\
    \hat{A}_5&: -\tfrac{7}{30} \\
    \hat{A}_9&: -\tfrac{65}{3528} \\ \vspace*{0.0002cm}
    \end{aligned}$ & $\begin{aligned}[t]
    \hat{A}_1&: 1 \\
    \hat{A}_7&: -\tfrac{272}{14175} \\
    \hat{A}_{13}&: -\tfrac{5256154}{6450283125} \\ 
    \end{aligned}$ & $\begin{aligned}[t]
    \hat{A}_1&: 1 \\
    \hat{A}_4&: -\tfrac{70}{9} \\
    \hat{A}_7&: -\tfrac{314432}{1148175} \\
    \end{aligned}$ & $\begin{aligned}[t]
    \hat{A}_1&: 1 \\
    \hat{A}_3&: \tfrac{40}{3} \\
    \hat{A}_5&: \tfrac{749}{5} \\ 
    \end{aligned}$ & $\begin{aligned}[t]
    \hat{A}_1&: 1 \\
    \hat{A}_5&: -\tfrac{89}{2560} \\
    \hat{A}_9&: -\tfrac{19407}{3670016} \\
    \end{aligned}$ & $\begin{aligned}[t]
    \hat{A}_1&: 1 \\
    \hat{A}_2&:  \tfrac{\xi (3125+6r_{1})}{250 \sqrt[3]{5}}  \\
    \end{aligned}$\\[0.5em] \hline
Instanton norm. & $ -\frac{1}{L(f_{32},1)} $ & $ \frac{2\;2^{1/3} \sqrt{3}  }{3^3 L(f_9,1)} $ & $-\frac{2^{5}}{3^{8}L(f_{108},1)}$ & $-\frac{i2\sqrt{2}}{L(f,1)}$ & $-\frac{i}{2\sqrt{2}L(f_{32},1)}$ & $\frac{5 \sqrt[6]{5}\xi }{L(f_{25},1)}$\\ \hline
modular form $f$ & $32.4.a.b$ & $9.4.a.a$ & $108.4.a.b$ & $8.4.a.a$ & $32.4.a.b$ & $25.4.b.a$\\ \hline
$L(f,1)$ & $\frac{\Gamma(1/4)^6\Gamma(1/2)^3}{16\sqrt{2}\pi^5}$ & $\frac{\Gamma(1/3)^9}{32\sqrt{3}\pi^5}$&$\frac{3\sqrt{3}\Gamma(1/3)^9}{16\, 2^{1/3}\pi^5}$ & $\simeq 4.01072$ & $\frac{\Gamma(1/4)^6\Gamma(1/2)^3}{16\sqrt{2}\pi^5}$ & $\simeq 1.62555$ \\ 
$L(f,2)$ & $\frac{\Gamma(1/4)^6\Gamma(1/2)^3}{64\sqrt{2}\pi^4}$ & $\frac{\Gamma(1/3)^9}{96\pi^4}$ & $\frac{\Gamma(1/3)^9}{32\, 2^{1/3}\pi^4}$ & $\simeq 1.9517$ & $\frac{\Gamma(1/4)^6\Gamma(1/2)^3}{64\sqrt{2}\pi^4}$ & $\simeq 1.56524$\\
$L(f,3)$ & $\frac{\Gamma(1/4)^6\Gamma(1/2)^3}{256\pi^{3}}$ & $\frac{\Gamma(\frac{1}{3})^9}{144\sqrt{3}\pi^3}$ & $\frac{2^{1/3}\Gamma(\frac{1}{3})^9}{192\sqrt{3}\pi^3}$ & $\simeq 1.23701$ & $\frac{\Gamma(1/4)^6\Gamma(1/2)^3}{256\pi^{3}}$ & \\ \hline
$q_C = (\rm{D6}, \rm{D4}, \rm{D2}, \rm{D0})$ & \multicolumn{3}{c|}{$(2,-1,0,-4)$} & $(1,0,0,0)$ & $(2,1,-5,6)$ & $(1,0,0,0)$\\ \hline
\end{tabular}}
\caption{\label{table:C} Summary of geometric data for conifold point examples, including both hypergeometric (the first three columns) and more generic models. We have given the rigid period $\tau$, extension data $\gamma,\delta$ and order $k$ for these examples. The instanton coefficients have been decomposed into a common factor times an (often) rational piece. The numerical values of this data can be specified by L-function values associated to the geometry in question.
} 
\end{table}

\subsubsection*{Physical couplings}
From the conifold prepotential in \eqref{eq:Fc} we can compute the relevant physical couplings for the vector multiplet sector of $\mathcal{N}=2$ supergravities arising from Type IIB Calabi-Yau compactifications. We focus here on the K\"ahler metric and gauge kinetic functions, since these encode the kinetic terms for the complex structure modulus and gauge fields in the theory. For the reader interested in moduli stabilization in flux compactifications we will also provide the scalar potential arising from turning on R-R and NS-NS three-form fluxes.

\paragraph{K\"ahler potential and metric.} Let us first write down the K\"ahler potential. We note that we use the periods given in \eqref{eq:summaryperiodsconifoldres}; these differ by a coordinate redefinition and rescaling from \eqref{eq:summaryperiodsconifold}, so the two K\"ahler potentials are related by a K\"ahler transformation. It is given by
\begin{equation}
K=-\log[K_{(0)}+ e^{-2\pi y}K_{(1)}+e^{-4\pi y}K_{(2)}+\ldots]\, .
\end{equation}
where the terms in the expansion of the K\"ahler potential in $|s|\ll 1$ read
\begin{equation}
    \begin{aligned}
        K_{(0)} &= 2\tau_2 \, , \\
        K_{(1)} &= 0\, , \\
        K_{(2)} &= -\frac{k}{\pi}A_1^2 (1+2\pi y)\, .
    \end{aligned}
\end{equation}
Using the prepotential periods \eqref{eq:summaryperiodsconifold} would have given a non-vanishing $K_{(1)}$ that depends on the axion $x$; however, this term is removed by a K\"ahler transformation to the periods \eqref{eq:summaryperiodsconifold}, as was already observed in \cite{Bastian:2021eom}. Indeed, one may verify the two K\"ahler potentials yield the same physical K\"ahler metric, which reads
\begin{equation}
    K_{t \bar{t}} =  \frac{4k\pi^2 y}{\tau_2}A_1^2 e^{-4\pi y} + \mathcal{O}(e^{-6\pi y})\, .
\end{equation}

\paragraph{Yukawa coupling.} By taking three derivatives of the periods in \eqref{eq:summaryperiodsconifoldres} we can compute the Yukawa coupling. We find that its expansion in $t$ takes the following form
\begin{equation}
    Y = \langle \Pi, \partial_t^3 \Pi \rangle = 4k\pi^2  \bigg(\sum_l l A_l e^{2\pi i t} \bigg)^2\, .
\end{equation}
In some examples the expansion coefficients $A_l$ take rational values (after extracting a common transcendental factor), cf.~table \ref{table:C}.

\paragraph{Gauge kinetic functions.} Next we compute the gauge kinetic functions \eqref{eq:KNIJ} for the vector sector of the 4d $\mathcal{N}=2$ supergravity. To leading order in the expansion in $s$ these couplings are given by
\begin{equation}\label{eq:Ccouplings}
\begin{aligned}
\mathcal{R}_{IJ} &= \begin{pmatrix} \tau_1 & 0 \\
 0 & k x
 \end{pmatrix}+\begin{pmatrix}
     0 & \gamma\tau_1-\delta \\
     \gamma \tau_1 - \delta & -\gamma (\gamma \tau_1 - \delta)
 \end{pmatrix}\, \\
\mathcal{I}_{IJ} &= \begin{pmatrix} \tau_2 & 0 \\
0 & ky
 \end{pmatrix}+\begin{pmatrix} 0 & -\gamma \tau_2 \\
-\gamma \tau_2  & \tau_2 \gamma^2
 \end{pmatrix} \, .
\end{aligned}
\end{equation}
These gauge couplings elucidate the physical interpretation of the model-dependent parameters, specifically the rigid period $\tau$ and also the extension data $\gamma,\delta$. The $\gamma,\delta$ parameterize mixing between the two vectors. Setting these to zero momentarily, we can read off the complexified gauge couplings of the two vectors straightforwardly as $\tau=\tau_1+i\tau_2$ and $k(x+iy)$. The first coupling is constant in the moduli and associated to the periods of the rigid conifold geometry associated to the moduli space singularity; on the other hand, the second coupling is dynamical and parametrized by an axio-dilaton-like field. For the rigid coupling it is also interesting to point out that in all examples we checked the theta angle takes the values $\tau_1=0,\frac{1}{2}$, which fits well with the proposal of \cite{Cecotti:2018ufg} that $\mathcal{N}=2$ supergravity theories with no vector multiplets can only have such theta angles.

\paragraph{Scalar potential.}  Finally we move on to the scalar potential \eqref{eq:LCSprepotential}. Similar to the large complex structure point, there is a rotation of the fluxes analogous to \eqref{eq:LCSlambda} that rotates out the extension data parameters $\gamma,\delta$ of the scalar potential. Recall from section \ref{ssec:conifoldconstruction} that there is an Sp$(4,\mathbb{Z})$ basis transformation \eqref{Eq:ExtShift} which shifts $\gamma,\delta$ independently by an integer. By relaxing the integrality condition, this means we can write down an Sp$(4,\mathbb{R})$ that effectively sets these parameters to zero (setting $b_1 = -\gamma, b_2 = -\delta$ in \eqref{Eq:ExtShift})
\begin{equation}\label{eq:Crot}
\Lambda = \left(
\begin{array}{cccc}
 1 & -\gamma  & 0 & 0 \\
 0 & 1 & 0 & 0 \\
 -\delta  & 0 & 1 & \gamma  \\
 0 & -\delta  & 0 & 1 \\
\end{array}
\right)\, .
\end{equation}
We also want to apply a rotation by $e^{xN}$ on the flux vector to define flux-axion polynomials, as this allows us to suppress the polynomial dependence on the axion in the scalar potential. Combining this with the Sp$(4,\mathbb{R})$ basis transformation above we find that we define the following flux-axion polynomials
\begin{equation}\label{eq:Crho}
    \rho = e^{x N}\Lambda^{-1} G_3 = \left(
\begin{array}{c}
 \gamma  g_2+g_1 \\
 g_2 \\
 -\gamma  g_4+\delta  g_1+g_2 k x+g_3 \\
 \delta  g_2+g_4 \\
\end{array}
\right)\, .
\end{equation}
Having set up these flux-axion polynomials, we are now in the position to compute the scalar potential. Let us first write down the leading part at polynomial order, which is given by
\begin{equation}\label{eq:Cpotential}
V_{(0)} = \bar{\rho}^T \begin{pmatrix}
\frac{(\tau_1)^2+(\tau_2)^2}{\tau_2} & 0 & 0 & -\frac{\tau_1+i\tau_2}{\tau_2} \\
0 & ky& -i & 0 \\
0 & i & 1/ky & 0 \\
-\frac{\tau_1-i\tau_2}{\tau_2} & 0 & 0 & \frac{1}{\tau_2}
\end{pmatrix} \rho\, .
\end{equation}
The components on the outside describe the coupling to the constant $\tau=\tau_1+i\tau_2$ part of the conifold periods, while the middle part corresponds to the period along the conifold cycle and its dual, as is indicated by the presence of the saxion $y$. For completeness, we also want to note that writing down the first corrections at order  $\mathcal{O}(e^{-2\pi y})$ can be written down relatively concisely as
\begin{equation}
    V_{(1)} =  \bar{\rho}^T
\scalebox{1}{$\left(
\begin{array}{cccc}
 0 & \frac{i A_1 k \text{Re}[e^{2\pi i x}(\tau_1-i\tau_2)]}{2 \pi  \tau _2} & \frac{i A_1 \text{Im}[e^{2\pi i x}(\tau_1-i\tau_2)]}{2 \pi  \tau _2 y} & 0 \\
 \frac{i A_1 k \text{Re}[e^{2\pi i x}(\tau_1-i\tau_2)]}{2 \pi  \tau _2} & \frac{4 A_2 k y \cos (2 \pi  x)}{A_1} & 0 & -\frac{i
   A_1 k \cos (2 \pi  x)}{2 \pi  \tau _2} \\
 \frac{i A_1 \text{Im}[e^{2\pi i x}(\tau_1-i\tau_2)]}{2 \pi  \tau _2 y} & 0 & \frac{4 A_2 \cos (2 \pi  x)}{A_1 k y} & -\frac{i
   A_1 \sin (2 \pi  x)}{2 \pi  \tau _2 y} \\
 0 & -\frac{i A_1 k \cos (2 \pi  x)}{2 \pi  \tau _2} & -\frac{i A_1 \sin (2 \pi  x)}{2 \pi  \tau _2 y} & 0 \\
\end{array}
\right)$} \rho\, ,
\end{equation}
where we suppressed the $e^{-2\pi y}$ factor. It is particularly interesting to note that the coupling between the middle two and outer two components does not involve the correction $A_2$, while in the middle block it plays an important role. Finally, also recall that these matrices are rotated by the Sp$(4,\mathbb{R})$ transformation \eqref{eq:Crot}, so in general the flux quanta are mixed by the extension data $\gamma,\delta$; however, note that there are examples where this mixing vanishes, see e.g.~the $X_{3,2,2}$ in table \ref{table:C}.

\subsection{K-points}
Here we summarize the results obtained for the periods near K-points. We give the prepotential with the corresponding coordinate, and include the resulting period vector. We also characterize the coefficients appearing in these periods, i.e.~which numbers are integer and which can be transcendental, and comment on their geometric origin. Similar to the other two boundaries, we conclude with a survey of the physical couplings in the asymptotic regime near this boundary.

\paragraph{Prepotential and periods.} We begin by writing down the prepotential and the resulting periods. For this boundary we take again, similar to the conifold point, a coordinate $s$ such that the boundary is located at $s=0$. In terms of this coordinate we can write an expansion for the prepotential as
\begin{equation}\label{eq:Kprepot}
    F = F_{(0)} + s F_{(1)}+s^2F_{(2)}+s^3 F_{(3)}+\mathcal{O}(s^4)\, ,
\end{equation}
where the terms at the various orders read
\begin{equation}\label{eq:Kprepotterms}
\begin{aligned}
    F_{(0)} &= \tfrac{1}{2}\delta+ \gamma \tau\, , \\
    F_{(1)} &= \gamma -\frac{c\tau_2}{2\pi}\bigg( \log\left[\frac{is}{2B_1\tau_2}\right]-1\bigg) \, , \\
    F_{(2)} &= \frac{ic}{4\pi^2}\bigg(\log\left[\frac{is}{2B_1\tau_2}\right] -2 + \frac{B_2}{4(B_1)^2}\bigg)\, , \\
    F_{(3)} &= \frac{c}{16\pi \tau_2} ( 1-\frac{B_2^2}{B_1^4} +\frac{2 B_3}{3B_1^3} )\, .
\end{aligned}
\end{equation}
We note that the logarithm $\log[s]$ only appears at first and second order $\mathcal{O}(s,s^2)$, while higher-order terms are free of any other moduli dependence. In order to compute the periods we must choose as projective coordinates $X^I=(1,\tau+s)$. We emphasize the presence of a constant term $\tau$, which means that -- unlike the conifold point -- there is no vanishing period in the limit $s \to 0$.\footnote{We will still encounter states that become massless at the K-point, which is driven by a diverging Planck mass $e^{K/2}$ at this infinite distance point.} We compute the periods to be
\begin{equation}
\Pi = \left(
\begin{array}{c}
 1 \\
 \tau \\
\delta+\gamma \tau -\frac{c\tau \tau_2}{2\pi} \log\big[\frac{is}{2B_1 \tau_2}\big] 
 \\
\gamma +\frac{c\tau_2}{2\pi}\log\big[\frac{is}{2B_1\tau_2}\big] \\
\end{array}
\right) +s\left(\begin{array}{c}
 0\\
 1 \\
\frac{ic\tau_1}{2\pi} \log\big[\frac{is}{2B_1 \tau_2}\big]+\gamma +\frac{3c\tau_1}{4\pi i} -\frac{cs\tau_2}{4\pi} +\frac{ics\tau B_2}{(B_1)^2}   \\
 \frac{c}{2\pi i} \log[\frac{is}{2B_1\tau_2}\big] +\frac{3ic}{4\pi}-\frac{icB_2}{4\pi(B_1)^2} \\
\end{array}
\right)+\mathcal{O}(s^2)\, ,
\end{equation}
where we used \eqref{eq:Ktau} to simplify some of the terms in this expression. While this parametrization of the periods in $s$ is the most easily accessible from the prepotential, for later applications to physical couplings it will prove to be helpful to introduce a covering coordinate for the asymptotic regime. To leading order this coordinate is given by
\begin{equation}\label{eq:Kmirror}
e^{2\pi i t} = -s/(2i \tau_2 B_1) + \mathcal{O}(s^2)\, ,
\end{equation}
such that the boundary is located at the large field limit $t \to i \infty$ and circling the boundary corresponds to $t \to t+1$. We refer to \eqref{eq:zs} for a description of the subleading terms in this expansion, which can be taken into account order-by-order similar to the mirror map at large complex structure. For completeness, let us write down the periods in terms of this covering coordinate $t$ as well
\begin{equation}\label{eq:Kperiodssummary}
    \Pi = \left(
\begin{array}{c}
 1 \\
 \tau  \\
 (a +b \tau ) t+\delta +\gamma  \tau  \\
 (b +c \tau ) t+\gamma  \\
\end{array}
\right)+2B_1 e^{2\pi i t} \left(
\begin{array}{c}
 0 \\
 -i\tau_2  \\
 - i b\tau_2 \,  t +\frac{\tau _2}{2\pi }(b+ic\tau_2)+ i \gamma \tau _2 \\ 
- i c\tau_2 \, t + \frac{c \tau _2}{2\pi }
\end{array}
\right)+ \mathcal{O}(e^{4\pi i t})\, ,
\end{equation}
where we note that certain terms at first in the expansion may have been altered by taking corrections to \eqref{eq:Kmirror} into account. Note in particular that this makes the form of them monodromy matrix more apparent under $t\to t+1$, as it now can be read off straightforwardly as
\begin{equation}\label{eq:KN}
    N = \begin{pmatrix}
        0 & 0 & 0 & 0 \\
        0 & 0 & 0 & 0 \\
        a & b & 0 & 0 \\
        b & c & 0 & 0 \\
    \end{pmatrix}
\end{equation}
with the monodromy given by $M = e^{N}$. Let us elaborate on the meaning of the model-dependent parameters appearing in these periods:
\begin{itemize}
    \item Log-monodromy coefficients $a,b,c$. From quantization of the monodromy matrix $M \in \mathrm{Sp}(4,\mathbb{Z})$ it follows that these parameters must be integers $a,b,c \in \mathbb{Z}$. Positivity of the K\"ahler metric \eqref{eq:Kmetric} will imply that the determinant is positive $d=b^2-ac >0$. It is worthwhile to briefly point out the geometrical interpretation of this data, since it encodes information about the singular CY geometry at the K-point. Assuming the K-point is realized by a Tyurin degeneration \cite{tyurin}, it can be understood as a stable degeneration limit where two threefolds intersect in a K3 surface. Special about one-modulus K-points is that the K3 surface is rigid, and the log-monodromy matrix specifies the intersection form on its $(2,0)+(0,2)$-part of its cohomology, see section \ref{sec:geometryinput} for more details.
    \item Rigid period $\tau=\tau_1+i\tau_2$. Similar to the conifold point, the K-point periods feature a model-dependent parameter living in the upper half plane $\tau_2 >0$, which is acted on by a subgroup $\mathrm{SL}(2,\mathbb{Z}) \subseteq \mathrm{Sp}(4,\mathbb{Z})$, see \eqref{eq:basischangeII} for a more detailed characterization. For K-points the monodromy coefficients $a,b,c$ completely specify (up to SL$(2,\mathbb{Z})$ transformations) this rigid period $\tau$ as
    \begin{equation}\label{eq:Ktau}
        \tau = \frac{-b+i\sqrt{b^2-ac}}{c} \, .
    \end{equation}
    \item Extension data $\gamma,\delta \in \mathbb{R}$. These parameters do not show up in the K\"ahler potential and metric, similar to the second Chern class at LCS. On the other hand, they do play a role in the precise quantization of the field strengths and three-form fluxes in the effective $\mathcal{N}=2$ or $\mathcal{N}=1$ supergravity theories: for the gauge fields these parameters correspond to non-vanishing theta angles when we set the axion to $x=0$, cf.~\eqref{eq:KNIJ}; for the scalar potential we can absorb the $\gamma,\delta$ in terms of the flux quanta by an Sp$(4,\mathbb{R})$ basis transformation \eqref{eq:Klambda}. In all examples we studied we find that $\gamma$ and $\delta$ are rational multiples of the same, possibly transcendental, number. We suspect this is a general constraint from geometry on the data, but it would be interesting to investigate this further, since the Hodge-theoretic constraints we used only impose they should be valued in $\mathbb{R}$. For special examples with monodromies with a finite order factor --- such as the hypergeometric cases --- this real number is rational, see table \ref{table:K} for a more complete overview.
    \item Instanton numbers $\hat{B}_1,\hat{B}_2,\ldots$. These coefficients are the series coefficients in the expansion in $s$ of the prepotential. Similar to the conifold points, these could in principle be arbitrary complex numbers, however, in geometrical examples there is an underlying arithmetic structure: all coefficients are rational, up to a common factor that can be expressed in terms of L-values associated to the geometry. In the physical couplings this rational series is displayed the most directly in the Yukawa coupling \eqref{eq:KYukawa}. In table \ref{table:K} we have summarized the findings for the $A_l$, giving both the common factor and the rational part of the first few instanton coefficients.
    \item Light states. At K-points there is a lattice of candidate BPS states that become light as we approach the singularity. In terms of the mirror D-brane basis we find that this lattice is spanned by D6-branes and D4-branes, bound to D2- and D0-branes.\footnote{The precise linear combinations of D-branes can be obtained by applying the transition matrices in section \ref{sec:examples} from the integral LCS basis to the integral K-point basis on the basis vectors $(0,0,1,0)$ and $(0,0,0,1)$ for the light states.} In light of the emergent string conjecture \cite{Lee:2019wij} the bound D4-brane state indicates a candidate 4-cycle that can be wrapped by an NS5-brane, signaling a light tower of string excitations for the distance conjecture \cite{Ooguri:2006in}.
\end{itemize}

\begin{table}[h!]
\centering
\renewcommand*{\arraystretch}{2.0}
\begin{tabular}{| c || c | c | c | c | c |}
\hline  Example & \hyperref[parX33]{$X_{3,3}$} & \hyperref[parX44]{$X_{4,4}$} & \hyperref[parX66]{$X_{6,6}$}  & \hyperref[par447]{4.47} & \hyperref[par37]{3.7}  \\[0.4em] \hline \hline 

{\renewcommand*{\arraystretch}{1}$\begin{pmatrix}
    a & b \\
    b & c
\end{pmatrix}$} & 
{\renewcommand*{\arraystretch}{1}$\begin{pmatrix}
    2 & 1 \\
    1 & 2
\end{pmatrix}$} & 
{\renewcommand*{\arraystretch}{1}$\begin{pmatrix}
    1 & 0 \\
    0 & 1
\end{pmatrix}$} & 
{\renewcommand*{\arraystretch}{1}$\begin{pmatrix}
    2 & 1 \\
    1 & 2
\end{pmatrix}$} & {\renewcommand*{\arraystretch}{1}$\begin{pmatrix}
    2 & 0 \\
    0 & 6
\end{pmatrix}$}  & {\renewcommand*{\arraystretch}{1}$\begin{pmatrix}
    1 & 1 \\
    1 & 2
\end{pmatrix}$}\\[0.4em] \hline
$\tau=\frac{-b+i\sqrt{ac-b^2}}{c}$ & $-\tfrac{1}{2}+i\frac{\sqrt{3}}{2}$  & $i$ & $-\tfrac{1}{2}+i\frac{\sqrt{3}}{2}$ & $\frac{i}{\sqrt{3}}$ & $-\frac{1}{2}+\frac{i}{2}$ \\ \hline
$(\gamma,\delta)$ & $(\tfrac{1}{6},\tfrac{1}{3})$ & $(\frac{1}{2},0)$ & $(\frac{1}{3},\tfrac{1}{6})$ & $(\tfrac{5}{12}, \tfrac{5}{6})$ & $0.355955 \cdot (1,1)$ \\ \hline
Instanton coef. & $\begin{aligned}[t]
\hat{B}_1&: 1\\
    \hat{B}_4&: -\tfrac{1}{2} \\
    \hat{B}_7&: \tfrac{501119}{196000} \\ 
    \end{aligned}$ & $\begin{aligned}[t]
    \hat{B}_1&: 1 \\
    \hat{B}_3&: 12 \\
    \hat{B}_5&: \tfrac{474122}{675} \\ 
    \end{aligned}$ & $\begin{aligned}[t]
    \hat{B}_2&: 1 \\
    \hat{B}_5&: -640 \\
    \hat{B}_8&: 1251305 \\ 
    \end{aligned}$ & $\begin{aligned}[t]
    \hat{B}_1&: 1 \\
    \hat{B}_4&: \tfrac{112}{3\ 3^{3/4}} \\
    \hat{B}_7&: \tfrac{5217112}{3375 \sqrt{3}} \\\\ 
    \end{aligned}$ &  $\begin{aligned}[t]
    \hat{B}_1&: 1\\
    \hat{B}_2&:  \tfrac{\left(  1269-16 \rho \right)}{48 \sqrt{3}} \\
    \\ 
    \end{aligned}$ \\[3em] \hline
Coef. norm. & $-\frac{1}{3L(f,1)^2}$ & $-\frac{1}{4L(f,1)^2}$ & $-\frac{9}{8\;2^{1/3} L(f,1)^2}$ & $\tfrac{8 \sqrt[3]{2} 3^{3/4} \pi^2}{L(f,1)^2}$ &  $-\frac{1}{4 \sqrt{3} L(f,1)^2}$\\ \hline
Modular form $f$ & 27.3.b.a & 16.3.c.a & 12.3.c.a & 108.3.c.a  & 36.3.d.a \\ \hline
$L(f,1)$ &  $\frac{\Gamma(\tfrac{1}{3})^6}{8\sqrt{3}\pi^3}$ & $\frac{\Gamma(\tfrac{1}{4})^4\Gamma(\tfrac{1}{2})^2}{32\pi^3}$ & $\frac{\sqrt{3}\Gamma(\frac{1}{3})^6}{32\; 2^{2/3}\pi^3}$ & $\tfrac{\sqrt{3}\Gamma(\tfrac{1}{3})^6}{8\sqrt[3]{2}\pi^3}$ & $\frac{\Gamma \left(\frac{1}{4}\right)^4\Gamma \left(\frac{1}{2}\right)^2}{12 \sqrt{2} \sqrt[4]{3} \pi ^3}$ \\ \hline
\end{tabular}
\caption{\label{table:K} Summary of geometric data for K-point examples. The first three rows specify the boundary data associated to the singularity: the K3 pairing, corresponding rigid period, and extension data. The next two rows give the first few terms in the instanton expansion, separated into a common factor and remainders $\hat{B}_k$. The last two rows specify the modular form associated to the geometry and its first critical L-value. }
\end{table}

\subsubsection*{Physical couplings}
Here we discuss the physical couplings in 4d supergravity theories near a K-point. We find that log-monodromy coefficients $a,b,c$ -- or equivalently the rigid period $\tau$ by \eqref{eq:Ktau} -- encode the leading perturbative couplings of the supergravity theory. On the other hand, the extension data $\gamma,\delta$ entails more refined information such as shifts \eqref{eq:KNIJ} to the theta angles of the gauge fields or shifts to the flux quanta by \eqref{eq:Krho}.

\paragraph{K\"ahler potential and metric.} Let us begin by computing the K\"ahler potential for K-points. It admits an expansion in exponential corrections in the saxion $y$ as
\begin{equation}\label{eq:Kpotential}
    K = -\log[ K_{(0)} + e^{-2\pi y}K_{(1)} +\ldots]
\end{equation}
where the leading term and first exponential correction read
\begin{equation}
    \begin{aligned}
        K_{(0)} & = 4|c| (\tau_2)^2 y\, , \\    
        K_{(1)} & = \frac{2 |c| (\tau_2)^2 e^{-2 \pi  y} \left(B_1^2 (4 \pi  y+1)-B_2\right) \cos (2 \pi  x)}{\pi  B_1}\, . \\
    \end{aligned}
\end{equation}
The first term is the usual term for a boundary at infinite distance. Note that it is linear in $y$ in contrast to the cubic term at large complex structure. We also want to point out that the extension data $\gamma,\delta$ does not show up in the K\"ahler potential, not even in the exponential corrections. This can be explained from the fact that this information can be rotated out of the periods by an Sp$(4,\mathbb{R})$ transformation \eqref{eq:Klambda}, which acts in the same way on $\Pi$ and $\bar\Pi$ and therefore cancels out. For this K\"ahler potential we compute the metric as an exponential expansion in $y$ as
\begin{equation}\label{eq:Kmetric}
    K_{t \bar{t}} = \frac{1}{4y^2} +\frac{\left(B_1^2-B_2\right) e^{-2 \pi  y} (2 \pi  y+1) \cos (2 \pi  x)}{4 \pi  B_1 y^3}+\mathcal{O}(e^{-4\pi y})\, .
\end{equation}
The leading term differs by a factor of three compared to the large complex structure point due to the different exponent in the K\"ahler potential. Also notice that the supergravity theory has an axionic shift symmetry near the K-point, broken by the exponential correction to the K\"ahler metric, i.e.~the kinetic term for the scalar field. This complements the recent analysis of \cite{Grimm:2022xmj} into the breaking of global symmetries at such boundaries by a generalized Witten effect.
 
\paragraph{Gauge kinetic functions.} Having characterized the kinetic terms for the scalars, we next turn to the vectors of the $\mathcal{N}=2$ supergravity theory. We compute the gauge kinetic functions to be
\begin{equation}\label{eq:KNIJ}
    \mathcal{I}_{IJ} = y\begin{pmatrix}
    a & b \\
    b & c
    \end{pmatrix}\, , \qquad 
    \mathcal{R}_{IJ} = -x\begin{pmatrix}
    a & b \\
    b & c
    \end{pmatrix}+\begin{pmatrix}
    \delta & \gamma \\
    \gamma & 0 
    \end{pmatrix}\, .
\end{equation}
This highlights how closely related the form of the geometrical period data is to that of the physical couplings. Indeed, the form of the coupling matrix is given precisely by the positive-definite $2\times 2$-matrix appearing in the log-monodromy matrix. The strengths of the couplings are given by the overall axion $x$ and saxion $y$ factors. Moreover, for the theta angle couplings $\mathcal{R}_{IJ}$ we note that the extension data $\gamma,\delta$ gives an overall shift to the background values. In special examples, i.e.~those with a finite order factor in the monodromy, these take rational values $\gamma,\delta \in \mathbb{Q}$; it would be interesting to investigate the vanishing of theta angles in quantum gravity in this context, extending the results of \cite{Cecotti:2018ufg}.

\paragraph{Yukawa coupling.} We next compute the Yukawa coupling by taking three derivatives of the periods in \eqref{eq:Kperiodssummary}. Taking just the polynomial part does not suffice to this end, since the periods are at most linear in $t$. By including the first exponential correction we find
\begin{equation}\label{eq:KYukawa}
    \langle  \Pi , \partial_t^3 \Pi \rangle  = 8 \pi ^2c \tau _2^2 \big( B_1 e^{2 i \pi  t} +\mathcal{O}(e^{4\pi i t}) \big)\, .
\end{equation}
From this perspective the essential instanton $B_1 \neq 0$ guarantees that the Yukawa coupling is non-vanishing (although exponentially small) near the K-point.

\paragraph{Scalar potential.} As last physical coupling we consider the scalar potential induced by three-form fluxes arising in Type IIB orientifold compactifications. We want to set up this scalar potential in the typical bilinear form in the flux quanta. The extension data $\gamma,\delta$ will yield some off-diagonal components, but these can be conveniently absorbed into a shift of the flux quanta. To this end, let us therefore first characterize this shift, which is generated by the Sp$(4,\mathbb{R})$ transformation
\begin{equation}\label{eq:Klambda}
    \Lambda = \left(
\begin{array}{cccc}
 1 & 0 & 0 & 0 \\
 0 & 1 & 0 & 0 \\
 -\delta  & -\gamma  & 1 & 0 \\
 -\gamma  & 0 & 0 & 1 \\
\end{array}
\right)\, .
\end{equation}
In turn, we apply this basis transformation on the flux vector $G_3=(g_1,g_2,g_3,g_4)$, with $g_i \in \mathbb{Z}$. We combine it with an axion-dependent rotation in order to remove the polynomial dependence on $x$ as well (exponential corrections still contain sine dependence on $x$). This yields a vector of flux-axion polynomials
\begin{equation}\label{eq:Krho}
    \rho =  e^{xN}\Lambda^{-1} G_3 = \left(
\begin{array}{c}
 g_1 \\
 g_2 \\
 g_1 (a x+\delta )+g_2 (b x+\gamma)+g_3 \\
 g_1 (b x+\gamma )+c g_2 x+g_4 \\
\end{array}
\right)\, .
\end{equation}
With these preparations in place, we can write down the scalar potential as
\begin{equation}
    V = \bar{\rho}^T\left(
\begin{array}{cccc}
 a y & b y & i & 0 \\
 b y & c y & 0 & i \\
 -i & 0 & \frac{c}{yd} & \frac{b}{yd} \\
 0 & -i & \frac{b}{yd} & \frac{a}{yd} \\
\end{array}
\right)\rho\,,
\end{equation}
where we used the shorthand $d=ac-b^2>0$ for the determinant. We see that, similar to the gauge kinetic functions \eqref{eq:KNIJ}, the subblock of the monodromy matrix \eqref{eq:KN} again encodes the coupling of the fluxes in the scalar potential. The rotation by \eqref{eq:Krho} of the flux quanta allowed us to ignore mixings between the heavy and light fluxes ($V$ scaling as $y$ or $1/y$ respectively), apart from the $i\langle G_3, \bar{G}_3 \rangle$ which gives the present off-diagonal $2\times 2$ blocks. 

\section{Construction of periods in an integral basis}\label{sec:construction}
In this section we derive the form of the periods near all possible boundaries in one-dimensional complex structure moduli spaces. We first construct the most general set of boundary data --- the limiting mixed Hodge structure and the log-monodromy matrix --- for each of these cases. Our procedure here closely follows the constructions in \cite{KerrLMHS}. Next we turn to assembling the period vector itself. We first write down the most general near-boundary expansion by using the instanton map formulation reviewed in section \ref{ssec:Hodge2}. We then bring these periods into the usual special geometry formulation, where we write down the prepotential and characterize the model-dependent coefficients.

\subsection{Large complex structure points}\label{ssec:LCS}
In order to ease into the subject, we will first review the well-established case of the large complex structure point. The reason for this is twofold: it will serve as a reminder about the necessary steps to bring the periods at large complex structure into the widely known form; and also, it highlights what part of the period structure originates from Hodge theory and what is encoded in the underlying geometry in a familiar setting. 

\subsubsection{Boundary Data}
Here we set up the boundary data for singularities of type $\mathrm{IV}_1$ --- LCS points --- following \cite{KerrLMHS}. For instructive purposes we start from the abstract point of view of asymptotic Hodge theory, where we argue for the most general form of the boundary data, and use Sp$(4,\mathbb{Z})$ basis transformations to reduce these to the most simple form. We also comment on the quantization conditions on the model-dependent coefficients in the boundary. Finally,  we perform the match with the geometric input from the prepotential \eqref{eq:LCSprepotential} such as intersection numbers.

\paragraph{Log-monodromy matrix.} Let us begin by writing down the log-monodromy matrix $N$. Recall from section \ref{ssec:Hodge2} that we chose our basis such that $e_i \in W_{2i}$ in the monodromy weight filtration \eqref{Wfiltr}. The log-monodromy matrix acts as $NW_{2i} \subseteq W_{2(i-1)}$ on these vector spaces. Therefore, the most general form for $N$ in this integral basis is given by
\begin{align}\label{eq:LCSN}
    N= \begin{pmatrix}
    0 & 0 & 0 & 0 \\
    a & 0 & 0 & 0 \\
    e & -b & 0 & 0 \\
    f & e & -a & 0 \\
    \end{pmatrix}\, ,
\end{align}
where we already imposed the symplectic property $N^T \Sigma + \Sigma N = 0$ with the pairing \eqref{eq:pairing} (hence the parameters $a$ and $e$ appear twice). The parameters $a,b,e,f \in \mathbb{Q}$ are some model-dependent coefficients, which are constrained by: (1) the integrality of the monodromy matrix $T=e^{N} \in \text{Sp}(4,\mathbb{Z})$, and (2) the polarization condition \eqref{eq:pol}. The former yields the following quantization conditions on the model-dependent parameters
\begin{equation}\label{eq:LCSintegral}
    a,b\in \mathbb{Z}, \quad e-ab/2 \in \mathbb{Z}, \quad f-a^2b/6 \in \mathbb{Z}\, ,
\end{equation}
while the polarization condition $e_3 \Sigma N^3 e_3 >0$ tells us that $b>0$. 

\paragraph{Basis transformations.} We can also use Sp$(4,\mathbb{Z})$ basis transformations to bound some of these parameters to a finite interval. We take any basis transformation that preserves the monodromy weight filtration, which is generated by the Lie algebra element
\begin{equation}
M = \begin{pmatrix}
    0 & 0 & 0 & 0 \\
    p & 0 & 0 & 0 \\
    r & q & 0 & 0 \\
    s & r & -p & 0 \\
    \end{pmatrix}\, ,
\end{equation}
where the parameters $p,q,r,s \in \mathbb{Q}$ satisfy the analogous constraints of \eqref{eq:LCSintegral}. Redefining the log-monodromy matrix by this transformation we find that
\begin{equation}\label{eq:shifts}
    N \to e^{M} N e^{-M}: \qquad e \to e-pb -qa\, , \quad f \to f +(2r-pq)a-p(pb+2e)\, ,
\end{equation}
while $a,b$ remain invariant. These shifts can be used to restrict the domains of the parameters to the intervals
\begin{equation}\label{eq:quantLCS}
\begin{aligned}
0 \leq e < \gcd(a,b)\, , \qquad  0\leq f < 2a\, ,
\end{aligned}
\end{equation}
which can be seen as follows. For the first interval we can choose suitable values of $p,q \in \mathbb{Z}$ to restrict to the domain $e \in [0,\gcd(a,b))$. After this redefinition, we can use the second shift in \eqref{eq:shifts} with $p=q=0$ to shift $f \to f+2ra$: picking a suitable $r \in \mathbb{Z}$ then restricts us to $f \in [0,2a)$.

\paragraph{Deligne splitting.} Having characterized the form of the log-monodromy matrix $N$, we next turn to the vectors that span the spaces $I^{p,q}$ of the Deligne splitting \eqref{Ipq}. A priori, the vector space $I^{3,3}$ is spanned by any complex linear combination of the basis vectors $e_i$ as
\begin{equation}
\omega_{3,3} =  (1,\, \pi_0 , \, \pi_1, \, \pi_2 )^{\rm T}\,   , 
\end{equation}
where we chose to rescale the first entry to one. We do not write down the other vector spaces here yet, as these can be obtained by application of the log-monodromy matrix $N$. By performing a coordinate redefinition $t \to t -\pi_0/a$ we can set the second entry to zero, i.e.~$\pi_0=0$. This transformation also shifts $\pi_1,\pi_2$, but we can simply absorb these by a redefinition. The next coefficient $\pi_1$ is constrained through the transversality condition $\Pi^T \Sigma \partial_t \Pi$, which requires us to impose
\begin{equation}
\omega_{3,3}^{\rm T} \Sigma N \omega_{3,3} = f -2a \pi_1 = 0\, \quad  \implies \quad \pi_1 = f/2a\, .
\end{equation}
The last complex coefficient $\pi_2$ is left unconstrained by asymptotic Hodge theory. To match with the notation of the other examples, let us write it out as $\pi_2=\gamma+i\delta$ into real model-dependent parameters $\gamma,\delta \in \mathbb{R}$. The other vectors $\omega_{p,p} \in I^{p,p}$ can be obtained straightforwardly by applying the log-monodromy operator $N$ as a lowering operator $3-p$ times. Altogether we find that
\begin{equation}\label{eq:LCSIpq}
\omega_{3,3} =  \begin{pmatrix}
1 \\
0 \\
f/2a \\
\gamma+i\delta
\end{pmatrix}, \quad \omega_{2,2} =  \begin{pmatrix}
0 \\
1 \\
e/a \\
f/a-af/2
\end{pmatrix},\quad \omega_{1,1} =  \begin{pmatrix}
0 \\
0 \\
1 \\
0 
\end{pmatrix}, \quad \omega_{0,0} =  \begin{pmatrix}
0 \\
0 \\
0 \\
1 
\end{pmatrix}.
\end{equation}

\paragraph{Geometric input.} Having exhausted the Hodge-theoretic constraints on the model-dependent data, we now perform the match with the prepotential data in \eqref{eq:LCSprepotential}. In this setting they can be understood geometrically from the topological data of the mirror Calabi-Yau threefold. The log-monodromy matrix for the LCS prepotential reads
\begin{equation}
    N=    \begin{pmatrix}
   0 & 0 & 0 & 0 \\
   1 & 0 & 0 & 0 \\
   -\sigma & -\kappa & 0 & 0 \\
   \frac{c_2}{12} & -\sigma & -1 & 0 \\
   \end{pmatrix}\, , \label{eq:LCSNtop}
\end{equation}
while the leading term in the periods is given by
\begin{equation}
    \omega_{3,3} = (1 , \, 0 , \, c_2/24, \, \frac{i \chi \zeta(3)}{8\pi^3} )^{\rm T} \, .
\end{equation}
We can now compare this log-monodromy matrix and period vector term to the expressions \eqref{eq:LCSN} and \eqref{eq:LCSIpq} we derived from asymptotic Hodge theory. Matching these results allows us to identify the model-dependent parameters as
\begin{equation}\label{eq:LCSmatch}
    a=1, \quad b=\kappa , \quad e= \sigma , \quad f = -c_2/12, \quad \gamma=0, \quad \delta =  \frac{ \chi \zeta(3)}{8\pi^3} \, .
\end{equation}
In particular, geometry thus imposes the stronger constraints $a=1$ and $\gamma=0$, while it also fixes the transcendentality of $\delta \in \mathbb{R}$ to be given by $\zeta(3)/\pi^3$. It is amusing to recall the quantization conditions \eqref{eq:LCSintegral} and domains \eqref{eq:quantLCS} we derived for the Hodge-theoretic parameters, and compare with the geometric data:
\begin{itemize}
\item First consider the parameter $e$: for $a=1$ its domain reduces to $0 \leq e < 1$, with the quantization condition $e-b/2 \in \mathbb{Z}$. This tells us that $e= b/2 \mod 1$, which is precisely the constraint that arises in mirror symmetry as $\sigma = \kappa/2 \mod 1$. 
\item Next consider the parameter $f$: for $a=1$ the Hodge-theoretic quantization constraints tell us that $f-b/6 \in \mathbb{Z}$, which in the geometric context reproduces the condition $c_2-\kappa/6 \in \mathbb{Z}$. 
\end{itemize}

\subsubsection{Periods and prepotential}
In this section we derive the form of the periods at large complex structure using the instanton map formulation reviewed in section \ref{ssec:Hodge2}. In particular, we show how this Hodge theoretic formulation relates to the prepotential formulation more familiar from the physics literature. For the sake of simplicity we take only the intersection number $\kappa$ into account, and set all other extension data to zero.\footnote{This may be achieved for instance by applying the basis transformation \eqref{eq:LCSlambda}, although we note that removing $\chi$ requires one to work over a complex rather than real basis.}

\paragraph{Instanton map formulation.} Let us begin by writing down the most general instanton map $\Gamma(z)$ for \ref{eq:instantonmap}. Recall that it must act on the Deligne splitting according to \eqref{eq:gamma-action} by lowering the first index of the vector spaces $I^{p,q}$. For the case of large complex structure points this simply means it must map $I^{p,p}$ into the spaces $I^{q,q}$ with $p>q$. In other words, it is simply a lower-triangular matrix with holomorphic coefficients given by
\begin{equation}\label{eq:instantonmapLCS}
\Gamma(z) = \begin{pmatrix}
0 & 0 & 0 & 0 \\
n(z) & 0 & 0 & 0 \\
B(z) & A(z)-\kappa n(z) & 0 & 0 \\
C(z) & -B(z) & -n(z) & 0 
\end{pmatrix}\, ,
\end{equation}
where we identified certain components in order to ensure a $\mathfrak{sp}(4)$-valued map. For later purposes it is useful to identify the holomorphic functions in this expression with the degree to which they lower the Deligne splitting following \eqref{eq:Gammadecomp}: the $\Gamma_{-1}$ part of the instanton map is comprised of the functions $A(z), n(z)$, the $\Gamma_{-2}$ part by $B(z)$, and $\Gamma_{-3}$ by $C(z)$. Griffith's transversality of the periods implies that the instanton map satisfies the conditions \eqref{eq:recursiongamma}, which reduces to differential relations on the functions given by
\begin{equation}\label{eq:LCStrans}
\begin{aligned}
    A(z)(1+2i\pi z \, n'(z) )&= \, 2\pi i z \, B'(z)+\pi i z \, \partial_z (n(z)A(z))\, ,\\ 
    6 B(z) (1+2i\pi z \, n'(z) ) &= i\pi z \, n(z) \big(n(z) A'(z)-n'(z) A(z) +B'(z)\big)+6\pi z\,  C'(z)\, ,
\end{aligned}
\end{equation}
Note that the factors of $1+2i\pi z\, n'(z)$ on the left-hand side can be understood as simply the Jacobian  factor $\partial_t \big(t+n(e^{2\pi it}) \big)$ for the derivatives on the right-hand sides. As we will see later, this corresponds precisely to the natural coordinate we introduce in \eqref{tildet} (the mirror map). For now let us write down the periods in terms of a generic coordinate, in which they can be expressed in terms of the functions in \eqref{eq:instantonmapLCS} as
\begin{equation}\label{eq:LCSGammaperiods}
    \Pi = \begin{pmatrix}
        1\\
        t+n(z) \\
        -\frac{1}{2}\kappa t^2 -\kappa t n(z)   +B(z) - \frac{1}{2} n(z) A(z) \\
        \frac{1}{6} \kappa t^3 +\frac{1}{2}\kappa t^2   n(z)- t B(z) +C(z) +\frac{1}{6} n^2(z) A(z)
    \end{pmatrix}.
\end{equation}
This expression together with the differential relations \eqref{eq:LCStrans} constitute the most general set of periods for a one-modulus large complex structure point. In the following we simplify these expressions by switching to a natural coordinate, which turns out to be the familiar mirror map. This helps us in elucidating the physical meaning of the quantities that appear in the instanton map formulation.

\paragraph{Natural coordinate.} Let us now first look at the role of the function $n(z)$: upon plugging the instanton map \eqref{eq:instantonmapLCS} into the expression for the periods \eqref{eq:instantonmap}, we find that the action on $\mathbf{a}_0 = (1,0,0,0)$ simply alters the second period from $t$ to $t+n(e^{2\pi i t})$ in \eqref{eq:LCSGammaperiods}. In this sense a different choice of $n(e^{2\pi i t})$ represents the freedom to change coordinates. We can therefore choose to set $n(z)$ to zero by redefining
\begin{equation}\label{tildet}
   \tilde{t}(t) = t+n(e^{2\pi i t})\, .
\end{equation}
Relating to a more familiar viewpoint from the physics literature, we can identify $\tilde{t}(t)$ with the mirror map, from which we find that $n(e^{2\pi i t})$ simply encodes the instanton expansion for the K\"ahler modulus. Let us denote the coordinate in which the LCS point is located at the origin by $q$ as
\begin{equation}
q(t) = \exp[2\pi i \tilde{t}(t)]  = \exp[2\pi i t + n(e^{2\pi i t})]\, ,
\end{equation}
Setting $n(e^{2\pi i t})=0$ by performing this change of coordinates, we next write down the differential equations relating $B(q)$ and $C(q)$ to $A(q)$. We find that \eqref{eq:LCStrans} reduces to
\begin{equation}\label{eq:LCStrans2}
    A(q) = 2\pi i q B'(q)\, , \quad B(q) = i \pi q C'(q)\, .
\end{equation}
where the primes denote derivatives with respect to $q$. In other words, we can write all functions in the instanton map in terms of derivatives of $C(q)$. Or vice versa, the function $A(q)$ determines the instanton map \eqref{eq:instantonmapLCS} completely after switching to the natural coordinate that sets $n(q)=0$. Either way, it allows us to rewrite the periods as
\begin{equation}
    \Pi = \begin{pmatrix}
    1 \\
    t \\
    \frac{1}{2}\kappa t^2 + \frac{1}{2}\partial_t  C(e^{2\pi i t}) \\ \frac{1}{6} \kappa t^3 + \frac{1}{2}t \partial_t C(e^{2\pi i t}) +C(e^{2\pi i t})
    \end{pmatrix}  \, ,
\end{equation}
where we dropped the tilde on $t$ for convenience.

\paragraph{Prepotential formulation.} The form of the derivatives here takes precisely the form needed to rewrite in terms of a prepotential. The periods then read
\begin{equation}
    \Pi = \begin{pmatrix}
    1\\
    t\\
    \partial_t F_{\rm LCS}\\
    2F_{\rm LCS}-t \partial_t F_{\rm LCS}
    \end{pmatrix},
\end{equation}
with the LCS prepotential is given by\footnote{Moreover, we can include the $\alpha'$-corrections by rotating to the Sp$(4,\mathbb{Z})$ basis by the rotation \eqref{eq:LCSlambda}. We will not bother to write down these additional polynomial terms here, but simply refer to the expression \eqref{eq:LCSprepotential}.}
\begin{equation}
F_{\rm LCS}(t) = \frac{1}{6}\kappa t^3 + \frac{1}{2}C(e^{2\pi i t})\, .
\end{equation}
We thus see that the other unconstrained functional degree of freedom in the instanton map encodes the instanton expansion of the prepotential, hence justifying the nomenclature.

In principle one can simply include the polynomial corrections to the periods by turning on the appropriate terms as given in \eqref{eq:LCSprepotential}. For our derivation here these are not as important, and the main takeaway message is that the two functional degrees of freedom from a Hodge-theoretic perspective are precisely the ones familiar from the physics literature: the mirror map and the instanton series in the prepotential.

\paragraph{Yukawa coupling.} To close of our discussion on the large complex structure periods, let us compute the corresponding Yukawa coupling. This allows us to elucidate further the physical meaning of the functional degree of freedom $C(q)$ in the instanton map \eqref{eq:instantonmapLCS}. We find by using the differential relations \eqref{eq:LCStrans2} that 
\begin{equation}
\langle   \Pi, \, (2\pi i q \partial_q)^3\Pi \rangle = \kappa -(2\pi i q \partial_q)^3 C(q) \, ,
\end{equation}
Thus the function $C(q)$ in the instanton map, or equivalently the function $A(q)$ through \eqref{eq:LCStrans2}, encodes the instanton corrections to the intersection number $\kappa$ in the Yukawa coupling.

\subsection{Conifold points}\label{ssec:conifoldconstruction}
In this section we derive the general form of the period vector for conifold points. Such singularities lie at finite distance in the K\"ahler metric and occur for instance in the moduli space of the mirror quintic \cite{Candelas:1990rm}. 

\subsubsection{Boundary data}
In this subsection we write down the most general boundary data --- the log-monodromy matrix and limiting mixed Hodge structure --- in an integral symplectic basis for one-parameter conifold points in complex structure moduli space. This data has already been constructed in \cite{KerrLMHS}, but we find it nevertheless instructive to briefly repeat their procedure here for completeness. 

\paragraph{Log-monodromy matrix.} First recall the integral basis we defined in section \ref{ssec:Hodge2} for conifold points: we chose basis vectors such that $e_3 \in W_2$, $e_1,e_2 \in W_3$ and $e_0 \in W_4$, pairing up $(e_3,e_0)$ and $(e_1,e_2)$ as described by the symplectic pairing \eqref{eq:pairing}. The action of the log-monodromy matrix is then described by $N W_4 \subseteq W_2$, which means that $N$ simply sends $e_0$ to a multiple of $e_3$ as $N e_0 = k e_3$. The log-monodromy matrix can thus be written out as
\begin{align}
N = \begin{pmatrix}
0 & 0 & 0 & 0 \\
0 & 0 & 0 & 0 \\
0 & 0 & 0 & 0 \\
-k & 0 & 0 & 0 
\end{pmatrix} \,, \label{Eq:ConiN}
\end{align}
with $k\in \mathbb{N}$ strictly positive, which follows from \eqref{eq:pol} as polarization condition $e_0\Sigma Ne_0>0$. From the analytic perspective this simple form of the log-monodromy matrix can be understood from the fact that the Picard-Fuchs operator has a single log-solution.

\paragraph{Deligne splitting.} As next step we write down the most general Deligne splitting \eqref{Ipq} for conifold singularities. For $\omega_{3,0} \in I^{3,0}$ we know that $I^{3,0} \subseteq W_3$, so this vector is given by a complex linear combination of $e_1,e_2,e_3 \in W_3$. On the other hand, for $\omega_{2,2} \in I^{2,2} \subseteq W_4$ we start from any complex linear combination of the four basis vectors. By normalizing the vectors suitably we find as starting point
\begin{equation}\label{eq:mhsI1start}
\omega_{3,0} = (0,1,\pi_1 , \pi_2)^{\rm T}\, , \quad \omega_{2,2} = (1,\pi_3, \pi_4, \pi_5)^{\rm T}\, .
\end{equation}
For the other vector spaces we have simply $\omega_{1,1}=e_3 \in I^{1,1}$ and $\omega_{0,3} = \bar{\omega}_{3,0} \in I^{0,3}$.

\paragraph{Transversality constraints.} Let us now try to constrain the parameters appearing in these vectors. The complex parameter $\pi_1$ in $\omega_{3,0}$ will turn out to be left unconstrained, so we set it to $\pi_1=\tau$. For the vector $\omega_{2,2}$ we must impose that it is real, up to an imaginary piece along  $e_3$, i.e.~the last component.\footnote{This follows from the fact that the Deligne splitting satisfies the complex conjugation rules $\bar{I}^{q,p} = I^{p,q} \mod_{r<p,s<q} I^{r,s}$, which in our case simply reads $\bar{I}^{2,2}=I^{2,2} \mod I^{1,1}$.} This means we must set $\pi_3 = \gamma$ and $\pi_4 = \delta$ with real model-dependent parameters $\gamma,\delta \in \mathbb{R}$. After these redefinitions it is a straightforward check that we now have the correct behavior under complex conjugation. By performing a coordinate transformation  $t \to t-\pi_5/k $ we find that $\omega_{2,2}$ transforms as 
\begin{equation}
\omega_{2,2} \to e^{-\pi_5/k N} \omega_{2,2} = (1,\gamma,\delta,0)\, ,
\end{equation}
thereby effectively setting $\pi_5 = 0$. This puts our LMHS in the $\mathbb{R}$-split, as $\omega_{2,2}$ is real now and coordinate transformations are the only possible deformations for conifold points, cf.~\cite{Bastian:2021eom, Grimm:2021ikg}. Next we turn to the polarization conditions \eqref{eq:pol}, which here demand an orthogonality condition between $\omega_{3,0}$ and $\omega_{2,2}$ that reads
\begin{equation}
    \omega_{2,2}^T \Sigma \omega_{3,0} = \pi_2 -\gamma +\delta \tau = 0\, .
\end{equation}
This requires us to fix $\pi_2 = \gamma-\delta \tau$. Altogether the above constraints exhaust the conditions we can impose on the parameters in the limiting mixed Hodge structure. To sum up, we find that the spanning vectors are given by
\begin{equation}\label{eq:mhsI1}
\omega_{3,0} = (0,\, 1,\, \tau,\,  \delta - \gamma \tau)^{\rm T}\, , \quad \omega_{2,2} = (1,\, \gamma , \, \delta,\,  0)^{\rm T}\, ,
\end{equation}
with $\tau \in \mathbb{C}$ and $\gamma,\delta \in \mathbb{R}$, and $\omega_{1,1}$ and $\omega_{0,3}$ given as described below \eqref{eq:mhsI1start}.

\paragraph{Basis transformations and coordinate redefinitions.} Next we characterize how we can shift the model-dependent parameters by coordinate redefinitions and symplectic basis transformations that preserve the weight filtration. Together these transformations are generated by the matrices
\begin{align}
M_1=\begin{pmatrix}
1 & 0 & 0 & 0 \\
0 & c_{11} & c_{12} & 0 \\
0 & c_{21} & c_{22} & 0 \\
0 & 0& 0 &1
\end{pmatrix}  , \qquad  M_2 = \begin{pmatrix}
1 & 0 & 0 & 0 \\
b_1 & 1 & 0 & 0 \\
b_2 & 0 & 1& 0 \\
b_1 \delta -b_2 \gamma & b_2 & -b_1 &1
\end{pmatrix} ,\label{Eq:ExtShift}
\end{align}
where $  c_{ij} \in \text{SL}(2,\mathbb{Z})$ and $b_i \in \mathbb{Z}$ for these transformations. The left-bottom component of $M_2$ corresponds to a coordinate shift $t \to t -(b_1\gamma-b_2 \delta)/k$, which is not quantized since $\gamma,\delta \in \mathbb{R}$ need not be; this shift has been included in to ensure that the last component of $\omega_{2,2}$ in \eqref{eq:mhsI1} remains zero. Also note that the log-monodromy matrix $N$ is invariant under these transformations $N \to M_i N M_i^{-1}$, and therefore we indeed leave the monodromy weight filtration unchanged. The first matrix can be thought of as generating $\text{SL}(2,\mathbb{Z})$ transformations on the vector $(1,\tau)^{\rm T}$, thereby bringing $\tau$ to the fundamental domain. 
The second matrix $M_2$ can be understood as a shift of the extension data parameters $\gamma,\delta$ by
\begin{align}\label{eq:Cshift}
    \gamma \mapsto \gamma + b_1 \,, \qquad \delta \mapsto \delta + b_2 \,,
\end{align}
which can be read off from the action of $M_2$ on the vector $\omega_{2,2}$ in \eqref{eq:mhsI1}. Thus from \eqref{eq:Cshift} we deduce that we can always bring these parameters into the interval $\delta,\gamma \in [0,1)$ by such shifts.

\subsubsection{Periods and prepotential}
In this section we derive the form of the periods at the conifold point, both in the Hodge-theoretic as well as the prepotential formulation. 

\paragraph{Instanton map formulation.} Let us begin by writing down the most general instanton map according to the properties described in section \ref{ssec:Hodge2}. For ease of presentation we work in the complex Hodge basis for the majority of our analysis. This basis is given by the vectors spanning \eqref{eq:mhsI1} as
\begin{equation}\label{eq:Cbasis}
m^{\rm LMHS} = \big( \omega_{2,2}, \, , \omega_{3,0} , \, \omega_{0,3} , \, \frac{1}{k} N\omega_{2,2} \big) = \begin{pmatrix}
    1 & 0 & 0 & 0 \\
    \gamma & 1 & 1 & 0 \\
    \delta & \tau & \bar\tau& 0 \\
    0 & \delta-\gamma \tau & \delta-\gamma \bar\tau & 1
\end{pmatrix}
\end{equation}
Performing rotations by the transition matrix $m^{\rm LMHS}$ and its inverse allows us to switch between this complex Hodge basis and the integral basis. For instance, by $N \to m^{\rm LMHS} N (m^{\rm LMHS})^{-1}$ we rotate the log-monodromy matrix from the Hodge to the integral basis. We use this complex basis up to equation \eqref{eq:ConifoldPeriod}, where we rotate our results back to the integral basis. For now, let us write down the most general instanton map in the complex Hodge basis \eqref{eq:Cbasis} as
\begin{equation}\label{eq:instantonmapconifold}
\Gamma(z) = m^{\rm LMHS}\left(
\begin{array}{cccc}
 0 & A(z) & 0 & 0 \\
 0 & 0 & 0 & 0 \\
 \frac{i B(z)}{2 \tau _2} & C(z) & 0 & -\frac{i A(z)}{2 \tau _2} \\
n(z) & B(z) & 0 & 0 \\
\end{array} 
\right) (m^{\rm LMHS})^{-1}\, ,
\end{equation}
where $A(z),B(z),C(z),n(z)$ are holomorphic functions that vanish at $z=0$. Note that the functions $A(z)$ and $B(z)$ appear twice, which ensures that $\Gamma(z)$ is an infinitesimal isomorphism of the symplectic pairing. One straightforwardly checks that this matrix indeed acts as $\Gamma(z) I^{p,q} \subseteq \oplus_{r<p-1} \oplus_{s} I^{r,s}$ in the basis \eqref{eq:Cbasis}. Moreover, according to the charge decomposition \eqref{eq:Gammadecomp} note that $A,n$ make up $\Gamma_{-1}$, $B$ gives $\Gamma_{-2}$ and $C$ gives $\Gamma_{-3}$. These holomorphic functions must further satisfy the differential equations \eqref{eq:recursiongamma}, which we find to be
\begin{equation}\label{eq:Ctrans}
\begin{aligned}
     A(z) \big(k+2\pi i z n'(z) \big) &= 2\pi i z \big( \partial_z(n(z) A(z))-2 B'(z) \big)  \, , \\
    6i B(z) A'(z) - 12 \tau_2 C'(z) &=-i A(z)(-n(z) A'(z)+A(z) n'(z))\, .
\end{aligned}
\end{equation}
In the first equation $k+2\pi i n'(z)$ can be understood as the Jacobian $\partial_t (kt+n(e^{2\pi i t}))$, corresponding precisely to the natural coordinate we introduce later through \eqref{eq:Cnatural}. By acting with the instanton map on the vector $\omega_{3,0}$ in \eqref{eq:mhsI1} we obtain the period vector as
\begin{equation}\label{eq:Cperiod}
    (m^{\rm LMHS})^{-1} \cdot  \Pi =  \begin{pmatrix}
    A(z) \\
    1 \\
    C(z) + \frac{i}{12\tau_2}A^2(z) n(z) \\
    A(z) \big(k t+\frac{1}{2} n(z)\big)+ B(z) \big)
    \end{pmatrix}.
\end{equation}
This period vector together with the differential relations \eqref{eq:Ctrans} constitutes the most general set of periods for one-modulus conifold points in the complex Hodge basis. In the following we simplify these expressions by introducing a natural coordinate, which elucidates the relation between this instanton map formulation and physical observables such as the Yukawa coupling and the prepotential.

\paragraph{Natural coordinate and Yukawa coupling.} Recall from the discussion in section \ref{ssec:Hodge2} that the piece of the instanton map \eqref{eq:instantonmapconifold} along $N$, i.e.~the function $n(z)$, can be set to zero by a coordinate redefinition. We can read off this function explicitly from the periods by using the first equation in the differential relations \eqref{eq:Ctrans} as
\begin{equation}\label{eq:Cnatural}
  n(z) = \frac{\partial_z ((m^{\rm LMHS})^{-1} \cdot\Pi)_4}{\partial_z ((m^{\rm LMHS})^{-1} \cdot\Pi)_1} - \frac{k \log(z)}{2 \pi i}\, .
\end{equation}
By redefining the coordinate as $z \to z \exp[2\pi i n(z)]$ we can effectively set $n(z)=0$. Denoting this new coordinate by $q$, the transversality conditions \eqref{eq:Ctrans} reduce to
\begin{equation}\label{eq:Ctrans2}
2\pi i q  B'(q) = -k A(q)\, , \qquad  B(q) A'(q) = -2i \tau_2 C'(q)\, .
\end{equation}
Using this coordinate, and rotating back to the integral monodromy weight frame by the basis transformation \eqref{eq:Cbasis}, we find as periods
\begin{equation}\label{eq:ConifoldPeriod}
    \Pi = \left(
\begin{array}{c}
 A(z) \\
 1+\gamma A(z)+C(z) \\
 \tau+\delta A(z)+\bar{\tau} C(z)  \\
\delta-\gamma \tau +A(z)  k \frac{\log (z)}{2\pi i}+ B(z)+(\delta-\gamma \bar{\tau} ) C(z) \\
\end{array}
\right).
\end{equation}
We now take these periods in the natural coordinate and compute the Yukawa coupling. By using the differential relations \eqref{eq:Ctrans2} we find that
\begin{equation}\label{eq:CYukawa}
\langle \Pi , \, (2\pi i\partial_q)^3 \Pi \rangle  = k  \left(2\pi i q\,  \partial_q A(q) \right)^2\, .
\end{equation}

\paragraph{Asymptotic periods.} We can now derive asymptotic models for the conifold periods by writing down a series expansion for the functions in the instanton map. This amounts to solving the differential relations \eqref{eq:Ctrans2} for $B(q),C(q)$ for a series ansatz for $A(q)$. We find that
\begin{equation}\label{eq:Cansatz}
\begin{aligned}
    A(q) &= \sum_{l=1}^\infty A_l q^l= A_1 q+ A_2 q^2 +A_3 q^3+A_4 q^4\ldots\, , \\
    \quad B(q) &= -\frac{k}{2\pi i} \sum_{l=1}^\infty \frac{1}{l}A_l q^l = -\frac{k}{2\pi i} ( A_1 q +\frac{1}{2} A_2 q^2 + \frac{1}{3} A_3 q^3 +\frac{1}{4} A_4 q^4+\ldots ) \, , \\
    C(q) &= -\frac{k}{8\pi \tau_2} \big( (A_1)^2 q^2 +\tfrac{5A_1 A_2}{3}  q^3 + \tfrac{3(A_2)^2+10 A_1 A_3}{6} q^4 + \tfrac{26 A_2 A_3+51 A_1 A_4}{30} q^5+\ldots \big)\, ,
\end{aligned}
\end{equation}
where we did not write down a general form for $C(q)$ since its derivative is equal to the product of two functions. The periods up to second order  then take the form
\begin{equation}\label{eq:Cpiasymp}
\Pi = \begin{pmatrix}
A_1 q + A_2 q^2 \\
1+ \gamma(A_1 q + A_2 q^2 ) - \frac{k}{8\pi \tau_2} (A_1)^2 q^2 \\
\tau+ \delta(A_1 q + A_2 q^2 ) - 
\frac{k\bar\tau}{8\pi \tau_2} (A_1)^2 q^2 \\
\delta-\gamma \tau + (A_1 q + A_2 q^2) k \frac{\log[z]}{2\pi i}-\frac{k}{2\pi i}(A_1 z+\frac{1}{2} A_2 q^2) + (\delta-\gamma \bar\tau) (A_1)^2 q^2
\end{pmatrix}+ \mathcal{O}(q^3)\, ,
\end{equation}
where we note that terms proportional to $A_1 A_2 q^3$ and $(A_2 q^2)^2$ have been dropped, but would be needed in order to satisfy the transversality conditions at subleading orders.

\paragraph{Prepotential coordinate.} We now want to recast these periods in terms of the usual prepotential formulation. A crucial difference to the LCS case is that the natural coordinate $q$, i.e.~the one in which the Yukawa coupling takes the simple form \eqref{eq:CYukawa}, does not correspond to the prepotential coordinate. For this formulation we introduce yet another coordinate
\begin{align}\label{eq:Cs}
    s(q)=\frac{A(q)}{1+\gamma  A(q)+C(q)} \,.
\end{align}
Similar to the mirror map at LCS, we now have to invert this relation order-by-order in the series expansion to solve for $q(s)$. We can take the series expansions for $A(q)$ and $C(q)$ in \eqref{eq:Cansatz} and invert the relation as a series in $s$ as
\begin{equation}\label{eq:Cz}
q(s) = \tfrac{1}{A_1} s +\tfrac{-A_2+(A_1)^2\gamma}{(A_1)^3}s^2+ \tfrac{16 (A_2)^2-8 A_1 A_3-16 A_1^2 A_2 \gamma+ A_1^4 (-\frac{k}{\pi \tau_2}+8\gamma^2 )}{8 (A_1)^5}s^3 +\mathcal{O}(s^4) \, .
\end{equation}

\paragraph{Prepotential.} Having established a coordinate for the prepotential with \eqref{eq:Cz}, we next turn to the prepotential itself. In order to get the periods in this frame we need to perform a rescaling and swap the first two and last two periods. From the periods in \eqref{eq:Cpiasymp} we then find
\begin{align}
    \Pi(s)= \begin{pmatrix}
    1 \\ s \\ 2 \, F_c - s \, \partial_s F_c \\ \partial_s F_c
    \end{pmatrix},
\end{align}
where the conifold prepotential is given by
\begin{equation}\label{eq:CF1}
    F_c = \frac{1}{2} \bar\tau + s\big(\delta-\gamma \bar \tau +\frac{i\tau_2}{A(q(s))}\big)+\frac{1}{2} s^2 \big(-\gamma (\delta-\gamma\bar\tau) + \frac{B(q(s))-2i \tau_2 \gamma}{A(q(s))} +\frac{k \log[q(s)]}{2\pi i} \big)\, ,
\end{equation}
where we chose to eliminate $C(q)$ by \eqref{eq:Cs}. Next we plug in the series solutions \eqref{eq:Cansatz} for $A,B$ and the expansion \eqref{eq:Cz} for $q(s)$ in terms of $s$.  Altogether we find that these expansions up to order $s^3$ yield\footnote{In particular, note that $1/A(q(s))$ scales asymptotically as $1/s$, so these denominators alter some of the lower-order terms in $s$ in \eqref{eq:CF1}, for instance shifting some factors of $\bar\tau$ to $\tau$.}
\begin{align}\label{eq:prepotentialconifold}
    F_c= \ &\frac{\tau}{2}+s(\delta-\gamma\tau)  + s^2\bigg( \frac{ k \log[s]}{4 \pi i} -\frac{k(3+2\log[A_1])}{8\pi i}-\frac{1}{2}\gamma (\delta-\gamma\tau) \bigg) \nonumber\\
    & - \frac{i k}{12 \pi A_1^2}(3 \gamma A_1^2-A_2)s^3 + \mathcal{O}(s^4)\, .
\end{align}
There are a few comments to be made about the form of this prepotential:
\begin{itemize}
\item Let us begin with the first two terms. The constant term involves the rigid period $\tau$ whose imaginary part is always strictly positive. On the contrary, the linear term arises purely due to the extension data parameters $\delta,\gamma$, and would not be there in their absence. Thus the linear term thereby carries information about the integral basis of the periods, as that is the role of the extension data. 
\item Next we move on to the terms at order $s^2$. The prefactor of the logarithmic term in $s$ is precisely given by the integer $k$ from  the monodromy matrix \eqref{Eq:ConiN}, as required for the quantization of the monodromies under $s\to e^{2\pi i}s$. On the other hand, the other terms at order $s^2$ are generically transcendental numbers involving logarithms and such.
\item Finally, the terms at higher orders in $s$ have only constant coefficients. In particular, note that no logarithmic terms in $s$ appear. This agrees with the fact that monodromies induce shifts by the second period, which is equal to $s$ and contains no higher order terms. Also note that these terms simplify when the extension data is dropped $\gamma,\delta=0$, i.e.~when examples where $\gamma,\delta=0$, or if one chooses to rotate to a convenient real basis by \eqref{eq:Crot}.
\item The expansion of the prepotential at higher orders in $s$ is parameterized by coefficients $A_l$. The first coefficient $A_1$ must be non-vanishing, as it corresponds to an essential exponential correction in $t=\log[q]/2\pi i$ to the periods in \eqref{eq:Cpiasymp}. On the other hand, the other coefficients $A_{l\geq 2}$ only correspond to subleading corrections to the physical couplings that are inessential.
\end{itemize}

\subsection{K-points}\label{ssec:Kconstruction}
In this section we derive the general form of the period vector for K-points, i.e.~one-modulus Type $\mathrm{II}_0$ singularities. These singularities lie at infinite distance in the K\"ahler metric and occur in the moduli spaces of some of the hypergeometric families in table \ref{table:hypergeom}. As before we begin from a Hodge-theoretic perspective, where we construct the most general form of the boundary data including quantization conditions by building upon the work in \cite{KerrLMHS}. We then write down the most general form of the exponential series, and subsequently bring these periods into the prepotential formulation. We close off with some remarks on the geometrical origin of the model-dependent data appearing in these periods.

\subsubsection{Boundary data}\label{II0:LMHS}
In this subsection we write down the most general form for the boundary data -- the log-monodromy matrix and the limiting mixed Hodge structure -- in an integral symplectic basis for one-modulus K-points. We note that in \cite{KerrLMHS} this data has already been constructed in the case that the log-monodromy matrix takes a simpler form (containing a diagonal sub-block), and here we extend this result to the general case (allowing any positive-definite sub-block).

\paragraph{Log-monodromy matrix.} We begin by writing down the log-monodromy matrix $N$. Recall from section \ref{ssec:Hodge2} that we chose basis vectors $e_2,e_3 \in W_2$ and $e_0,e_1 \in W_4$. The action of $N$ is summarized by $NW_4\subseteq W_2$, which means that $N$ sends $e_0$ and $e_1$ into linear combinations of $e_2$ and $e_3$. The log-monodromy matrix can thus be written out as
\begin{align}\label{Eq:IIN}
N = \begin{pmatrix}
0 & 0 \\ 
B & 0
\end{pmatrix} =
\begin{pmatrix}
0 & 0 & 0 & 0 \\
0 & 0 & 0 & 0 \\
a & b & 0 & 0 \\
b & c & 0 & 0 
\end{pmatrix} \,, 
\end{align}
where $a,b,c \in \mathbb{Z}$.
The subblock $B$ is symmetric because the log-monodromy matrix needs to be an infinitesimal isomorphism of the symplectic pairing \eqref{eq:pairing}, i.e.~$N^T \Sigma+\Sigma N = 0$. Moreover, it has to be a positive-definite matrix by the polarization conditions\footnote{To be precise, we must impose that $\Sigma N$ is a positive-definite matrix on the subspace spanned by $e_0$ and $e_1$.}, implying in particular a positive discriminant $d=ac-b^2>0$. 

\paragraph{Deligne splitting.} Next we write down the most general Deligne splitting \eqref{Ipq} for K-points. For $\omega_{3,1} \in I^{3,1}$ we have that $I^{3,1} \subseteq W_4$, so we can rescale the first entry to one, but are otherwise left with 3 arbitrary complex coefficients. On the other hand, for the basis vector $\omega_{2,0} \in I^{2,0}$ we know that $I^{2,0} \subseteq W_2$, so it is a linear combination of $e_2$ and $e_3$. Altogether this means we can write down as starting point
\begin{align}\label{eq:Kmhs}
\omega_{3,1}= \begin{pmatrix}
1 \\ \tau\\ \pi_0 \\  \pi_1 
\end{pmatrix} , \qquad \omega_{2,0} = \begin{pmatrix}
0 \\ 0 \\ \pi_2 \\ 1
\end{pmatrix} ,
\end{align}
with $\tau, \pi_0 ,\pi_1,\pi_2 \in \mathbb{C}$. 
We now constrain these coefficients by using the conditions imposed by the limiting mixed Hodge structure. We begin with the transversality condition that $\omega_{3,1}$ and $\omega_{2,0}$ should be orthogonal to each other under the symplectic pairing, which yields
\begin{equation}\label{eq:Kdelta}
    \omega_{3,1} \Sigma \omega_{2,0} =\pi_2+\tau = 0 \quad \implies \quad \pi_2=-\tau\, .
\end{equation}
Next we constrain the parameter $\tau$ itself by the requiring $\omega_{3,1}$ and $N \omega_{3,1}$ to be orthogonal under the symplectic pairing. Equivalently, this constraint is obtained by demanding $N I^{3,1} = I^{2,0}$, i.e.~parallel $N \omega_{3,1}$ and $\omega_{2,0}$. In either case, the resulting constraint reads
\begin{equation}\label{eq:Ktau2}
    c\tau^2 +2b \tau +a =0 \,  \quad \implies \quad \tau = \frac{-b \pm i\sqrt{d}}{c}\, , 
\end{equation}
where we used again the shorthand $d=ac-b^2>0$ for the determinant of $B$. Note that there is a choice of sign here, which we fix by demanding $\text{Im}(\tau)>0$.

\paragraph{Basis transformations.} The above constraints on the parameters constitute all constraints imposed by the limiting mixed Hodge structure. We may, however, still use coordinate redefinitions and basis transformations to fix some of the redundancies in this description. Let us begin with the basis transformations, for which we take integral symplectic matrices that preserve the weight filtration, i.e.~the fact that $e_2,e_3 \in W_2$. This is achieved by setting its top-right $2\times 2$-subblock to zero. There are two different sorts of transformations, given by
\begin{equation}\label{eq:basischangeII}
M_1 = \begin{pmatrix}
 C & 0 \\
 0 & C^{-\rm T}
\end{pmatrix}, \qquad M_2 = \begin{pmatrix}
1 & 0 \\
D & 1 
\end{pmatrix} = \begin{pmatrix}
1 & 0 & 0 & 0 \\
0 & 1 & 0 & 0 \\
d_1 & d_2 & 1 & 0 \\
d_2 & d_3 & 0 & 1 
\end{pmatrix}\, .
\end{equation}
where $C,D$ are integer matrices, with $\det C=\pm 1$ to ensure integrality of $M_1$, and $D=D^T$ to ensure the symplectic property of $M_2$. The matrix $M_1$ can be interpreted as a transformation of the period vector $(1,\tau)$ by $C \in \text{SL}(2,\mathbb{Z})$. Therefore we assume in the following that $M_1$ has readily been used to bring the block $B$ in the log-monodromy matrix \eqref{Eq:IIN} and $\tau$ in \eqref{eq:Ktau2} to a convenient form (e.g.~restrict $\tau$ to the fundamental domain). On the other hand, the matrix $M_2$ can be used to shift the parameters $\pi_1,\pi_2 \in \mathbb{C}$. A convenient way to parametrize these is given by
\begin{equation}\label{eq:alphabeta}
\pi_0 = \delta + \tau \gamma\, , \qquad \pi_1 = \alpha + \tau \beta \, ,
\end{equation}
with $\alpha,\beta,\gamma,\delta \in \mathbb{R}$. The matrix $M_2$ then simply shifts these parameters under $\omega_{3,1} \to M_2 \, \omega_{3,1}$ by
\begin{equation}\label{eq:alphabetashift}
    \delta \to \delta+d_1\, , \quad \gamma \to \gamma+d_2\, , \quad \alpha \to \alpha+d_2 \, , \quad \beta \to \beta+d_3\, .
\end{equation}
This allows us to bring the parameters $\beta,\delta$ to the finite intervals $[0,1)$. On the other hand, the parameters $\alpha,\gamma$ are shifted by the same integer $d_2$, so in general only one of these can be brought to this interval. We will see shortly, however, that by using coordinate redefinitions we can restrict both of these remaining parameters.

\paragraph{Coordinate redefinitions.} We next use coordinate redefinitions to reduce the number of arbitrary parameters in the boundary data. Let us first characterize how such a redefinition shifts the parameters in the limiting mixed Hodge structure \eqref{eq:Kmhs}. Under a shift of the coordinate $t$ we find that $\omega_{3,1}$ changes by a piece along $\omega_{2,0}$ as
\begin{equation}
t \to t+\frac{\lambda}{b+c \tau}: \qquad   \omega_{3,1} \to \omega_{3,1}+\frac{\lambda}{b+c \tau} N \omega_{3,1} = \omega_{3,1}+\lambda \, \omega_{2,0},
\end{equation}
for $\lambda \in \mathbb{C}$, and where we included a factor of $b+c \tau$ for convenience. We parametrize the coordinate shift as $\lambda = \lambda_1 + \tau \lambda_2$ with $\lambda_i \in \mathbb{R}$, similar to $\pi_0,\pi_1$ in \eqref{eq:alphabeta}. Then we find by using \eqref{eq:Ktau2} that the extension data parameters in the limiting mixed Hodge structure are shifted as
\begin{equation}
    \delta \to \delta + \frac{a}{c}\lambda_2 \, , \quad \gamma \to \gamma -\lambda_1+\frac{2b}{c} \lambda_2\, , \quad \alpha \to \alpha +\lambda_1\, , \quad \beta \to \beta +\lambda_2\, .
\end{equation}
Note in particular that the shift induced by $\lambda_1$ on $\gamma,\alpha$ is orthogonal compared to the shift induced by $d_2$ in \eqref{eq:alphabetashift}. Hence this coordinate redefinition will allow us to identify the parameters $\gamma,\alpha$ and thereby bring them to a bounded domain. More explicitly, we now choose to use $\lambda$ to fix $\beta$ by setting $\beta=0$ and $\alpha=\gamma$. This is achieved by setting
\begin{equation}\label{eq:shiftfix}
\lambda_1 = \frac{1}{2}\gamma - \frac{1}{2}\alpha + \frac{b}{c} \beta \, , \qquad \lambda_2 = -\beta\, ,
\end{equation}
and subsequently relabeling $\gamma,\delta$ accordingly.\footnote{For completeness, these relabelings are given by $\delta'=\delta+\frac{a}{c} \lambda_2$ and $\gamma'=\gamma-\lambda_1+\frac{2b}{c}\lambda_2$, after which we drop the primes.} It is amusing to note that our choice of coordinate here brings us to the $\mathbb{R}$-split mixed Hodge structure.\footnote{As explained in \cite{Bastian:2021eom, Grimm:2021ikg} and similar to conifold points, coordinate transformations are the only possible deformations away from the $\mathbb{R}$-split for K-points.} This can be seen straightforwardly from the fact that now $\omega_{3,1} \Sigma  \bar{\omega}_{3,1}=0$, and thus the vector spanning $I^{1,3}$ can be taken to be simply $\omega_{1,3} = \bar{\omega}_{3,1}$.

\paragraph{Basis transformations revisited.} We look again at the action of the basis transformation $M_2$ in \eqref{eq:basischangeII}, but now combined with fixing $\alpha=\gamma$ and $\beta=0$ by the coordinate shift \eqref{eq:shiftfix}. The coefficients $c_1,c_2$ can be absorbed directly into $\gamma,\delta$ by the shifts in \eqref{eq:alphabetashift}, so these can be set to zero. They allow us to bring $\alpha=\delta+\tau \gamma$ to the standard domain of the torus in the quotient $\mathbb{C}/(\mathbb{Z}+\tau \mathbb{Z})$, which amounts to $\gamma,\delta \in [0,1)$. Then remains the last coefficient $c_3$, which induces a non-vanishing $\beta$. This piece can subsequently be set to zero again by performing the shift described by \eqref{eq:shiftfix} with $\alpha=\gamma$ and $\beta=c_3$. The resulting action on the parameters $\gamma,\delta$ induced by $M_2$ and this coordinate change is found to be
\begin{equation}\label{eq:gen3}
\delta \to \delta - \frac{a}{c} c_3\, , \qquad \gamma \to \gamma -\frac{3 b}{c} c_3\, .
\end{equation}
In other words, this means that, taking also \eqref{eq:alphabetashift} into account, $\alpha$ takes values in the complex plane quotiented by the lattice generated by $\{1, \, \tau\, , \frac{1}{c}(a+3b \tau)\}$.

\paragraph{Summary of boundary data.} To close off, let us summarize our findings for the boundary data of K-points. We find that the log-monodromy matrix takes the form
\begin{equation}
    N = \begin{pmatrix}
0 & 0 & 0 & 0 \\
0 & 0 & 0 & 0 \\
a & b & 0 & 0 \\
b & c & 0 & 0 
\end{pmatrix}\, , 
\end{equation}
with $a,b,c \in \mathbb{Z}$ such that $d=ac-b^2>0$. The limiting mixed Hodge structure is determined by the vectors 
\begin{equation} \label{eq:KLMHSbasis}
\omega_{3,1}= \begin{pmatrix}
1 \\ \tau\\ \delta+\gamma \tau \\  \gamma  
\end{pmatrix} , \qquad \omega_{2,0} = \begin{pmatrix}
0 \\ 0 \\ -\tau\\ 1
\end{pmatrix},
\end{equation}
where $\tau = \frac{-b \pm i\sqrt{d}}{c}$. The extension data is parametrized by one complex parameter $\alpha = \gamma+\delta \tau \in \mathbb{C}/(\mathbb{Z} \oplus \tau \mathbb{Z} \oplus  \frac{1}{c}(a+3b \tau) \mathbb{Z})$ with $\alpha_i \in \mathbb{R}$: the first two bring $\alpha$ to the standard domain for an elliptic curve with complex structure $\tau$; the third vector can make this domain slightly smaller for particular values of $a,b,c \in \mathbb{Z}$. We also note that this limiting mixed Hodge structure has been brought to an $\mathbb{R}$-split form, so the other two vectors are given by $\omega_{1,3}=\bar{\omega}_{3,1}\in I^{1,3} $ and $\omega_{0,2}=\bar{\omega}_{2,0} \in I^{0,2}$. Finally, we stress that we fixed a convention choice for the last two components of $\omega_{3,1}$, i.e.~$(\delta+\gamma \tau, \gamma)$, by using coordinate transformations and basis changes; while this worked well for the examples considered in this work, for other models slightly different conventions might fit better.

\subsubsection{Periods and Prepotential}
In this section we derive the form of the periods for K-points. We start from the instanton map formulation reviewed in section \ref{ssec:Hodge2}, which allows us to characterize the full expansion of the periods around this singularity compatible with the boundary structure. We then recast these expressions for the periods into the usual prepotential formulation of the periods.

\paragraph{Instanton map formulation.} Let us first construct the most general instanton map and write down the resulting form of the periods. For expository reasons we choose to work in the complex Hodge basis for this analysis, since this simplifies the expressions significantly. This basis is given by
\begin{equation}\label{eq:Kbasis}
m^{\rm LMHS} = \big( \omega_{3,1} , \, N\omega_{3,1} ,\, \omega_{1,3}, \, N \omega_{1,3} \big) = \begin{pmatrix}
 1 & 0 & 1 & 0 \\
 \tau & 0 & \bar\tau & 0 \\
 \delta+\gamma \tau & a+b\tau & \delta + \gamma \bar\tau & a+b\bar\tau \\
 \gamma & b+c\tau & \gamma & b+c\bar\tau
\end{pmatrix}\, .
\end{equation}
Performing basis rotations by $m^{\rm LMHS}$ allows us to switch between the complex Hodge basis and the integral basis of the periods. We will use this complex basis up to \eqref{eq:Kperiods2}, where we switch back to the integral basis. At that stage we have simplified the periods sufficiently by introducing a natural coordinate \eqref{eq:Kcoor} to write down asymptotic models for the periods in full. For now, let us write down the most general form of the instanton map in the complex Hodge basis as
\begin{equation}
    \Gamma(z) = m^{\rm LMHS} \begin{pmatrix}
    0 & 0 & 0 & 0 \\
    n(z) & 0 & 0 & 0 \\
    B(z) & A(z) & 0 & 0 \\
    C(z) & -B(z) & n(z) & 0
    \end{pmatrix} (m^{\rm LMHS})^{-1}\, ,
\end{equation}
where we identified the (4,2)- and (4,3)-components of this matrix to ensure that $\Gamma$ is an element of $\mathfrak{sp}(4)$. Transversality of the periods implies that these holomorphic functions must satisfy the differential equations
\begin{equation}\label{eq:Ktrans}
\begin{aligned}
A(z) \left(1+i \pi  z n'(z)\right)&=i \pi  z \left(n(z) A'(z)+2 B'(z)\right) \, ,\\
B(z) \left(1+ i \pi  z n'(z)\right)
   +i \pi  z C'(z) &= -\frac{1}{6} n(z) \left(A(z)-2 i \pi  z B'(z)\right)\, .
\end{aligned}
\end{equation}
The periods in the basis \eqref{eq:Kbasis} then take the form
\begin{equation}\label{eq:Kperiods}
(m^{\rm LMHS})^{-1} \cdot\Pi =  \left(
\begin{array}{c}
 1 \\
 n(z)+t \\
 \frac{1}{2} n(z) A(z)+B(z) \\
 (\frac{1}{2}n(z) A(z)+B(z) )t +\frac{1}{6} n^2(z)A(z)+C(z)\\
\end{array}
\right)\, .
\end{equation}
This expression together with the differential relations \eqref{eq:Ktrans} constitutes the most general set of periods at K-points in the complex Hodge basis \eqref{eq:Kbasis}. We now comment on the choice of coordinate and the form of the prepotential.

\paragraph{Natural coordinate and Yukawa coupling.} We can take the second period as our coordinate parametrizing the asymptotic regime. This amounts to a redefinition of $t$ by
\begin{equation}\label{eq:Kcoor}
t'(t)= t+n(e^{2\pi it})\, .
\end{equation}
This relation can be inverted order-by-order when expanding $n(e^{2\pi i t})$, similar to the mirror map at large complex structure. In the following we assume that this coordinate change has been performed. This allows us to drop the prime and set $n(e^{2\pi i t})=0$. In particular, in this coordinate, which we will denote by $q$, the differential relations \eqref{eq:Ktrans} simplify considerably to
\begin{equation}\label{eq:Ktrans2}
A(q) =i \pi  q 2 B'(q) \, ,\qquad B(q) = -i \pi  q C'(q) \, .
\end{equation}
The advantage of the use of this coordinate becomes most clear when computing the Yukawa coupling. By using the differential relations \eqref{eq:Ktrans2} we can bring it to the simple form
\begin{equation}\label{eq:KYukawa2}
\langle \Pi, (2\pi i \partial_q)^3 \Pi \rangle = c (\tau_2)^2\,  (2\pi i q\partial_q)^2 B(z)\, .
\end{equation}

\paragraph{Asymptotic periods.} We can also straightforwardly write down the asymptotic form of the periods in terms of this natural coordinate \eqref{eq:Kcoor}. It amounts to solving the differential equations \eqref{eq:Ktrans2} for $B(z)$ and $C(z)$ order-by-order for a series ansatz for $A(z)$. The combined set of solutions read
\begin{equation}\label{eq:Ksol}
    A(z) = \sum_{k=1}^\infty 2\pi i k \,  B_k z^k\, , \qquad  B(z) = \sum_{k=1}^\infty B_k z^k\, , \qquad C(z) = \sum_{k=1}^\infty \frac{i B_k}{k\pi} z^k \, ,
\end{equation}
with $B_k \in \mathbb{C}$ model-dependent coefficients, which for later convenience we normalized by a factor of $2\pi i k$. Let us next substitute these expressions into the formula for the periods \eqref{eq:Kperiods}. We take only the essential exponential corrections into account, meaning we keep $B_1$ and set the other $B_{k} = 0$ ($k\geq 2$). We also rotate to the integral symplectic basis by using \eqref{eq:Kbasis}. The resulting periods then read
\begin{equation}\label{eq:Kperiods2}
\begin{aligned}
    \Pi &= \begin{pmatrix}
    1+B \\
    \tau + \bar\tau B\\
    (a+b\tau)t  +(a+b \bar{\tau}) \big(t B+C\big)+ \delta+\gamma \tau +(\delta+\gamma \bar\tau )B\\
     (b+c\tau)t  +(b+c \bar{\tau}) \big(t B+C\big) +\gamma (1+B )\\   
    \end{pmatrix} \\
    & = \left(
\begin{array}{c}
 1+B_{1} e^{2 i \pi  t} \\
 \tau +B_{1} e^{2 i \pi  t} \bar\tau \\
 (a+b \tau )t +B_{1} e^{2 i \pi  t} (a+b\bar\tau) (t+\frac{i}{\pi})
+\delta+\gamma \tau +(\delta+\gamma \bar\tau) B_1 e^{2 i \pi t} \\
(b+c\tau)t + 
 B_{1} e^{2 i \pi  t} (b+c\bar\tau) (t+\frac{i}{\pi})
+\gamma (1+ B_{1} e^{2 i \pi  t} )
   \\
\end{array}
\right)\, ,
\end{aligned}
\end{equation}
where we first wrote the general form in terms of the functions $B(e^{2\pi i t})$ and $C(e^{2\pi i t})$, and subsequently plugged in the first order term in the exponential expansion. We note that one may rewrite the monodromy-related coefficients in the last period by using $b+c\tau = \pm i\sqrt{d}$ for convenience, as follows from \eqref{eq:Ktau2}.

\paragraph{Prepotential coordinate.} Let us now recast the above expressions for the periods in terms of the usual prepotential formulation. For the prepotential we have to pick a different coordinate compared to the natural coordinate \eqref{eq:Kcoor}. We define this coordinate through the ratio of the first two periods in \eqref{eq:Kperiods2} as
\begin{equation}\label{eq:Kcoorprep}
    s+\tau  = \frac{\tau + \bar\tau B(q)}{1+B(q)}\, ,
\end{equation}
with the boundary now located at $s=0$. This relation can easily be solved for $B(q)$ as
\begin{equation}\label{eq:KBsol}
B(q(s)) =  - \frac{s}{s+2i\tau_2}\, .
\end{equation}
In order to express the prepotential in terms of the coordinate $s$ we have to solve this equation for $q(s)$. This can be achieved by inverting $B(q)$, which in practice is most easily carried out order-by-order in $|s|\ll 1$. Notice that this is reminiscent of the inversion of the mirror map at large complex structure, which also is necessary in order to describe the (instanton contributions to the) prepotential. Let us carry out this inversion explicitly for the general form of $B(q)$ given in \eqref{eq:Ksol}, yielding
\begin{equation}\label{eq:zs}
    B_1 q(s) = - \frac{s}{i \tau_2} -  (1-\frac{B_2}{(B_1)^2}) (\frac{s}{i\tau_2})^2 +\mathcal{O}(s^3)\, ,
\end{equation}
where notice that we manipulated the expression into an expansion is $s/i\tau_2$.

\paragraph{Prepotential.} We next move on to the prepotential itself. We first compute it in terms of $q(s)$ and $C(q(s))$, where we eliminate $B(q(s))$ by using \eqref{eq:KBsol}. In this formulation it takes a particularly simple form
\begin{equation}\label{eq:KF}
\begin{aligned}
F &= (\Pi_1\Pi_3 + \Pi_2 \Pi_4)/(2\Pi_1) \\
&=  (c i \tau_2 s+\tfrac{1}{2} cs^2)\frac{ \log [q(s)]}{2\pi i  }+ c \tau _2 C(q(s))\left(\tau _2-i s\right) + \frac{1}{2}\delta+(s+\tau)\gamma \, ,
\end{aligned}
\end{equation}
with the corresponding periods given by
\begin{equation}
    \Pi = \begin{pmatrix}
        1\\
        s+\tau \\
        2F-(s+\tau) \partial_s F  \\
        \partial_s F
    \end{pmatrix}\, .
\end{equation}
Note that the second period is related to the coordinate $s$ in a linear way, so this does not change the derivative structure of the other periods. The above expression for the prepotential \eqref{eq:KF} is, however, not the most practical expression, as $q(s)$ still has to be evaluated in terms of $s$. By using the series solution \eqref{eq:zs} we can write down the prepotential up to second order explicitly as
\begin{equation}\label{eq:KF2}
\begin{aligned}
    F &= \tfrac{1}{2}\delta+\tau \gamma-s F_{(1)} +s^2 F_{(2)}+\ldots \, , \\
    F_{(1)} &= \frac{c\tau_2}{2\pi} (\log[\frac{is}{2\tau_2 B_1}]-1)+\gamma \, , \\
    F_{(2)} &= \frac{c}{4\pi i}\log[\frac{is}{2 \tau_2 B_1}]+\frac{ic}{8\pi } (4-\frac{B_2}{(B_1)^2})  \, .
\end{aligned}
\end{equation}
Let us now make some comments regarding the above form of the prepotential:
\begin{itemize}
\item The extension data parameters $\gamma,\delta$ appear only at constant and linear order in $s$. Moreover, in absence of these parameters --- for instance when set to zero by an Sp$(4,\mathbb{R})$ basis transformation \eqref{eq:Klambda} --- then the constant part of the prepotential would vanish. 
\item The logarithmic terms in $s$ only appear at linear and square order, and are absent at higher orders. This is consistent with the fact that we shift by the first two periods --- which are constant and linear in $s$ --- under monodromies $s \to e^{2\pi i}s$, cf.~\eqref{Eq:IIN}. Moreover, notice that the coefficients of the logarithms are quantized in the sense that $c\tau_2, c/2 \in \mathbb{Q}(\sqrt{d})$, which precisely agrees with the integrality of the monodromy matrix.
\item Finally, the higher-order terms in the prepotential arise from expanding $\log[z(s)]$ and $C(q(s))$ in terms of $s$ by using \eqref{eq:zs} and \eqref{eq:Ksol}. An example of such a term is the second term of $F_{(2)}$, and at higher orders the prepotential is made up of just such terms, i.e.~ratios of the coefficients $A_l$.
\item Among the higher-order terms, a special role is reserved for the coefficient $A_1$. This model-dependent parameter cannot vanish, as it is \textit{essential} for the leading higher-order terms at $\mathcal{O}(s)$ and $\mathcal{O}(s^2)$. On the other hand, the parameters $A_{l\geq 2}$ do not play such an important role and could in principle vanish in a given example.
\end{itemize}

\section{Observations and interpretation of boundary data}\label{sec:geometryinput}
In the previous section, we have established general expressions for the periods in an integral basis for each singularity type. It is natural to ask whether the various coefficients that appear are arbitrary or if they contain meaningful information about the underlying geometry. Recall that this data may be divided into a perturbative and a non-perturbative part: the first encodes the leading asymptotic periods through discrete boundary data, while the latter captures the full expansion of the period functions. For LCS points this data is of course well-known: the topological numbers of the mirror Calabi-Yau are encoded in the boundary data, while the coefficients of the holomorphic series from the instanton map are related to BPS state counting, or equivalently the numbers of holomorphic curves on the mirror CY. For the conifold singularities, such a correspondence has not yet been established in full generality. Great progress has been made recently in \cite{Bonisch:2022mgw} for the case of ordinary conifolds. For K-point singularities, the authors are not aware of any concrete results in the literature.

\subsection{Semisimple monodromy factors}\label{ssec:semisimple}
Before we treat conifold and K-points individually, let us first make some general remarks about the semisimple part of monodromies around these singularities. The purpose of this discussion is to illustrate how the presence of semisimple factors constrains the allowed boundary data. Namely, instead of generic transcendental data, such a factor restricts the boundary data to take algebraic values, i.e.~roots to polynomial equations with rational coefficients.\footnote{Closely related observations have been made recently in \cite{chmiel2019computing, chmiel2021coefficients} for the transition matrices of conifold points with semisimple monodromies.} Here we argue for this result from the perspective of the LMHS associated to the singularity.

First we discuss briefly the action of semisimple monodromy factors on LMHS in general, and then specialize to conifold and K-points. Recall from our discussion around equation \eqref{Tdecomp} that the monodromy around any singularity may be decomposed into two commuting factors $T_u,T_{ss} \in \text{Sp}(4,\mathbb{Q})$: a unipotent part $T_u=e^{N}$, and a semisimple part $T_{ss}$ of some finite order $n \in \mathbb{N}$. Usually one removes these semisimple factors by performing a coordinate transformation of the form $z \to z^n$; however, for our purposes it is worthwhile to pay closer attention to this semisimple part, as it imposes certain symmetries on the boundary data. The precise statement is that $T_{ss}$ is an automorphism of the LMHS, in the sense that it acts on the vector spaces $I^{p,q}$ of the Deligne splitting \eqref{Ipq} as
\begin{equation}
    \omega_{p,q} \in I^{p,q}: \qquad T_{ss}\omega_{p,q} \in I^{p,q}\, .
\end{equation}
Given a non-trivial semisimple monodromy matrix $T_{ss}$, this yields a non-trivial set of constraints on the vectors $\omega_{p,q}$ spanning the Deligne splitting, and thereby also on the boundary data that make up these vectors. In order to see this in more detail, we specialize to the two classes of boundaries of interest to our work.

\paragraph{Conifold points.} First we consider conifold points. Here the Deligne splitting is made up of four one-dimensional vector spaces, where the semisimple monodromy can act non-trivially on the subspaces $I^{3,0}\oplus I^{0,3}$ and/or $I^{2,2}\oplus I^{1,1}$.  A non-trivial action on the former implies that the spanning vector $\omega_{3,0} \in I^{3,0}$ is rotated by a root of unity as
\begin{equation}\label{eq:T30}
    T_{ss} \, \omega_{3,0} = e^{2\pi i a_1}\omega_{3,0}\, ,
\end{equation}
where $a_1$ is the first index in the Riemann symbol of the singularity, i.e.~the local exponent of the first period. On the other hand, the vectors spanning $I^{2,2}\oplus I^{1,1}$ are real (assuming we are in the $\mathbb{R}$-split basis), so there is only a potential minus sign
\begin{equation}\label{eq:T22}
    T_{ss} \omega_{2,2} = \pm \omega_{2,2}\, , \qquad T_{ss} \omega_{1,1} =  \pm \omega_{1,1}\, ,
\end{equation}
corresponding to indices  $a_2=a_3=0,\tfrac{1}{2} \mod 1$ respectively. How \eqref{eq:T30} and \eqref{eq:T22} precisely constrain the boundary data becomes more apparent when considering the explicit form of the Deligne splitting given in \eqref{eq:mhsI1}, which we recall here for completeness
\begin{equation}\label{eq:sec5mhsI1}
    \omega_{3,0} = (0,1,\tau,\delta-\gamma \tau)^T\, , \qquad \omega_{2,2} = (1,\gamma,\delta,0)^T\, , \qquad \omega_{1,1} = (0,0,0,1)^T\, ,
\end{equation}
and the conjugate $\omega_{0,3}=\bar{\omega}_{3,0}$. With \eqref{eq:T30} and \eqref{eq:T22} we know the action of $T_{\rm ss}$ on the complete integral basis, so we may write out this matrix explicitly as
\begin{equation}\label{eq:TssC1}
    T_{ss} = \scalebox{0.8}{$\left(
\begin{array}{cccc}
 \pm 1 & 0 & 0 & 0 \\
 \frac{s(a_1) \left(\gamma  \tau _1-\delta \right)}{\tau _2}-\gamma  \left(c(a_1) \mp 1\right) & c(a_1)-\frac{\tau _1 s(a_1)}{\tau _2} & \frac{s(a_1)}{\tau _2} & 0 \\
 \frac{s(a_1) \left(\gamma  |\tau|^2-\delta  \tau _1\right)}{\tau _2}-\delta  (c(a_1)\mp 1) &
   -\frac{|\tau|^2s(a_1)}{\tau _2} & c(a_1)+\frac{\tau _1 s(a_1)}{\tau _2} & 0 \\
 -\frac{s(a_1) \left(\gamma ^2 \tau _2^2+\left(\delta -\gamma  \tau _1\right){}^2\right)}{\tau _2} & \frac{s(a_1)
   \left(\gamma  |\tau|^2-\delta  \tau _1\right)}{\tau _2}+\delta \left( c(a_1)\mp  1\right) & \frac{s(a_1)
   \left(\delta -\gamma  \tau _1\right)}{\tau _2}-\gamma  \left(c(a_1)\mp 1\right) & \pm 1 \\
\end{array}
\right)$}\, ,
\end{equation}
where we used the shorthands $c(a_1)=\cos(2\pi a_1)$ and $s(a_1)=\sin(2\pi a_1)$, and the $\pm$-signs refer to whether the monodromy acts respectively trivially or non-trivially in \eqref{eq:T22} on $I^{2,2}$ and $I^{1,1}$. Notice that, if we put this together with the unipotent factor $T_u = e^{N/n}$, with $N$ given by \eqref{Eq:ConiN}, only the lower-left element changes by adding a $k/n$. For all other components the quantization condition $T \in \mathrm{Sp}(4,\mathbb{Z})$ thus directly carries over to the semisimple factor $T_{ss}$ given in \eqref{eq:TssC1}. In particular, let us point out the middle subblock
\begin{equation}\label{eq:sl2block}
    \begin{pmatrix}
        c(a_1)-\frac{\tau_1 s(a_1)}{\tau_2} & \frac{s(a_1)}{\tau_2} \\
        -\frac{|\tau|^2 s(a_1)}{\tau_2} & c(a_1)+\frac{\tau_1s(a_1)}{\tau_2}
    \end{pmatrix}\in \mathrm{SL}(2,\mathbb{Z})\, .
\end{equation}
From the condition \eqref{eq:T30} we know that this element fixes the rigid period $\tau$ by the usual $\mathrm{SL}(2,\mathbb{Z})$ transformation. The only possible values (on the fundamental domain) for $\tau$ that are fixed by an $\mathrm{SL}(2,\mathbb{Z})$ transformation are
\begin{equation}
    \tau = i \, , \qquad \tau = -\frac{1}{2}+\frac{i\sqrt{3}}{2}\, .
\end{equation}
Notice that this only applies when the SL$(2,\mathbb{Z})$-matrix acts non-trivially, i.e.~for $c(a_1) \neq \pm 1$; otherwise it is proportional to $\pm \mathbb{I}$ and any $\tau$ is allowed. The element fixing $\tau=i$ has order $2$, while for $\tau = -\tfrac{1}{2}+\tfrac{i\sqrt{3}}{2}$ it is $3$. These order constraints restrict the exponent as
\begin{equation}\label{eq:Ca1}
    4a_1 \in \mathbb{Z} \, , \quad \text{or} \quad 6 a_1 \in \mathbb{Z}\, ,
\end{equation}
i.e.~whenever $\cos(2\pi a_1) = 0 , \pm \tfrac{1}{2}, \pm 1$. In turn, we can also constrain the total order of the semisimple monodromy, as we only have to take an additional $\mathbb{Z}_2$ action \eqref{eq:T22} into account. Altogether this gives for $T_{ss}$ as possible orders
\begin{equation}
    n=2,3,4,6\, .
\end{equation}
Note that this is stronger than the Euler totient constraint $\phi(n)\leq 4$ mentioned below \eqref{Tdecomp}, which yielded $n\leq 12$.  As a cross-check of our analysis, for conifold points these orders --- or equivalently the allowed values for $a_1<a_2=a_3<a_4$ in \eqref{eq:Ca1} --- are indeed the only possibilities in the database \cite{almkvist2005tables}.

With these preparations in place, let us next explicitly go through the semisimple monodromy factors at all possible orders $n$. We first consider $n=2$. There are two choices, either it acts with a minus sign in \eqref{eq:T30} or in \eqref{eq:T22} (one has $T_{ss}=-\mathbb{I}$ when it acts on both with a minus sign). In both cases there is no constraint on $\tau$ imposed, as the middle SL$(2,\mathbb{Z})$ block is simply $\pm \mathbb{I}$. Taking $c(a_2)=-c(a_1)=\pm 1$ for the signs, we find that \eqref{eq:TssC2} reduces to 
\begin{equation}\label{eq:TssC2}
    T_{ss} = \pm \begin{pmatrix}
    1 & 0 & 0 & 0 \\
    2\gamma & - 1 & 0 & 0 \\
    2\delta & 0 & - 1 & 0 \\
    0 & -2\delta & 2\gamma & 1
    \end{pmatrix}\, ,
\end{equation}
For order $n=3$ we are constrained to $\tau=(-1+i\sqrt{3})/2$. The middle block \eqref{eq:sl2block} is given by an SL$(2,\mathbb{Z})$-element that fixes this point, corresponding to $c(a_1)=- 1/2$ and $s(a_1)=\pm \sqrt{3}/2$. Additionally we fix $c(a_2)=1$, as otherwise our semisimple monodromy would have order 6. We find that \eqref{eq:TssC1} then reads
\begin{equation}\label{eq:TssC3}
    T_{ss} = \left(
\begin{array}{cccc}
 1 & 0 & 0 & 0 \\
 \frac{3 \gamma }{2} & -\frac{1}{2} & 0 & 0 \\
 \frac{3 \delta }{2} & 0 & -\frac{1}{2} & 0 \\
 0 & -\frac{3 \delta }{2} & \frac{3 \gamma }{2} & 1 \\
\end{array}
\right) \pm \left(
\begin{array}{cccc}
 0 & 0 & 0 & 0 \\
 -\frac{\gamma }{2}-\delta  & \frac{1}{2} & 1 & 0 \\
 \gamma +\frac{\delta }{2} & -1 & -\frac{1}{2} & 0 \\
 -\gamma ^2-\gamma  \delta -\delta ^2 & \gamma +\frac{\delta }{2} & \frac{\gamma }{2}+\delta  & 0 \\
\end{array}
\right)\, ,
\end{equation}
The order $n=4$ constrains us to $\tau=i$. The action \eqref{eq:T30} on $I^{3,0}$ has phase $\pm i$, corresponding to $c(a_1)=0$ and $s(a_1)=(-1)^{m_1}$ for some $m_1 \in \mathbb{Z}$. The sign in the other action \eqref{eq:T22} may be chosen independently as $c(a_2)=(-1)^{m_2}$ for some $m_2 \in \mathbb{Z}$. The semisimple monodromy factor \eqref{eq:TssC1} then reads
\begin{equation}
    T_{\rm ss} = (-1)^{m_1} \left(
\begin{array}{cccc}
 0 & 0 & 0 & 0 \\
 -\delta  & 0 & 1 & 0 \\
 \gamma  & -1 & 0 & 0 \\
 -\gamma ^2-\delta ^2 & \gamma  & \delta  & 0 \\
\end{array}
\right) + (-1)^{m_2} \left(
\begin{array}{cccc}
 1 & 0 & 0 & 0 \\
 \gamma  & 0 & 0 & 0 \\
 \delta  & 0 & 0 & 0 \\
 0 & -\delta  & \gamma  & 1 \\
\end{array}
\right),
\end{equation}
The order $n=6$ constrains us to $\tau=(-1+i\sqrt{3})/2$. We leave out cases where $(T_{ss})^3=- \mathbb{I}$, as these may be obtained from order $n=3$ in \eqref{eq:TssC3} by including an overall minus sign; this requires us to pick $c(a_1)$ and $c(a_2)$ with the same sign, i.e.~$c(a_1)=(-1)^{m_1}/2$ and $c(a_2)=(-1)^{m_1}$. On the other hand, we can choose the sign independently in $s(a_1)=(-1)^{m_2}\sqrt{3}/2$. Altogether, this reduces the semisimple monodromy factor \eqref{eq:TssC1} to
\begin{equation}\label{eq:TssC6}
    T_{ss} = (-1)^{m_1}\left(
\begin{array}{cccc}
 1 & 0 & 0 & 0 \\
 \frac{\gamma }{2} & \frac{1}{2} & 0 & 0 \\
 \frac{\delta }{2} & 0 & \frac{1}{2} & 0 \\
 0 & -\frac{\delta }{2} & \frac{\gamma }{2} & 1 \\
\end{array}
\right)+ (-1)^{m_2}\left(
\begin{array}{cccc}
 0 & 0 & 0 & 0 \\
 -\frac{\gamma }{2}-\delta  & \frac{1}{2} & 1 & 0 \\
 \gamma +\frac{\delta }{2} & -1 & -\frac{1}{2} & 0 \\
 -\gamma ^2-\gamma  \delta -\delta ^2 & \gamma +\frac{\delta }{2} & \frac{\gamma }{2}+\delta  & 0 \\
\end{array}
\right) .
\end{equation}
With this exhaustive set of semisimple monodromy factors, let us return to the quantization condition $T \in \mathrm{Sp}(4,\mathbb{Z})$ and how it constrains the extension data $\gamma,\delta$. Recall that the unipotent factor \eqref{Eq:ConiN} only alters the bottom-left component, so the quantization conditions directly carry over to the other components of $T_{ss}$. We find that
\begin{equation}
    \begin{aligned}
        n=2,4&: \qquad 2\gamma, 2\delta \in  \mathbb{Z}\, , \\
        n=3,6&: \qquad \gamma, \delta \in \mathbb{Z}\, ,
    \end{aligned}
\end{equation}
The former follows by inspecting $(T_{ss})^{n/2}$ for $n=2,4$, which is given by \eqref{eq:TssC2}. The latter by looking directly at entries of \eqref{eq:TssC3} and \eqref{eq:TssC6}. Recall from section \ref{ssec:conifoldconstruction} that integer shifts of the extension data may be used to set $\gamma,\delta \in [0,1)$: for $n=2,4$ this yields $\gamma,\delta =0,1/2$, while for $n=3,6$ we can always set $\gamma=\delta=0$.

Before we move onto K-points, let us comment on the case that only $T \in \mathrm{Sp}(4,\mathbb{Q})$, as such monodromies might arise when the Calabi--Yau threefold for instance has singularities or a higher-dimensional moduli space. In this case we lose the precise quantization conditions obtained above over $\mathbb{Z}$, but suitable statements still apply over $\mathbb{Q}$. For the rigid period $\tau$ this means that, for $n=3,4,6$, it still lies in $\tau \in \mathbb{Q}(i)$ or $\tau \in \mathbb{Q}(i\sqrt{3})$, just no longer constrained to the cusps. Similarly, for the extension data we have at all orders $n=2,3,4,6$ that still $\gamma,\delta \in \mathbb{Q}$, but we lose the sharp quantization conditions $\gamma,\delta = 0,\tfrac{1}{2}$.

\paragraph{K-points.} Next we consider K-points. Here again the Deligne splitting is made up of one-dimensional vector spaces, where now the action of the semisimple monodromy factorizes over the subspaces $I^{3,1}\oplus I^{1,3}$ and $I^{2,0}\oplus I^{0,2}$. Similar to the action \eqref{eq:T30} on $\omega_{3,0}$ for conifold points, the action on $\omega_{3,1}\in I^{3,1}$ is a rotation by a root of unity
\begin{equation}\label{eq:T31}
        T_{ss} \, \omega_{3,1} = e^{2\pi i a_1}\omega_{3,1}\, ,
\end{equation}
with the order $n$ of the semisimple factor given by the denominator of $a_1 \in \mathbb{Q}$. The action on $\omega_{2,0}$ is identical, and is inferred from the fact that $\omega_{2,0}=N\omega_{3,1}$ and the commuting of monodromy factors $T_{ss}N=NT_{ss}$. The action of $\omega_{1,3}$ and $\omega_{0,2}$ follows simply by complex conjugation.

The constraints on the boundary data imposed by rationality of $T_{ss}$ through \eqref{eq:T31} become more apparent when considering the explicit form of the Deligne splitting \eqref{eq:KLMHSbasis}. Let us recall the precise expressions for completeness
\begin{equation}
    \omega_{3,1} = (1,\tau,\delta+\gamma \tau,\gamma)^T \, ,\qquad \omega_{2,0} = (0,0,-\tau,1)^T\, ,
\end{equation}
and conjugates $\bar{\omega}_{3,1}=\omega_{1,3}$ and $\bar{\omega}_{2,0}=\omega_{0,2}$. For the rigid period it was already established that $\tau \in \mathbb{Q}(\sqrt{-d})$, with $d \in \mathbb{N}$ the determinant of the pairing matrix in the monodromy \eqref{eq:KN}, even in the absence of semisimple monodromy factors. In order to see the constraints on $\gamma,\delta$ it is convenient to write $T_{ss}$ in the integral basis as
\begin{equation}\label{eq:TssK}
    T_{ss} = \scalebox{1}{$\left(
\begin{array}{cccc}
 c(a_1)-\frac{\tau _1 s(a_1)}{\tau _2} & \frac{s(a_1)}{\tau _2} & 0 & 0 \\
 -\frac{|\tau|^2 s(a_1)}{\tau _2} & \frac{\tau _1 s(a_1)}{\tau _2}+ c(a_1) & 0 & 0 \\
 -\frac{2 s(a_1) \left(\tau _1 \left(\delta+\gamma  \tau _1\right)+\gamma  \tau _2^2\right)}{\tau _2} & \frac{\delta  s( a_1)}{\tau _2} & \frac{\tau _1 s(a_1)}{\tau _2}+c(a_1) & \frac{|\tau|^2 s(a_1)}{\tau _2} \\
 \frac{\delta  s(a_1)}{\tau _2} & \frac{2 \gamma s(a_1)}{\tau _2} & -\frac{s(a_1)}{\tau _2} & c(a_1)-\frac{\tau _1 s(a_1)}{\tau _2} \\
\end{array}
\right)$}\, ,
\end{equation}
where we used the shorthands $c(a_1)=\cos(2\pi a_1)$ and $s(a_1)=\sin(2\pi a_1)$. Notice that, similar to the conifold point, the SL$(2,\mathbb{Z})$-matrix \eqref{eq:sl2block} appears in this monodromy matrix. Again the unipotent factor does not affect this subblock, as \eqref{eq:KN} only alters the lower-left part. Thus we are analogously led to the conclusion that
\begin{equation}
    \tau = i \, , \qquad \tau=-\frac{1}{2}+\frac{i\sqrt{3}}{2}\, .
\end{equation}
This is even stronger than the constraint $\tau \in \mathbb{Q}(i\sqrt{d})$ (for any $d \in \mathbb{N}$)  for generic K-points without semisimple part obtained before in section \ref{ssec:Kconstruction}. We may therefore choose the log-monodromy matrix \eqref{eq:KN} to be fixed to $a=c=1$ and $b=0$ in the first case, and $a=c=2$ and $b=1$ in the second. Similar to the conifold point, the order of the semisimple factor is constrained to
\begin{equation}
    n=2,3,4,6\, ,
\end{equation}
which is again stronger than $n\leq 12$ obtained from the Euler totient bound $\phi(n)\leq 4$. As a cross-check of our analysis we consider K-points in the database \cite{almkvist2005tables}, where we find that all but two example obey this constraint. The outliers are operators 2.42 and 2.43 with Riemann symbols $\tfrac{11}{12},\tfrac{11}{12},\tfrac{13}{12},\tfrac{13}{12}$ and $\tfrac{7}{8}, \tfrac{7}{8}, \tfrac{9}{8}, \tfrac{9}{8}$; for both no mirror Calabi-Yau threefold is known, so we cannot fix an integral basis through their topological data and investigate these examples further, but likely these geometries possess singularities or have higher-dimensional moduli spaces.

Let us now go through these orders for $n$ and reduce \eqref{eq:TssK} in each of these cases. We include the unipotent factor \eqref{eq:KN} with the corresponding values for $a,b,c$ for later reference. For convenience we regard semisimple factors with $(T_{ss})^{n}=-\mathbb{I}$ as order $n$, even though then only $(T_{ss})^{2n}=\mathbb{I}$; this is in agreement with our example study in section \ref{sec:examples}. For order $n=2$, leaving out the case $T_{ss}=-\mathbb{I}$, we have $\tau=i$ with $c(a_1)=0$ and $s(a_1)=\pm 1$, which yields
\begin{equation}\label{eq:TssK2}
    T = T_u T_{ss} = \pm 
\left(
\begin{array}{cccc}
 0 & 1 & 0 & 0 \\
 -1 & 0 & 0 & 0 \\
 -2 \gamma  & \delta +\frac{1}{2} & 0 & 1 \\
 \delta -\frac{1}{2} & 2 \gamma  & -1 & 0 \\
\end{array}
\right)\, ,
\end{equation}
where we note that this semisimple factor squares to minus the identity. For order $n=3$ we have instead $\tau=(-1+i\sqrt{3})/2$, with $c(a_1)=-1/2$ and $s(a_2)=\pm \sqrt{3}/2$, which yields
\begin{equation}\label{eq:TssK3}
    T = T_u T_{ss} = 
\left(
\begin{array}{cccc}
 -\frac{1}{2} & 0 & 0 & 0 \\
 0 & -\frac{1}{2} & 0 & 0 \\
 -\frac{1}{3} & -\frac{1}{6} & -\frac{1}{2} & 0 \\
 -\frac{1}{6} & -\frac{1}{3} & 0 & -\frac{1}{2} \\
\end{array}
\right) \pm \left(
\begin{array}{cccc}
 \frac{1}{2} & 1 & 0 & 0 \\
 -1 & -\frac{1}{2} & 0 & 0 \\
 \delta -2 \gamma  & \delta +\frac{3}{8} & -\frac{1}{2} & 1 \\
 \delta -\frac{3}{8} & 2 \gamma  & -1 & \frac{1}{2} \\
\end{array}
\right)\, ,
\end{equation}
For order $n=4$ any semisimple factor with $(T_{ss})^4=\mathbb{I}$ also has $(T_{ss})^2=-\mathbb{I}$, so this reduces to the analysis in \eqref{eq:TssK2}. For order $n=6$ one simply includes an overall minus sign in the semisimple factor of the $n=3$ case, so this reduces to the analysis in \eqref{eq:TssK3}.

To conclude our discussion on semisimple factors for K-points, let us investigate the quantization condition $T = T_u T_{ss} \in \mathrm{Sp}(4,\mathbb{Z})$ in detail. From the monodromies given in \eqref{eq:TssK2} and \eqref{eq:TssK3} we directly read off the conditions
\begin{equation}\label{eq:Kgammadelta}
\begin{aligned}
   n=2 &: \qquad     2\gamma \in \mathbb{Z}\, , \quad \delta + \frac{1}{2} \in \mathbb{Z} \, , \\
    n=3 &: \qquad     2\gamma \mp \frac{1}{3} \in \mathbb{Z}\, , \quad \delta \pm  \frac{1}{3} \in \mathbb{Z} \, , 
\end{aligned}
\end{equation}
Recall from section \ref{ssec:Kconstruction} that we can perform integer shifts on the extension data $\gamma,\delta$ by integral basis transformations. With the above quantization conditions this thus allows us to pick $\gamma = 0,\tfrac{1}{2}$ and $\delta=\tfrac{1}{2}$ for $n=2$, while for $n=3$ we can set $\gamma = \pm  \tfrac{1}{6}, \tfrac{1}{2}\pm \tfrac{1}{6}$ and $\delta = \mp \frac{1}{3}$.

Similar to the conifold point analysis, many of the observations made above can be generalized to the case of rational monodromies $T \in \mathrm{Sp}(4,\mathbb{Q})$. For the rigid period $\tau$ it follows that it is no longer restricted to the cusps, but still lies in the number fields $\tau \in \mathbb{Q}(i)$ or $\tau \in \mathbb{Q}(i\sqrt{3})$. This is still a remarkably stronger condition than $\tau \in \mathbb{Q}(i\sqrt{d})$ which holds for generic K-points. Similarly for the extension data $\gamma,\delta$ we cannot make precise statements about the rational piece as before, but it still must take rational values $\gamma,\delta \in \mathbb{Q}$.

\subsection{Conifold points}\label{ssec:geometryI1}
In the previous subsection we have discussed the boundary data of conifold points in the presence of semisimple monodromy factors. Here we turn to the general situation and review connections between the boundary information and the degenerate conifold variety associated to this singularity.

\subsubsection*{Degenerate geometry and monodromy}
Ordinary conifolds are among the most well studied types of degeneration \cite{Candelas:1989ug,Candelas:1989js,Strominger:1995cz}. Locally the geometry can be understood as the vanishing of an $S^3$. One might naturally ask what the vanishing cycle is for a general conifold point. It can be inferred from the general form of the periods \eqref{eq:ConifoldPeriod}, that there is still a single vanishing cycle, i.e.~one period goes strictly to zero at the singular point. We can thus analyze the situation with a standard Picard-Lefschetz monodromy argument as was done in \cite{vanEnckevort:2003zz}. There the authors observe that in many examples one does not encounter the usual expression,
as would apply for ordinary conifolds, but rather a more general transformation. Given a three-cycle $\alpha$ in our geometry, the monodromy transformation from encircling the conifold point should be described by 
\begin{align}
    \alpha \mapsto \alpha - \lambda \langle \beta , \alpha \rangle \beta \,,
\end{align}
where $\beta$ denotes the homology class of the vanishing cycle and $\lambda \in \mathbb{Z}$. For $\lambda=1$, one recovers the usual Picard-Lefshetz formula corresponding to the ordinary conifold. As explained in \cite{vanEnckevort:2003zz}, the cases $\lambda \neq 1$ could be understood as orbifolds of the ordinary conifold, where the vanishing fiber corresponds to $S^3/K$, where $K$ is a finite subgroup of $SU(2)$ and $|\lambda|$ corresponds to the order of $K$. Using the form of the conifold periods in the monodromy weight frame immediately identifies $\lambda=k$, i.e. it is given by the only integer that specifies the log-monodromy matrix \eqref{Eq:ConiN}. The physics of such more general degenerations has been studied in \cite{Gopakumar:1997dv,Davies:2013pna}.  

\subsubsection*{Graded pieces and modularity}
We now turn to the pure Hodge structures lying on the graded pieces of the weight filtration. The only interesting piece for the discussion that follows (as the others are just one-dimensional) is given by
\begin{align}
    Gr_3 \cong H^{3,0} \oplus H^{0,3} \,, \label{eq:RigidThreefoldHS}
\end{align}
resembling the middle cohomology of a rigid Calabi-Yau threefold, i.e.~no complex structure deformations. This does not really come as a surprise. At least for ordinary conifolds, it was first shown for the quintic by Schoen \cite{Schoen1986OnTG}, that one can blow-up the conifold singularity to recover a smooth Calabi-Yau threefold that is rigid. If this resolution can be done over $\mathbb{Q}$, then the threefold is said to be modular in the usual sense, i.e. there is a correspondence between its middle cohomology and a certain classical weight-four modular form. On a general level, proving modularity for rigid threefolds over $\mathbb{Q}$ was the outcome of a series of works culminating in \cite{Gouva2009RigidCT}. Thus, as soon as we see an object like \eqref{eq:RigidThreefoldHS}, it is reasonable to look for modularity. 

How does modularity in the above sense manifest itself at the level of the periods? An answer to this question was recently given in \cite{Bonisch:2022mgw} for ordinary conifold points. Our empirical data in section \ref{sec:examples} suggests that the story for generalized conifold points follows the same pattern. The modularity shows up through critical values of the L-functions that are naturally associated to the relevant weight-four modular form $f$, see appendix \ref{app:Lvalues} for definitions relating to L-functions. At the level of the boundary data, these L-values show up on the period piece that is associated with the pure Hodge strucutre \eqref{eq:RigidThreefoldHS}. More concretely, restricting the vector \eqref{eq:mhsI1} that spans $I^{3,0}$ to the entries that lie in $Gr_3$, we observe the general form
\begin{align}
  \omega_{3,0}|_{Gr_3} = \begin{pmatrix}
      1 \\ \tau_1 + i \mathbb{Q} \frac{L(f,2)}{\pi L(f,1)}
  \end{pmatrix} \sim \begin{pmatrix}
     \mathbb{Q} \pi^2 L(f,1) \\ \mathbb{Q} \tau_1 \pi^2L(f,1) + i \mathbb{Q} \pi L(f,2)
  \end{pmatrix} \,, \quad \tau_1=0,\frac{1}{2} \, , \label{eq:ConiRigidPeriod}
\end{align}
where the $\mathbb{Q}$ represent rational factors. In the second step we have used the scaling freedom to get a more natural normalization from an arithmetic point of view.
Actually, it is this second normalization (up to possibly an algebraic number) that naturally comes out when computing the periods from the Picard-Fuchs equations in examples. We rather do an ad-hoc rescaling of the LMHS pieces to set some entries to one according to our conventions. As \eqref{eq:ConiRigidPeriod} stands, it agrees with the results from \cite{Bonisch:2022mgw} for ordinary conifolds, only reformulated in the language of LMHS, and also with the worked out examples of generalized conifolds of section \ref{sec:examples}. Currently, we do not have a good general understanding of the parameter $\tau_1$. For all conifolds that are close to a LCS point, in the sense that both regions of convergence overlap, it is identified with $\sigma$ appearing in the classical LCS prepotential \eqref{eq:LCSprepotential}. For all the examples we have studied in this paper that are not, in the above sense, next to an LCS point, we find that $\tau_1=\frac{1}{2}$. 
The two-dimensional vector \eqref{eq:ConiRigidPeriod} should allow the interpretation as period vector associated to the rigid Hodge structure \eqref{eq:RigidThreefoldHS}. Making this statement more precise is beyond the scope of this paper, but it seems to be in line with conjectures about critical L-values \cite{Deligne1979ValeursDF,Beilinson1985}.

Further observations can be made for generalized conifolds that have a finite order part to their monodromy, i.e.~fractional Riemann indices. In section \ref{ssec:semisimple} we provided a thorough analysis of this semisimple factor. When there are three distinct eigenvalues, i.e.~it has order $n=3,4,6$, it restricts the rigid period to the cusps $\tau=i$ or $\tau = (-1+i\sqrt{3})/2$. In light of \eqref{eq:ConiRigidPeriod} this implies that the modular form associated with the singularity enjoys a special property, namely complex multiplication in  $\mathbb{Q}(\sqrt{-1} )$ or $\mathbb{Q}(\sqrt{-3})$. It is observed from the examples that the transcendental part of the critical L-values is expressible in terms of conventional Gamma functions, which is in line with the study in \cite{li2018computing}. In addition, under the assumption that none of the critical L-values is zero, there is an algebraic relation between the two critical L-values of the form
\begin{align}
      L(f,2) = \mathbb{Q} \sqrt{D} \pi L(f,1) \,,
\end{align}
where $D$ is the CM discriminant. This behavior seems to hold more generally for classical CM modular forms of weight $2w$ for $w>1$, although the authors were not able to find an underlying statement in the literature.

When rotating the Frobenius solution into the integral monodromy weight frame, there is another ingredient that appears in the transition matrix, namely the quasiperiods $\eta^{+}$ and $\eta^{-}$ that are related to the modular forms associated with the geometry. Without going into any details about their origin, we simply want to share some observations when the modular form in question has CM. We refer the interested reader to \cite{Bonisch:2022mgw} for a nice introduction to quasiperiods. For weight-four CM modular forms, we observe in examples that the quasiperiods can be expressed in terms of the critical L-values as
\begin{align}
    \eta^{+} =  \mathbb{Q}\frac{\sqrt{D} \pi}{L(f,1)} + \mathbb{Q}  \pi^2 L(f,1)\,,\label{eq:ConiQuasiP}
\end{align}
The second term is well understood and is inherent to the definition of the quasiperiods as they are only defined up to shifts. However, the simplicity of this expression should be understood as a consequence of the CM property. For all the conifold cases with CM analysed in \ref{sec:examples}, the relation \eqref{eq:ConiQuasiP} is even in its most simple form, meaning that the quasiperiods appearing in the transition matrices are simply given by the inverse of the critical L-value without shift term. A more non-trivial example can be found in the geometry $X_{4,3}$ for the ordinary conifold at $z=1$ analysed in \cite{Bonisch:2022mgw}, which happens to have a CM modular form of level 9 associated with it\footnote{Although, it seems like an accident that there CM in the picture as it is not expected as the monodromy has no apparent finite order part.} Using their notation from table 2 on page 27, one can numerically verify the following relation
\begin{align}
    \eta^+=\frac{2 \sqrt{3}\pi}{3 L(f_9,1)}-63 \pi^2 L(f_9,1) \,.
\end{align} 
We want to end this part with a note about the occurrence of the same modular forms (or twists thereof) in different varieties. According to the formulation of the Tate conjecture in \cite{Yui}, this should imply a correspondence between geometries. Such geometries were referred to as relatives in \cite{Cynk2005MODULARCT}. In our case, the relevant geometries should be the rigid ones obtained after resolutions of the conifold geometries. Whereas a detailed study of this correspondence is beyond the scope of this paper, we just point out the occurrences we observe. We find the same modular form of level $9$ for the point $z=\infty$ in $X_{3,2,2}$ that is also associated with the point $z=2^{-6}3^{-3}$ of $X_{4,3}$. For the geometry related to the CY operator 2.17, we find a modular form of level $64$ at $z=2^{-8}$ which is a twist of the modular form of level $8$ that occurs at $z=2^{-8}$ in $X_{2,2,2,2}$. For the geometry related to CY operator 2.62, we find the same modular form of level $32$ at $z=\infty$ which occurs also at $z=\infty$ in $X_{4,2}$.

\subsubsection*{Extension data}
There are two relevant extensions to look at. First, there is the extension $\text{Ext}^1(Gr_4,Gr_2)$, which is a-priori a generic complex number. Its imaginary part parametrizes how far away we are from the $\mathbb{R}$-split LMHS, i.e.~$\bar{I}^{p,q}=I^{q,p}$; it corresponds to the Beilinson-Bloch height as recently discussed in \cite{bloch2023computing}. One may remove this complex number in the extension data by performing a coordinate rescaling. In this work we explicitly kept track of this extension, and observed that in all examples its real part vanishes. For the coordinate transformation this means that one simply has to apply a real rescaling. For the imaginary part we observed that in examples with a semisimple monodromy factor the scaling parameter is an algebraic number; on the other hand for more generic examples, such as the quintic in \eqref{quintic:etah}, it is transcendental and may be expressed in terms of derivatives of L-functions.

Second, there is the extension $\text{Ext}^1(Gr_4,Gr_3)$, which is dual to $\text{Ext}^1(Gr_3,Gr_2)$ by the isomorphism property of the weight graded pieces under the map $N$, and hence the latter contains the same information. We have seen in section \ref{ssec:conifoldconstruction}, following the work of \cite{KerrLMHS}, that it is
given by two real numbers, defined up to integer shifts
\begin{align}\label{eq:ConiExt}
    \text{Ext}^1(Gr_4,Gr_3) \sim \begin{pmatrix}
        \gamma \\
        \delta
    \end{pmatrix} \in \mathbb{R}^2/\mathbb{Z}^2 \,. 
\end{align} 
In all the cases we checked, including the ordinary conifolds of the 14 hypergeometric families, there is only a single real number specifying this extension, with $\gamma$ and $\delta$ being simply rational multiples of each other. This constraint does not follow from the linear algebra construction in section \ref{ssec:conifoldconstruction}, but it seems to follow rather from geometry. 

Something drastic happens when the monodromy has a finite order part to it, namely the extension parameters become rational numbers, as we argued in detail in section \ref{ssec:semisimple}. In most mathematics applications this is already quite special as one often works over $\mathbb{Q}$, and one says that the Hodge-structure splits over the rational numbers. As physicists we are to some extend forced to work over $\mathbb{Z}$ by charge quantization. Among the cases we checked, there is a single example, the geometry \hyperref[parX322]{$X_{3,2,2}$}, where both extensions parameters turn out to be an integer, so one can set them to zero using integer shifts. As a result the LMHS completely splits over the integers. 

We still have not said anything about the origin of these extension parameters. In the following, we try to sketch part of what is known. This presentation is by no means meant to be rigorous and the interested reader should consult the relevant literature. In work by Carlson \cite{Carlson1980ExtensionsOM} on extensions of Hodge structures, the following isomorphism was established (specializing to the case at hand) 
\begin{align}
    \text{Ext}^1(H^3,\mathbb{Z}(1)) \cong J^3 = \frac{H^3_{\mathbb{C}}}{F^2 + H^3_{\mathbb{Z}}} \,.
\end{align}
The extension class on the left is exactly of the form \eqref{eq:ConiExt} with $H^{3} \cong Gr_{3}$ looking like the pure Hodge structure of a rigid Calabi-Yau threefold and $\mathbb{Z}(1)=Gr_4$ being a one-dimensional Hodge-Tate structure\footnote{The integer in parenthesis specifies the Tate twist, which effects the weights of the Hodge structure. We shall think of this as a bookkeeping device on where the pure Hodge structures sit with respect to each other in the LMHS. At the level of the periods Tate twists will introduce relative factors of $2 \pi i$.}. The object on the right hand side of the isomorphism is the so-called Griffith's intermediate Jacobian \cite{zbMATH03272123}, which can be thought of as an appropriate generalization of the Jacobian variety for elliptic curves. This correspondence suggests that the extension class \eqref{eq:ConiExt} captures some information about cycles. In order to say more, we need to introduce some relevant notions. Given a variety $X$, one can define the group of codimension $m$ subvarieties $Z^m(X)$ on which one can pose several equivalence relations, such as rational, homological and algebraic equivalence. The Chow group $CH^m(X)=Z^m(X)/Z^{m}(X)_{rat}$ captures the codimension $m$ cycles that are rationally equivalent. By means of the other equivalences, there is a natural filtration on the Chow group $ CH^m(X)_{alg} \subset CH^m(X)_{hom} \subset CH^m(X) $. The Griffith's Abel Jacobi map takes the homologically trivial cycles $Z \in CH^m(X)_{hom}$ to the intermediate Jacobian:
\begin{align}
    aj: \, CH^m(X)_{hom} \to J^m(X)\,. 
\end{align}
By the story above we thus get that $aj(Z) \in \text{Ext}^1(H^3,\mathbb{Z}(1))$, i.e.~the extension data is capturing the image of an homologically trivial cycle in the intermediate Jacobian. 
The gist of the story lies in the fact that homologically trivial cycles need not be algebraically trivial. An early example of this occurence in the case of threefolds was given in \cite{pjm/1102723329}. At this point one might wonder what this cycle is. A natural candidate arises when considering the singular geometry at the conifold point as was discussed by Bloch for the quintic threefold in \cite{10.1215/S0012-7094-85-05219-6}. Upon blowing up the double point singularities, one gets an exceptional quadric surface. The latter has two rulings whose difference of lines defines a cycle that has the desired properties, i.e.~homologically trivial but not necessarily algebraically trivial. The obstruction to algebraic triviality should be measured by the Abel-Jacobi map of the cycle and thus the corresponding extension class. Unfortunately, to the authors knowledge, there is to this date no closed form expression for this quantity for conifold degenerations of Calabi-Yau threefolds. 

For physics purposes, the important take-away of the above story is that this information enters into the period data --- and thus the physical couplings --- through the coefficients $\gamma,\delta$. It is important information for the quantization of the basis, for instance in model building applications when stabilizing moduli with fluxes. This extension data may be conveniently encoded in terms of an Sp$(4,\mathbb{R}$) basis rotation \eqref{eq:Crot} that shifts the flux quanta. To this end, it is also interesting to point out that there are examples where these numbers are rational, and even if this is not the case we find they are rational multiples of each other.

\subsubsection*{$\Gamma$-map}
As explained in section \ref{ssec:Hodge2}, the instanton map $\Gamma$ is encoded in the single holomorphic series $A(q)$ when using the canonical coordinate. Based on examples it seems to be of the form 
\begin{align}
    A(q)=\sum_{k=1}^{\infty}A_k q^k= \frac{\overline{\mathbb{Q}}}{L(f,1)} \sum_{k=1}^{\infty} \hat{A}_k q^k \,,
\end{align}
where we extracted a common factor involving an algebraic piece and an L-value. It should be noted that the L-value appears due to our normalization of the period vector, and taking the arithmetic normalization would get rid of the L-value. The coefficients $\hat{A}_k$ are complex numbers from abstract Hodge-theoretic considerations, but in all geometrical examples studied we find they take real values. Moreover, when the boundary is algebraic (as happens for examples with semisimple monodromies), these $\hat{A}_k$ become rational. In contrast to the LCS case, where the analogue of this series is integral and has a nice interpretation in terms of instanton numbers, it seems that for the conifold singularities this structure is more intricate. It would be interesting to see if this series allows for a more sophisticated notion of integrality along the lines of \cite{Schwarz:2017gnp,Jefferson:2013vfa}.

\subsection{K-points}
Having reviewed the connections between the boundary data and the degenerate variety for conifold points, we now proceed and provide a similar analysis for K-points. In the absence of a semisimple monodromy factor these points correspond to so-called Tyurin degenerations \cite{Doran2016MirrorST}. In the presence of a finite order part we can say in hindsight that the general K-point geometry (at least at the level of the LMHS) still shares the same qualitative features.

\subsubsection*{Degenerate geometry and monodromy matrix}
The geometry of a Tyurin degeneration for a Calabi-Yau threefold with $h^{2,1}=1$, is given by two (quasi) Fano-varieties intersecting in a rigid K3 surface. For a more thorough and also general discussion the reader can consult \cite{Doran2016MirrorST}. For the broader class of  singularities we expect the geometry to be similar, but an explicit description from basic principles is beyond the scope of this paper and we rather focus on structure of the periods near the boundary. One might be curious if the log-monodromy matrix contains any information about the pieces that make up the singular geometry. A very natural invariant attached to an isomorphism class of a rigid K3 surface $X$ is its transcendental lattice $T(X)$ . The latter can be uniquely identified with its positive-definite intersection form
\begin{align} \label{eq:K3transL}
    T(X) \longleftrightarrow \begin{pmatrix}
        a & b \\
        b & c
    \end{pmatrix} \,, \quad a,b,c \in \mathbb{Z}\,,
\end{align}
which has discriminant $d=ac-b^2>0$. In each SL$(2,\mathbb{Z})$ equivalence class, there is always a representative for which $-2a\leq b<2a\leq c$, so we will assume this form in the following. Furthermore, such matrices are in direct correspondence with imaginary quadratic fields $\mathbb{Q}(\sqrt{-D})$, where $d=r D$, with $r$ being a perfect square. These number fields will play a role when we look at the modularity for the K-points. 

Looking at action of the monodromy operator $N$ when restricted to the $Gr_4$ part, we can recognize it as candidate for an intersection form on a rigid K3 surface. More precisely, by using \eqref{eq:KN} for $N$ we find the matrix
\begin{align}
   \left( Q . N \right) |_{Gr_4}=\begin{pmatrix}
        a & b\\
        b & c
    \end{pmatrix} \,, \label{eq:RigidK3IntMat}
\end{align}
which enjoys the correct properties under SL$(2,\mathbb{Z})\subset \text{Sp}(4,\mathbb{Z})$. We will elaborate on the empirical evidence we find for this correspondence in the next part when we discuss the modularity associated with the K-points. In the meantime, we can turn to    \cite{Katzarkov2009GeneralizedHM,Doran2016MirrorST} for an independent check. They identify the intersection matrix for the rigid K3 occurring at the K-point of $X_{3,3}$ and agrees with what we find in \eqref{eq:BoundaryDataX33}. This correspondence could probably be made more precise by tracking the intersection form of the rigid K3 through the Clemens-Schmid exact sequence \cite{Morrison+1984+101+120}, which formally relates the singular geometry at the degeneration point to the LMHS. 
\subsubsection*{Graded pieces and modularity}
When looking at the pure Hodge structures on the graded pieces, one realizes that they look exactly like what one would expect for a K3 manifold which is rigid, also known as exceptional K3 surface. More precisely, we have 
\begin{align}
    Gr_{4} \cong H^{3,1} \oplus H^{1,3} \,, \nonumber \\
    Gr_{2} \cong H^{2,0} \oplus H^{0,2} \,,
\end{align}
where the indices of the Hodge structure on $Gr_4$ are shifted compared to the usual indices (which we have on $Gr_2$). This is just the Tate twist we have encountered already for the conifold. It is a well established result that rigid K3 surfaces over $\mathbb{Q}$ are modular. So there is a precise way in which a weight-three modular form is attached to their middle cohomology \cite{Livne1995MotivicOT}.

At the level of the LMHS, the modularity manifests itself again through the presence of critical L-values of a weight-three modular form $f$. More concretely, projecting the vector that spans $I^{3,1}$ in \eqref{eq:KLMHSbasis} onto $Gr_4$ we have 
\begin{align}
    \omega_{3,1}|_{Gr_4}= 
    \begin{pmatrix}
        1 \\ \frac{-b+i \sqrt{d}}{c}
    \end{pmatrix}=\begin{pmatrix}
        1 \\ - \frac{b}{c}+ i \mathbb{Q} \frac{ L(f,2)}{ \pi L(f,1)}
    \end{pmatrix} \sim \begin{pmatrix}
        \mathbb{Q} \pi L(f,1) \\ \mathbb{Q}  \pi L(f,1) + i \mathbb{Q}  L(f,2)\end{pmatrix} \,,  
\end{align}
where in the second step we used the scaling freedom to have a normalization that is more natural from the arithmetic point of view. Again, it is actually the latter normalization that comes out naturally when computing the periods from the Picards-Fuchs equation.  Similar to conifold points, the real period in all the examples we checked is given (up to some conventional factor of $\pi$) by a rational multiple of a critical L-value. As before, this occurence is in line with modern conjectures about critical L-values and periods.    

A major difference to the conifold case is that weight-three modular forms always enjoy the CM property. It can already be seen from the functional relation \eqref{eq:LFuncRel} obeyed by the associated L-function, that for a weight-three modular form $f_N$ of level $N$, there is always an algebraic relation relating the two critical L-values
\begin{align}
    \pi L(f_N,1) = \frac{1}{2} \sqrt{N} L(f_N,2) \, .
\end{align}
In all the examples we checked, the CM field for the occurring modular forms is given by $\mathbb{Q}( \sqrt{-D})$, where $D$ is the non-square part of the discriminant mentioned below equation \ref{eq:K3transL}, which is related to the intersection form of the candidate rigid K3 occurring in the degenerate geometry. According to discussions in \cite{Livne1995MotivicOT, ELKIES2013106} this would be expected and provides a nice consistency check. 

The transition matrix that rotates the Frobenius solution for K-points into the monodromy weight frame also has the quasi-periods associated with the weight-three modular forms among its entries. As there is always CM, we expect the quasi-periods to be completely expressible in terms of the critical L-values. We find indeed evidence for a relationship of the form
\begin{align}
    \eta_+ = \mathbb{Q}\frac{ \pi}{L(f,1)} + \mathbb{Q} \pi L(f,1)\,.
\end{align}
Up to shifts, they are also inverses of the critical L-values. For all the K-point examples treated in section \ref{sec:examples}, except the operator 3.7, the quasiperiods showing up in the transition matrices are actually simply given by the inverses of their respective L-values without shift.

\subsubsection*{Extension data}
From the Hodge-theoretic construction in section \ref{ssec:Kconstruction}, it follows that the extension data for type $\text{II}_0$ singularities is given by two real numbers, defined up to integer shifts
\begin{align}
    \text{Ext}^1(Gr_4,Gr_2) \sim \begin{pmatrix}
        \gamma + i \delta
    \end{pmatrix} \in \mathbb{C}/\left(\mathbb{Z} \oplus \tau \mathbb{Z} \right)  \,,
\end{align}
where the quotient just restricts to the usual fundamental domain in the lattice quotient of the torus. The presence of a non-trivial semisimple monodromy factor with two distinct eigenvalues\footnote{This excludes a non-trivial $T_{ss}=-\mathbb{I}$ corresponding to local exponents $a_i=\tfrac{1}{2}\mod 1$, which would yield $s(a_1)=0$ in \eqref{eq:TssK} and therefore no constraints on $\gamma,\delta$.} constrains these numbers to specific rational values $\gamma,\delta \in \mathbb{Q}$, see below \eqref{eq:Kgammadelta}

\subsubsection*{$\Gamma$-map}
Finally we turn to the formulation in terms of the instanton map $\Gamma(q)$ reviewed in section \ref{ssec:Hodge2}. For the K-points the holomorphic series $A(q)$ does not explicitly appear in the period vector when using the canonical coordinate. It is thus more convenient to formulate everything in terms of the second function $B(q)$, which takes the form
\begin{align}
    B(q)=\sum_{k=1}^{\infty}B_k q^k = \frac{\overline{\mathbb{Q}}}{L(f,1)^2} \sum_{k=1}^{\infty} \hat{B}_k q^k \,, \label{eq:KpointSeries}
\end{align}
where we extracted a common factor involving an algebraic piece $\overline{\mathbb{Q}}$ and a critical L-value piece. Similar to conifold points, from abstract Hodge-theoretic considerations the $\hat{B}_k$ are complex numbers, but take real values in all geometrical examples. Furthermore, in the presence of a semisimple monodromy factor the $\hat{B}_k$ are rational. Again, it would be interesting to investigate if some more sophisticated notion of integrality holds for these series. For the K-points occurring in hypergeometric examples, the series \eqref{eq:KpointSeries} agrees with the recent computation of the Givental J-function in the context of gauged linear sigma models associated with the same geometries \cite{Erkinger:2022sqs}. We explain this match in precise detail for the \hyperref[parX33]{$X_{3,3}$} example in the next section.

\section{Geometric examples of integral periods}\label{sec:examples}
In this section we collect explicit geometric examples that serve as complementary data to the more abstract discussion laid out so far. We record the transition matrices 
rotating from the respective Frobenius solutions into the monodromy weight frame. We then extract the boundary data and obtain the beginnings of the series expansions for coordinate changes and other relevant (coupling) functions.  

Let us briefly elaborate on the procedure to extract this information, in particular the series expansions. To this end, it is convenient to rotate the period vector to the LMHS basis, which yields \eqref{eq:Cperiod} for conifold points and \eqref{eq:Kperiods} for K-points. We note that this period vector has been rescaled by \eqref{eq:fC} and \eqref{eq:fK} for conifold and K-points respectively. The canonical coordinate change for conifold points is then given by inverting \eqref{eq:Cnatural}, while for K-points one takes the (rescaled) second period in \eqref{eq:Kperiods}. Similarly the holomorphic series $A(q)$ and $B(q)$ (specifying the instanton map $\Gamma(q)$) in this coordinate $q$ may then be read off from period entries in this LMHS basis. Finally, the Yukawa coupling computed in terms of this coordinate $q$ is then straightforwardly obtained from \eqref{eq:CYukawa} and \eqref{eq:KYukawa2} respectively.

The logic behind the order of examples in the following presentation is to gradually increase the complexity of the objects involved in the computation. This is ultimately tied to the finite order part of the monodromy: the more distinct eigenvalues it has, the more algebraic relations one has among the different components of the periods, making it more easily solvable. The first class of examples we describe occur in the hypergeometric models for which computations can be done analytically due to their rather special nature. Then we pick individual examples where we computations can only be done numerically and quantities have to be matched afterwards. All non-hypergeometric examples are labeled by their operator number according to the Calabi-Yau differential operator database \cite{almkvist2005tables}.

In order to prepare for our example analysis, let us first set up the transition matrices from the complex Frobenius basis to the integral monodromy weight basis at conifold and K-points. To this end, it is convenient to separate the transition matrix into two factors as 
\begin{equation}\label{mfactorized}
    \hat{m}_{s} = \hat{m}^{\rm LMHS}_{s} \hat{m}^{\rm Frob}_{s}\, ,
\end{equation}
where the first $m_{s}^{\rm LMHS}$ encodes the defining data of the LMHS, while the second $m_{s}^{\rm Frob}$ arranges the complex Frobenius periods into suitable linear combinations with convenient normalizations. Based off of \eqref{eq:Cbasis}, we take for conifold points as LMHS transition matrix
\begin{equation}\label{mC}
    \hat{m}^{\rm LMHS}_{\rm C} = \begin{pmatrix}
    0 & 1 & 0 & 0 \\
    1 & \gamma & 0 & 1 \\
    \tau & \delta & 0 & \bar{\tau} \\
    \delta-\gamma \tau & 0 & 1 & \delta-\gamma \bar\tau
\end{pmatrix}\, ,
\end{equation}
where $\tau$ is the rigid period of the conifold, and $\gamma,\delta \in \mathbb{R}$ the extension data specifying the integral basis. Based on \eqref{eq:Kbasis} we take for K-points
\begin{equation}\label{mK}
    \hat{m}^{\rm LMHS}_{\rm K} = \left(\begin{array}{cccc}
 1 & 0 & 1 & 0 \\
 \tau & 0 & \bar{\tau} & 0 \\
 \delta+\gamma \tau  & a+b\tau & \delta+\gamma \bar{\tau} & a+b\bar{\tau} \\
\gamma & b+c\tau & \gamma & b+c\bar{\tau} \\
\end{array}
\right) \, ,
\end{equation}
where $a,b,c \in \mathbb{Z}$ specifies the monodromy matrix \eqref{eq:KN}, $\tau = (-b+i\sqrt{b^2-ac})/c$ the rigid K3 period, and $\gamma,\delta \in \mathbb{R}$ extension data fixing the integral basis. We find that this decomposed form of the transition matrix $\hat{m}_s$ is instructive in the study of examples, as the remaining factor $\hat{m}_s^{\rm Frob}$ mostly has vanishing off-diagonal components. The non-vanishing entries can be given precise interpretations as L-function values and other arithmetic data associated to the degenerate variety.

\subsection{Hypergeometric examples}
A class of examples where computations can be done explicitly is found in the hypergeometric models which are conveniently summarized in table \ref{table:hypergeom}. The complex structure moduli space of these models has three singular points, it is simply $\mathbb{P}^1/ \{ 0,1,\infty \} $, where the LCS point is located at $0$ by convention. In the appropriate formulation, the periods centered around the LCS point of these models can be expressed completely in terms of hypergeometric functions and their derivatives. For these one can use inversion formulas to get analytic expression for the periods centered around the point at infinity. Three of these models have conifold singularities at the infinite point, while other three have K-points at infinity. This makes these examples a natural starting point for our investigation. Explicit expressions for the transition matrices from the Frobenius basis to the integral symplectic basis have already been obtained in \cite{Joshi:2019nzi}. However, the chosen integral symplectic frame was not adapted to the weight filtration structure as dictated by Hodge theory, which makes it difficult to connect with the results of \cite{Bonisch:2022mgw} in the case of the conifold point. Once in the monodromy weight frame, it will be clear from the expressions that the conifold points follow the same story as \cite{Bonisch:2022mgw}. For the reader interested in reproducing the results for these hypergeometric models, we want to mention that we have attached two notebooks to the submission detailing the computations for these conifold and K-points.

\subsubsection{Conifold singularities in hypergeometric examples}
Among the fourteen hypergeometric examples, three have a so called generalized conifold point, i.e. there is a finite order monodromy on top of the usual conifold monodromy. Their Riemann indices are given by $a,\tfrac{1}{2},\tfrac{1}{2},1-a$, where $a=\tfrac{1}{4},\tfrac{1}{3},\tfrac{1}{6}$ depending on the specific example. Due to the fractional nature of the Riemann indices, the finite order part of the monodromy will have three different eigenvalues. The modular forms associated to the L-values appearing in the transition matrix will have the CM property and the extension data will be rational. 

\paragraph{$\mathbf{X_{4,2}}$ geometry.}\label{parX42} We start with the geometry $X_{4,2}$, which has a conifold point with indices $\tfrac{1}{4},\tfrac{1}{2},\tfrac{1}{2},\tfrac{3}{4}$ at infinity.  The topological numbers of its mirror are
\begin{equation*}
    \kappa=8\, ,\qquad c_2=56\, ,\qquad \chi=-176\,.
\end{equation*}
The singularity we are interested in has $a=\frac{1}{4}$, so one has to go to the $4$-fold cover of moduli space to make the monodromy unipotent. With the methods explained in the appendix \ref{appendix:transition} we can compute the transition matrix from the complex Frobenius basis at the conifold point to the integral LCS basis analytically with the result 
\begin{equation}\label{eq:m42}
   m_{\rm C}= \left(
\begin{array}{cccc}
 \frac{\left(\frac{1}{32}-\frac{i}{32}\right) \Gamma \left(\frac{1}{4}\right)^6}{\pi ^{9/2}} & -\frac{\pi -4 i (1+\log (8))}{8 \pi ^2} & -\frac{1}{4 \pi} & \frac{\left(\frac{1}{4}+\frac{i}{4}\right) \pi ^{3/2}}{\Gamma \left(\frac{1}{4}\right)^6} \\
 \frac{i \Gamma \left(\frac{1}{4}\right)^6}{32 \pi ^{9/2}} & -\frac{i (1+\log (8))}{4 \pi ^2} & -\frac{1}{8 \pi} & -\frac{i \pi ^{3/2}}{4 \Gamma \left(\frac{1}{4}\right)^6} \\
 \frac{\Gamma \left(\frac{1}{4}\right)^6}{8 \pi ^{9/2}} & -\frac{1}{4 \pi } & 0 & \frac{\pi ^{3/2}}{\Gamma \left(\frac{1}{4}\right)^6} \\
 \frac{i \Gamma \left(\frac{1}{4}\right)^6}{16 \pi ^{9/2}} & -\frac{i (1+\log (8))}{\pi ^2} & -\frac{1}{2 \pi } & -\frac{i \pi ^{3/2}}{2 \Gamma \left(\frac{1}{4}\right)^6} \\
\end{array}
\right)\, .
\end{equation}
For the conifold point at infinity there is a more convenient integral frame, the monodromy weight frame, that respects the local weight filtration, which is obtained by an additional  Sp$(4,\mathbb{Z})$ matrix
\begin{equation}\label{eq:conifoldmatrix}
    m^{\rm C}_{\rm LCS}  = \left(
\begin{array}{cccc}
 4 & 0 & -1 & -2 \\
 2 & 0 & 0 & -1 \\
 1 & 2 & 0 & -1 \\
 0 & -1 & 0 & 0 \\
\end{array}
\right)\;,
\end{equation}
which rotates us from the integral LCS to the integral conifold frame. The first row of this matrix --- corresponding to the vanishing cycle at the conifold point --- turns out to be universal for the three hypergeometric examples. From the perspective of the LCS frame, this means that the same bound state of D-branes, consisting of 2 D6, $-1$ D4 and $-4$ D0-branes (after a multiplication by the pairing \eqref{eq:pairing}), becomes massless at this conifold point in all these models. 

In order to compare with the conventions of \cite{Bonisch:2022mgw}, let us also write out the transition matrix from the local Frobenius basis to the integral monodromy weight basis for completeness, in such a way that $\sqrt{\kappa}(2 \pi)^3$ appears as coefficient of the vanishing cycle. Combining \eqref{eq:m42} and \eqref{eq:conifoldmatrix} we find
\begin{align}\label{eq:mhat42}
(2 \pi)^3 4 \sqrt{2} \, \hat{m}_{\rm C} = &\left(
\begin{array}{cccc}
 0 & -8 \sqrt{2} \pi ^2 & 0 & 0 \\
 \frac{2 \sqrt{2} \Gamma \left(\frac{1}{4}\right)^6}{\pi ^{3/2}} & -8 \sqrt{2} \pi ^2 & 0 & \frac{16 \sqrt{2} \pi ^{9/2}}{\Gamma \left(\frac{1}{4}\right)^6} \\
 \frac{(1+i) \sqrt{2} \Gamma \left(\frac{1}{4}\right)^6}{\pi ^{3/2}} & -4 \sqrt{2} \pi ^2 & 0 & \frac{(8-8 i) \sqrt{2} \pi ^{9/2}}{\Gamma \left(\frac{1}{4}\right)^6} \\
 -\frac{i \sqrt{2} \Gamma \left(\frac{1}{4}\right)^6}{\pi ^{3/2}} & 8 i \sqrt{2} \pi  (1+\log (8)) & 4 \sqrt{2} \pi ^2 & \frac{8 i \sqrt{2} \pi ^{9/2}}{\Gamma \left(\frac{1}{4}\right)^6} \\
\end{array}
\right)
\end{align}
In line with our discussion below \eqref{mfactorized}, it proves to be convenient to decompose this transition matrix into two factors, one based on LMHS information, and another minimally rotating and rescaling the complex Frobenius periods. Let us repeat the LMHS transition matrix \eqref{mC} for this first example explicitly as
\begin{equation}
    \hat{m}^{\rm LMHS}_{\rm C} = \begin{pmatrix}
    0 & 1 & 0 & 0 \\
    1 & \gamma & 0 & 1 \\
    \tau & \delta & 0 & \bar{\tau} \\
    \delta-\gamma \tau & 0 & 1 & \delta-\gamma \bar\tau
\end{pmatrix}\, ,
\end{equation}
which is specified by the boundary data
\begin{equation}
    \tau=\frac{1}{2}+\frac{i}{2}\, , \qquad k=2\, , \quad n=4\, , \qquad  \gamma = 1, \quad \delta=\frac{1}{2}\, .
\end{equation}
On the other hand, the matrix acting directly on the Frobenius periods reads
\begin{equation}\label{mFrob42}
    (2 \pi)^3 4 \sqrt{2} \, \hat{m}_{\rm C}^{\rm Frob} = \begin{pmatrix}
       64  \pi^2 L(f_{32},1) & 0 & 0 & 0 \\
       0 & (2\pi i)^2 \sqrt{\kappa} & 0 & 0 \\
       0 & (2\pi i)^2 \sqrt{\kappa} \frac{k}{n}h & -(2\pi i)^2 \sqrt{\kappa}\frac{k}{n} & 0 \\
       0 & 0 & 0 & \frac{\pi}{L(f_{32},1)}
    \end{pmatrix}\, .
\end{equation}

We want to emphasize that the form of the transition matrix given by \eqref{mC} and \eqref{mFrob42} will be the same for all conifold points, up to some model-dependent LMHS data $\tau, k,\gamma, \delta$ and Frobenius data $L(f,1), m, h$.\footnote{Going beyond the hypergeometric cases requires us to introduce quasi-periods $\theta$, which is just additional geometrical information that needs to be specified.} These numbers can all be determined efficient, for instance numerically, in examples. Or, on the flip side, given this geometrical data, one can straightforwardly rotate the Frobenius solution to the integral basis. This latter perspective is similar to the transition matrix \eqref{eq:mLCS} at LCS, which is specified in terms of the topological data $\kappa,c_2,\chi,\sigma$. Let us briefly go over the parameters appearing in \eqref{eq:mhat42}:
\begin{itemize}
    \item The integer $k=2$ specifies the unipotent part of the monodromy \eqref{Eq:ConiN}, while $n=4$ refers to the $n$-fold cover of moduli space on which the monodromy is unipotent.
    \item The extension data $\gamma,\delta$ takes rational values $\gamma=1$ and $\delta=\tfrac{1}{2}$, where we note that one may set $\gamma=0$ by an additional Sp$(4,\mathbb{Z})$ transformation \eqref{Eq:ExtShift}.
    \item The L-function values $L(f_{32},1)$ are associated to the weight-four modular form $f_{32}$ of level $32$ labeled by $32.4.a.b$ in the database \cite{lmfdb}. This modular form has complex multiplication (CM) by $d=-4$. The numerical values of the first two L-function values are
    \begin{equation}\label{L42}
        L(f_{32},1) = \frac{1}{16\sqrt{2}\pi^5} \Gamma(\tfrac{1}{4})^6 \Gamma(\tfrac{1}{2})^3\, ,\qquad L(f_{32},2) = \frac{1}{64\sqrt{2}\pi^4} \Gamma(\tfrac{1}{4})^6 \Gamma(\tfrac{1}{2})^3\, .
    \end{equation}
    Seemingly as a consequence, there is a simple algebraic relation between the L-values explaining why we can write the transition matrix solely in terms of $L(f_{32},1)$. Such a relation is also observed for L-values of other weight-four modular forms with CM, although the authors were not able to find a general statement in the literature about this. Another observation is that, instead of the quasiperiods of the relevant modular form as outlined in \cite{Bonisch:2022mgw}, there is simply the inverse of the L-value appearing in the transition matrix. This seems to hint that at least in the observed cases, CM also introduces a relation between the periods and quasi-periods of a modular form. However, we did not explicitly compute the quasi-periods from the modular form. 
    \item Finally, the parameter $h$ induces an off-diagonal component in the Frobenius matrix \eqref{mFrob42}. It adds a piece along $(0,0,0,1)$ (corresponding to the conifold cycle) to the column vector $(1,\gamma,\delta,0)$ in the LMHS matrix \eqref{mC}. This can be understood as a shift $t \to t+kh$ for the covering coordinate (or equivalently a rescaling of the normal coordinate). The numerical value of $h$ can be given in a uniform way for all hypergeometric examples
\begin{align}
   h=\frac{H_{a-1/2}+H_{1/2-a}+\log(256 \mu)}{2\pi i} \,,
   \label{eq:ConiFrobCombi}
\end{align}
with $H_n$ being the harmonic numbers, and $\mu$ and $a=a_1$ are given in table \ref{table:hypergeom}. For the example under consideration recall that $a=\tfrac{1}{4}$ and $\mu = 2^{10}$. This special number originates from our choice of the normalization in the complex Frobenius bases. It describes the rotation from the MeijerG function solution to the Frobenius solution and this normalization. It is an interesting question how this generalizes to the non-hypergeometric cases where this interpretation is missing. 
\end{itemize}  
Let us next record the boundary data associated to this singularity, using the integral monodromy weight frame. The semi-simple factor of the monodromy matrix is given by
\begin{equation}\label{eq:Tss42}
   T_{ss} = \left(
\begin{array}{cccc}
 -1 & 0 & 0 & 0 \\
 -1 & -1 & 2 & 0 \\
 0 & -1 & 1 & 0 \\
 -\frac{1}{2} & 1 & -1 & -1 \\
\end{array}
\right) \, .
\end{equation}
Going subsequently to the covering space we find,  in the normalized form of section \ref{ssec:conifoldconstruction}, as log-monodromy matrix and LMHS
\begin{align}
  N= \begin{pmatrix}
   0 & 0 & 0 & 0 \\
   0 & 0 & 0 & 0 \\
   0 & 0 & 0 & 0 \\
   -2 & 0 & 0 & 0 \\
   \end{pmatrix}\,, \quad \omega_{3,0} = \begin{pmatrix}
   0 \\ 1 \\ \frac{1}{2}+  \frac{i}{2} \\ -\frac{i}{2}
   \end{pmatrix} \,, \quad \omega_{2,2}= \begin{pmatrix}
   1 \\ 1 \\ \frac{1}{2} \\  \frac{3 \log(2)}{i \pi}
   \end{pmatrix} \,.
\end{align}
Note that the eigenvectors of the semisimple factor \eqref{eq:Tss42} are $\omega_{3,0}$ and $\omega_{0,3}$ with eigenvalues $\pm i$, and $\omega_{2,2}$ and $\omega_{1,1}=N\omega_{2,2}$ with eigenvalue $-1$. Using this boundary data and the procedure outlined in section \ref{ssec:conifoldconstruction}, we can make the coordinate change to the natural Hodge-theoretic coordinate $q$ by
\begin{align}
    z =  8 q \left(1-\frac{4}{9}q^4+\frac{709}{3969}q^8-\frac{373424}{12966723}q^{12}+ \frac{23219329}{5718324842} q^{16}\dots \right)\, ,
\end{align}
where $z$ denotes the local normal coordinate used in the Picard-Fuchs equation. The first few terms in the holomorphic series $A(q)$ specifying the instanton map are given by 
\begin{align}
    A(q)= -  \frac{ \sqrt{2}}{ L(f_{32},1)} \cdot \left( q- \frac{7}{30}q^5-\frac{65}{3528}q^{9}+\frac{7762663}{385296912}q^{13}+ \frac{929036467}{84646767744} q^{17}\dots \right)\,.
\end{align}
From this series we compute the Yukawa coupling, either directly or through \eqref{eq:CYukawa}, to be
\begin{align}
    Y(q)= \frac{16 \pi^2}{ L(f_{32},1)^2} \left(q^{2}-\frac{7}{3}q^6+\frac{454}{441}q^{10} + \frac{1687031}{1852389} q^{14}- \frac{133727507}{635369427} q^{18}+ \dots \right)\, .
\end{align}
Note that the coefficients in this series expansion take just rational values.

\paragraph{$\mathbf{X_{3,2,2}}$ geometry.}\label{parX322} For our next example we consider the $X_{3,2,2}$, for which we study its conifold point at infinity with local exponents $\tfrac{1}{3},\tfrac{1}{2}, \tfrac{1}{2}, \tfrac{2}{3}$, i.e.~$a=\frac{1}{3}$. To make the monodromy unipotent, one has to go to the $6$-fold cover of moduli space. The topological numbers of its (mirror) manifold are
\begin{equation*}
    \kappa=12\qquad c_2=60\qquad \chi=-144\;.
\end{equation*}
Using this information we can fix the integral basis at the conifold point; the transition matrix from the local Frobenius basis to the integral monodromy weight basis is given by
\begin{align}
(2\pi)^3 2 \sqrt{3} \, \hat{m}_{\rm C} = &\left(
\begin{array}{cccc}
 0 & -8 \sqrt{3} \pi ^2 & 0 & 0 \\
 \frac{9 \sqrt{3} \Gamma \left(\frac{1}{3}\right)^9}{4 \pi ^3} & 0 & 0 & \frac{128 \pi ^6}{\Gamma \left(\frac{1}{3}\right)^9} \\
 \frac{9 \left(\sqrt{3}+i\right) \Gamma \left(\frac{1}{3}\right)^9}{8 \pi ^3} & 0 & 0 & \frac{64 \left(3-i \sqrt{3}\right) \pi ^6}{3 \Gamma
   \left(\frac{1}{3}\right)^9} \\
 0 & 8 i \sqrt{3} \pi  (3+\log (16)) & 8 \sqrt{3} \pi ^2 & 0 \\
\end{array}
\right)\, ,
\end{align}
which may be factorized into the LMHS transition matrix \eqref{mC} and a transition matrix acting on the Frobenius basis
\begin{equation}
(2\pi)^3 2 \sqrt{3} \,     \hat{m}_{\rm C}^{\rm Frob} = \begin{pmatrix}
        216 \pi^2 L(f_{9},1) & 0 & 0 & 0 \\
        0 & (2\pi i)^2 \sqrt{\kappa} & 0 & 0 \\
        0 & (2\pi i)^2 \sqrt{\kappa} \frac{k}{n} h & -(2\pi i)^2 \sqrt{\kappa} \frac{k}{n}& 0 \\
        0 & 0 & 0 & \frac{4\pi}{\sqrt{3} L(f_{9},1)}
    \end{pmatrix}\, ,
\end{equation}
with the boundary data given by
\begin{equation}
    \tau = \frac{1}{2}+\frac{i}{2\sqrt{3}}\, , \qquad k=6\, , \quad n=6\, , \qquad \gamma =\delta= 0\, .
\end{equation}
where $h$ was defined in \eqref{eq:ConiFrobCombi} (with $a=\tfrac{1}{3}$ and $\mu = 2^3 3^3$), The modular form $f_9$ denotes $9.4.a.a$\footnote{The same modular form is associated to the ordinary conifold point in the $X_{4,3}$ geometry.} which has CM by $d=-3$. Its L-function values are given by
\begin{equation}
    L(f_9,1) = \frac{1}{32\sqrt{3}\pi^5} \Gamma(\tfrac{1}{3})^9\, , \qquad L(f_9, 2) = \frac{1}{96\pi^4}\Gamma(\tfrac{1}{3})^9\, .
\end{equation}
As for the previous example, only $L(f_9,1)$ appears in the transition matrix due to additional relation among the L-values. Before going to the covering space, we extract the finite order monodromy matrix 
\begin{align}\label{eq:Tss322}
   T_{ss} = \begin{pmatrix}
        -1 & 0 & 0 & 0 \\
        0 & -2 & 3 & 0 \\
        0 & -1 & 1 & 0 \\
       0 & 0 & 0 & -1 \end{pmatrix}\, .
\end{align}
In the $n=6$-fold cover we extract the boundary data and normalize it as in section \ref{ssec:conifoldconstruction}, which gives 
\begin{align}
  N= \begin{pmatrix}
   0 & 0 & 0 & 0 \\
   0 & 0 & 0 & 0 \\
   0 & 0 & 0 & 0 \\
   -6 & 0 & 0 & 0 \\
   \end{pmatrix}\,, \quad \omega_{3,0} = \begin{pmatrix}
   0 \\ 1 \\ \frac{1}{2}+  \frac{i}{2\sqrt{3}} \\ 0
   \end{pmatrix} \,, \quad \omega_{2,2}= \begin{pmatrix}
   1 \\ 0 \\ 0 \\ -i \frac{4 \log(2)}{\pi}
   \end{pmatrix} \,.
\end{align}
Note that the semisimple factor indeed acts as an automorphism on the LMHS, having $\omega_{3,0}$ and $\omega_{0,3}$ as eigenvectors with eigenvalues $e^{2\pi i/3}$ and $\omega_{2,2}$ and $\omega_{1,1}=N\omega_{1,1}$ with eigenvalue $-1$. Also observe that the boundary data of this example is rather special as $\gamma=\delta=0$. This means that after the appropriate coordinate change that makes the boundary data $\mathbb{R}$-split, the LMHS will actually be completely split over $\mathbb{Z}$, i.e.~it is a direct sum of pure Hodge structures. 

Using this boundary data and the procedure outlined in section \ref{ssec:conifoldconstruction}, we can make the coordinate change to the natural Hodge-theoretic coordinate $q$ by
\begin{align}
    z = 2^{4/3} q \left(1-\frac{28}{675}q^6+\frac{80458}{55130625}q^{12}+\frac{64027928}{2933074546875}q^{18}+ \dots \right)\, ,
\end{align}
where $z$ is the local normal coordinate for the conifold point used in the Picard-Fuchs equation. In turn, we obtain the holomorphic series $A(q)$ specifying the instanton map
\begin{align}
    A(q)=  -\frac{2\; 2^{1/3}\sqrt{3}}{27 L(f_9,1)} \cdot \left( q- \frac{272}{14175}q^7-\frac{5256154}{6450283125}q^{13}+\frac{82545304736}{16551339668015625}q^{19}+\dots \right)\, . \nonumber
\end{align}
Next we compute the Yukawa coupling, either directly or through \eqref{eq:CYukawa}, to be
\begin{align}
    Y(q)=   \frac{32 \;2^{2/3}\pi^2}{ 81  L(f_9,1)^2}\left(  q^{2}-\frac{544}{2025}q^8-\frac{1560244}{496175625}q^{14}+ \dots \right)\, .
\end{align}
\paragraph{$\mathbf{X_{6,2}}$ geometry.}\label{parX62} For our final generalized conifold point in the hypergeometric examples we consider the $X_{6,2}$, which has Riemann indices $\tfrac{1}{6},\tfrac{1}{2},\tfrac{1}{2},\tfrac{5}{6}$, i.e.~$a=\frac{1}{6}$. To make the monodromy unipotent, one has to go to the $3$-fold cover of moduli space. The topological numbers of this manifold are
\begin{equation*}
    \kappa=4\qquad c_2=52\qquad \chi=-256\;.
\end{equation*}
The transition matrix from the complex Frobenius basis to the integral monodromy weight frame reads
\begin{align}
(2 \pi)^3 16 \, \hat{m}_{\rm C} = &\left(
\begin{array}{cccc}
 0 & -8 \pi ^2 & 0 & 0 \\
 \frac{6 \sqrt[3]{2} \sqrt{3} \Gamma \left(\frac{1}{3}\right)^9}{\pi ^3} & -\frac{16 \pi ^2}{3} & 0 & \frac{4\ 2^{2/3} \pi ^6}{9 \Gamma \left(\frac{1}{3}\right)^9} \\
 -\frac{6 (-1)^{2/3} \sqrt[3]{2} \sqrt{3} \Gamma \left(\frac{1}{3}\right)^9}{\pi ^3} & -\frac{8 \pi ^2}{3} & 0 & \frac{4 \sqrt[3]{-1} 2^{2/3} \pi ^6}{9 \Gamma \left(\frac{1}{3}\right)^9} \\
 \frac{6 i \sqrt[3]{2} \Gamma \left(\frac{1}{3}\right)^9}{\pi ^3} & -\frac{4}{3} i \pi  (3+16 \log (2)) & \frac{8 \pi ^2}{3} & -\frac{4 i 2^{2/3} \pi ^6}{9 \sqrt{3} \Gamma \left(\frac{1}{3}\right)^9} \\
\end{array}
\right),
\end{align}
which may be decomposed according to \eqref{mfactorized} into an LMHS factor $\hat{m}_{\rm C}^{\rm LMHS}$ and a matrix acting on the Frobenius solution
\begin{equation}
    (2 \pi)^3 16 \, \hat{m}^{\rm Frob}_{\rm C} = \begin{pmatrix}
       64 \pi^2 L(f_{108},1) & 0 & 0 & 0 \\
       0 & (2\pi i)^2 \sqrt{\kappa} & 0 & 0 \\
       0 & (2\pi i)^2 \sqrt{\kappa}\frac{k}{n}h & -(2\pi i)^2 \sqrt{\kappa}\frac{k}{n} & 0 \\
       0 & 0 & 0 & \frac{\pi}{4 \,L(f_{108},1)}
    \end{pmatrix}
\end{equation}
where $\hat{m}_{\rm C}^{\rm LMHS}$ is fixed through \eqref{mC} by the boundary data
\begin{equation}
    \tau = \frac{1}{2}+\frac{i\sqrt{3}}{2}\, , \qquad k = 1 \, , \quad n=3\, , \qquad \gamma = \frac{2}{3} \, , \quad \delta = \frac{1}{3} \, ,
\end{equation}
with $h$ defined in \eqref{eq:ConiFrobCombi} (with $a=\tfrac{1}{6}$ and $\mu =2^8 3^3$). The modular form $f_{108}$ denotes $108.4.a.b$, which has CM by $d=-3$. Its L-function values are
\begin{equation}
    L(f_{108},1) = \frac{3\sqrt{3}}{16 2^{1/3}\pi^5}\Gamma(\tfrac{1}{3})^9\, , \qquad L(f_{108},2) =\frac{1}{32 2^{1/3}\pi^4}\Gamma(\tfrac{1}{3})^9 \, , 
\end{equation}
Before going to the covering space, we extract the finite order monodromy matrix for completeness
\begin{align}\label{eq:Tss62}
   T_{ss} = \begin{pmatrix}
        -1 & 0 & 0 & 0 \\
        -1 & 0 & 1 & 0 \\
        0 & -1 & 1 & 0 \\
       -\frac{1}{3} & 1 & -1 & -1 \end{pmatrix}\, .
\end{align}
Upon going to the covering space we extract the boundary data and normalize it as described in section \ref{ssec:conifoldconstruction}, which gives 
\begin{align}
  N= \begin{pmatrix}
   0 & 0 & 0 & 0 \\
   0 & 0 & 0 & 0 \\
   0 & 0 & 0 & 0 \\
   -1 & 0 & 0 & 0 \\
   \end{pmatrix}\,, \quad \omega_{3,0} = \begin{pmatrix}
   0 \\ 1 \\ \frac{1}{2}+  \frac{i\sqrt{3}}{2} \\ -\frac{i}{\sqrt{3}}
   \end{pmatrix} \,, \quad \omega_{2,2}= \begin{pmatrix}
   1 \\ \frac{2}{3} \\ \frac{1}{3} \\ -i \frac{8 \log(2)}{\pi}
   \end{pmatrix} \,.
\end{align}
Notice that this LMHS indeed has $T_{ss}$ in \eqref{eq:Tss62} as an automorphism, with eigenvalues $e^{\pm 2\pi i/3}$ for $\omega_{3,0}$ and $\omega_{0,3}$, and eigenvalue $-1$ for $\omega_{2,2}$ and $\omega_{1,1} = N\omega_{2,2}$. Using this boundary data and the procedure outlined in section \ref{ssec:conifoldconstruction}, we can make the canonical coordinate change 
\begin{align}
    z = 32 \cdot 2^{1/3} q \left(1-\frac{40}{27}q^3+\frac{36856}{18225}q^6-\frac{544612}{295245}q^{9}+ \dots \right)\, .
\end{align}
The first few terms in the holomorphic series $A(z)$ specifying the instanton map are given by 
\begin{align}
    A(q)= -\frac{4 \; 2^{1/3}}{L(f_{108},1)} \cdot \left(q- \frac{70}{81} q^4-\frac{314432}{1148175}q^{7}+\frac{2713286}{7971615}q^{10}+\dots \right)\, .
\end{align}
In turn, the Yukawa coupling is obtained through \eqref{eq:CYukawa} to be
\begin{align}
    Y(q)=\frac{62 \; 2^{2/3} \pi^2}{ L(f_{108},1)^2} \left( q^{2}-\frac{560}{81}q^5+\frac{2588864}{164025}q^{8} + \dots \right)\, .
\end{align}

\subsubsection{K-points}
In this section we study --- in similar fashion as the above conifold points --- the three K-points arising in the hypergeometric models. We start with the $X_{3,3}$, where we display our methodology in the greatest detail. We then repeat our analysis more briefly for the other two K-points.

\paragraph{$\mathbf{X_{3,3}}$ geometry.}\label{parX33} We begin our investigation of K-point examples with the Calabi-Yau threefold $X_{3,3}$. Its periods are obtained from the hypergeometric Picard-Fuchs equation \eqref{eq:PF} with indices $\tfrac{1}{3}, \tfrac{1}{3}, \tfrac{2}{3}, \tfrac{2}{3}$. This first set of solutions at the K-point is given in the complex Frobenius basis \eqref{Kfrobansatz}. In order to fix an integral basis we use the topological data of its mirror
\begin{equation}
    \kappa= 9\, ,\qquad c_2=54\,, \qquad \chi=-144\, .
\end{equation}
By using this information we can fix an integral basis in the LCS regime. In turn, by analytic continuation of the periods we can then determine an integral basis in the K-point phase. Without performing these steps here explicitly, let us directly state the transition matrix from the complex Frobenius basis \eqref{Kfrobansatz} at the K-point to this integral basis as
\begin{equation}
    m_{\rm K} = \scalebox{0.77}{$\left(
\begin{array}{cccc}
 \frac{\left(\pi -3 i \sqrt{3} \pi +9 \left(\sqrt{3}+3 i\right) \log (3)\right) \Gamma \left(\frac{1}{3}\right)^6}{64 \pi ^5} & \frac{\left(3-i \sqrt{3}\right)
   \Gamma \left(\frac{1}{3}\right)^6}{32 \pi ^4} & \frac{2 (-1)^{2/3} \pi  \left(3 \sqrt{3} (2+3\log (3))+\left(2-i \sqrt{3}\right) \pi \right)}{27 \Gamma
   \left(\frac{1}{3}\right)^6} & \frac{4 \sqrt[6]{-1} \pi ^2}{9 \sqrt{3} \Gamma \left(\frac{1}{3}\right)^6} \\
 -\frac{i \left(\sqrt{3} \pi -27 \log (3)\right) \Gamma \left(\frac{1}{3}\right)^6}{96 \pi ^5} & \frac{\Gamma \left(\frac{1}{3}\right)^6}{16 \pi ^4} & \frac{2 i \pi
    \left(\sqrt{3} \pi +9 (2+3\log (3))\right)}{81 \Gamma \left(\frac{1}{3}\right)^6} & \frac{4 \pi ^2}{27 \Gamma \left(\frac{1}{3}\right)^6} \\
 -\frac{3 i \left(\sqrt{3} \pi -9 \log (3)\right) \Gamma \left(\frac{1}{3}\right)^6}{32 \pi ^5} & \frac{3 \Gamma \left(\frac{1}{3}\right)^6}{16 \pi ^4} & \frac{2 i
   \pi  \left(6+\sqrt{3} \pi +9 \log (3)\right)}{9 \Gamma \left(\frac{1}{3}\right)^6} & \frac{4 \pi ^2}{9 \Gamma \left(\frac{1}{3}\right)^6} \\
 \frac{\left(\left(27+i \sqrt{3}\right) \pi +27 \sqrt{26-6 i \sqrt{3}} \log (3)\right) \Gamma \left(\frac{1}{3}\right)^6}{192 \pi ^5} & \frac{\left(-1-3 i
   \sqrt{3}\right) \Gamma \left(\frac{1}{3}\right)^6}{32 \pi ^4} & \frac{\pi  \left(\left(27-i \sqrt{3}\right) \pi -9 \sqrt{26+6 i \sqrt{3}} (2+3\log
   (3))\right)}{81 \Gamma \left(\frac{1}{3}\right)^6} & \frac{\left(-2+6 i \sqrt{3}\right) \pi ^2}{27 \Gamma \left(\frac{1}{3}\right)^6} \\
\end{array}
\right)$} .
\end{equation}
For our purposes, however, this integral basis coming from LCS is not the most convenient frame to study the K-point. Instead, we implement a further Sp$(4,\mathbb{Z})$ basis change that rotates us to the integral monodromy weight frame specialized to the K-point. It is given by
\begin{equation}
    \hat{m}_{\rm K} (m_{\rm K})^{-1} = \left(
\begin{array}{cccc}
 0 & 3 & -1 & 0 \\
 3 & -2 & -1 & -1 \\
 -1 & 1 & 0 & 0 \\
 0 & -1 & 0 & 0 \\
\end{array}
\right)\, ,
\end{equation}
where $\hat{m}_{\rm K}$ denotes the transition matrix from the complex Frobenius basis directly to this integral monodromy weight basis. Already on a first impression $\hat{m}_{\rm K}$ appears more adapted to the singularity, as can be seen from the fact that certain entries now vanish, i.e.
\begin{equation}\label{mK33}
\hat{m}_{\rm K}=\scalebox{0.8}{$\left(
\begin{array}{cccc}
 \frac{\sqrt{3} \Gamma \left(\frac{1}{3}\right)^6}{4 \pi ^2} & 0 & -\frac{16 \pi ^4}{9 \sqrt{3} \Gamma \left(\frac{1}{3}\right)^6} & 0 \\
 \frac{(-1)^{2/3} \sqrt{3} \Gamma \left(\frac{1}{3}\right)^6}{4 \pi ^2} & 0 & \frac{8 \left(\sqrt{3}+3 i\right) \pi ^4}{27 \Gamma \left(\frac{1}{3}\right)^6}
   & 0 \\
 \frac{\left(\left(7 \sqrt{3}+3 i\right) \pi +27 i \left(\sqrt{3}+i\right) \log (3)\right) \Gamma \left(\frac{1}{3}\right)^6}{48 \pi ^3} & \frac{\sqrt[6]{-1}
   \Gamma \left(\frac{1}{3}\right)^6}{4 \pi ^2} & -\frac{4 \pi ^3 \left(18+18 i \sqrt{3}+\left(7 \sqrt{3}+13 i\right) \pi +\log \left(27^{9+9 i
   \sqrt{3}}\right)\right)}{81 \Gamma \left(\frac{1}{3}\right)^6} & -\frac{8 \left(\sqrt{3}-i\right) \pi ^4}{27 \Gamma \left(\frac{1}{3}\right)^6} \\
 \frac{\left(\sqrt{3} \pi -27 \log (3)\right) \Gamma \left(\frac{1}{3}\right)^6}{24 \pi ^3} & \frac{i \Gamma \left(\frac{1}{3}\right)^6}{4 \pi ^2} & -\frac{8
   \pi ^3 \left(\sqrt{3} \pi +9 (2+\log (27))\right)}{81 \Gamma \left(\frac{1}{3}\right)^6} & \frac{16 i \pi ^4}{27 \Gamma \left(\frac{1}{3}\right)^6} \\
\end{array}
\right)$}.\\
\end{equation}
In order to make the underlying geometrical data more manifest, it proves to be convenient to perform a decomposition $\hat{m}_{\rm K} = \hat{m}^{\rm LMHS}_{\rm K}  \hat{m}^{\rm Frob}_{\rm K} $ as discussed below \eqref{mfactorized}. The first factor $\hat{m}^{\rm LMHS}_{\rm K}$ given by \eqref{mK} is based on LMHS data, while the matrix $\hat{m}^{\rm Frob}_{\rm K}$ rotates the Frobenius periods into convenient linear combinations. Let us state both these factors here explicitly: the LMHS transition matrix is given by
\begin{equation}\label{mLMHS33}
    \hat{m}^{\rm LMHS}_{\rm K} = \left(\begin{array}{cccc}
 1 & 0 & 1 & 0 \\
 \tau & 0 & \bar{\tau} & 0 \\
 \gamma+\delta \tau  & a+b\tau & \gamma+\delta \bar{\tau} & a+b\bar{\tau} \\
\delta & b+c\tau & \delta & b+c\bar{\tau} \\
\end{array}
\right) \, ,
\end{equation}
where the relevant boundary data reads
\begin{equation}
  \begin{pmatrix}
      a & b \\
      c & d
  \end{pmatrix}  = \begin{pmatrix}
      2 & 1 \\
      1 & 2
  \end{pmatrix} \, , \qquad \tau = -\frac{1}{2}+\frac{i\sqrt{3}}{2}\, , \qquad (\delta\, , \ \gamma) = \big(\tfrac{2}{3}\, , \ \tfrac{1}{6}\big)\, .
\end{equation}
The remaining matrix acting on the complex Frobenius basis is given by
\begin{equation}\label{frob33}
    \hat{m}^{\rm Frob}_{\rm K} =\begin{pmatrix} 6\pi L(f_{27},1) & 0 & 0 & 0 \\
    6\pi h L(f_{27},1)  &  2\pi L(f_{27},1) & 0 & 0 \\
    0 & 0 &  -\frac{2\pi}{27L(f_{27},1)} & 0 \\
    0 & 0 & -\frac{2\pi}{27L(f_{27},1)} (h+\frac{1}{\pi i}) & -\frac{2\pi}{81L(f_{27},1)}
    \end{pmatrix}\, ,
\end{equation}
where $L(f_{27},1)$ denotes an L-function value we will elaborate upon shortly. The off-diagonal entries are parametrized by the number
\begin{equation}\label{h33}
    h = -\frac{3\log[3]}{2\pi i} \, .
\end{equation}
These components correspond to a rescaling of the local normal coordinate by $z \to 3^{-3} z$, similar to the off-diagonal components we encountered for the conifold points in \eqref{eq:ConiFrobCombi}.\footnote{For K-points the log-monodromy matrix \eqref{eq:KN} acts on the first and third column of LMHS \eqref{mLMHS33} by shifts along $(0,0,a+b\tau,b+c\tau)$ and $(0,0,a+b\bar\tau,b+c\bar\tau)$. These are precisely the pieces that the off-diagonal components in \eqref{frob33} add to these columns in the transition matrix.}

We want to emphasize that the above decomposition of the transition matrix applies not just to this example, but more broadly to all K-point examples that will follow. In fact, the LMHS matrix \eqref{mLMHS33} takes the same form, just the model-dependent data ($\tau, a, b, c, \gamma , \delta$) changes. The only minor caveat is that the Frobenius matrix \eqref{frob33} requires a slight generalization for non-hypergeometric examples: in addition to L-value $L(f,1)$ also certain quasi-periods for other off-diagonal components appear. Let us elaborate briefly on all of the parameters occurring in this example:
\begin{itemize}
    \item The integers $(a,b,c)=(2,1,2)$ specify the monodromy matrix in \eqref{eq:KN}. In fact, this bilinear form coincides with the quadratic form on the transcendental lattice of the rigid K3 surface that appears in the degenerated geometry at the K-point \cite{Doran2016MirrorST}. 
    \item The period $\tau = -\frac{1}{2}+i\frac{\sqrt{3}}{2} $ corresponds to the rigid K3 period, fixed in terms of the coefficients of the above quadratic form by \eqref{eq:Ktau}.
    \item The extension data is rational $\gamma=\tfrac{2}{3}$, $\delta=\tfrac{1}{6}$. It is therefore crucial for fixing the integral basis. However, we note that one may set them to zero by an Sp$(4,\mathbb{Q})$-basis, as can be seen from \eqref{eq:basischangeII}.
    \item The L-function values are associated to the weight-three modular form $f_{27}$ of level 27 labeled by $27.3.b.a$ in the database \cite{lmfdb}. The first two values are given by
    \begin{equation}
    L(f_{27},1)=\frac{3\sqrt{3}}{2\pi} L(f_{27},2)=\frac{\Gamma \left(\frac{1}{3}\right)^6}{8 \sqrt{3} \pi ^3}\;.
    \end{equation}
    Notice that the ratio of these two, up to a factor of $\pi$, is given by $\sqrt{3}$, which matches precisely with the non-rational part of our period $\tau$.
\end{itemize}
Let us close our discussion of this data with giving the monodromies and LMHS associated to this K-point. We begin with the semisimple part of the monodromy, which reads
\begin{equation}\label{eq:Tss33}
    T_{ss} = \left(
\begin{array}{cccc}
 0 & 1 & 0 & 0 \\
 -1 & -1 & 0 & 0 \\
 \frac{1}{3} & \frac{2}{3} & -1 & 1 \\
 \frac{2}{3} & \frac{1}{3} & -1 & 0 \\
\end{array}
\right)\, .
\end{equation}
In turn, we go to the 3-fold cover to remove any of these semisimple monodromies due to cube roots. The remaining part of the boundary data is then given by the log-monodromy of the unipotent part and the vectors spanning the LMHS
\begin{align}
 N=\left(
\begin{array}{cccc}
 0 & 0 & 0 & 0 \\
 0 & 0 & 0 & 0 \\
 2 & 1 & 0 & 0 \\
 1 & 2 & 0 & 0 \\
\end{array}
\right) \,, \quad
\omega_{3,1}=\left(
\begin{array}{c}
 1 \\
 -\frac{1}{2}+i \frac{\sqrt{3}}{2} \\
 \frac{5}{12}+i \frac{\sqrt{3}}{12}\\
 \frac{1}{6} \\
\end{array}
\right) \, , \quad \omega_{2,0}=\left(
\begin{array}{c}
 0 \\
 0 \\
 1 \\
 \frac{1}{2}+i \frac{\sqrt{3}}{2} \\
\end{array}
\right). \label{eq:BoundaryDataX33}
\end{align}
Notice that the semisimple monodromy in \eqref{eq:Tss33} indeed acts as an automorphism on the LMHS, with the vectors $\omega_{3,1},\omega_{2,1}$ as its eigenvectors with eigenvalues $e^{2\pi i/3}$.

Let us now turn to the series expansions around this K-point. We begin by switching to the natural coordinate $q$ from the local normal coordinate $z$ used in the Picard-Fuchs equation. The relation between these two is given by
\begin{align}
    z = 27q(1-\frac{7}{4}q^3-\frac{11779}{4000}q^6-\frac{194463}{8000}q^{9}-\frac{10139207770507}{42592000000}q^{12}-\dots)\,, \label{P33CC}
\end{align}
As discussed in section \ref{ssec:Kconstruction}, the utility of this is twofold. First, the overall rescaling by 27 assures that the LMHS is $\mathbb{R}$-split, matching precisely with the expectation from \eqref{h33}. Second, the non-constant part of the coordinate change is exactly what brings the instanton map $\Gamma$ into reduced form, i.e.~it is completely determined by a single holomorphic function. In that sense the definition of \ref{P33CC} is canonical. It is also remarkable that the same coordinate change has been encountered in the GLSM calculation of \cite{Erkinger:2022sqs}.\footnote{To be precise, the series for the coordinate change is equation 4.18 in \cite{Erkinger:2022sqs}, with our coordinates related by $q_{\text{there}} = q_{\text{here}}/27$.}

We can identify this remaining function $B(q)$ from the periods through \eqref{eq:Kperiods}, which yields
\begin{align}
    B(q)=-\frac{1}{3 L(f_{27},1)^2} \left(q+\frac{1}{2}q^4+\frac{501119}{196000}q^7+\frac{33777}{1600}q^{10} + \dots \right) \,.
\end{align}
Remarkably, these coefficients also match with the GLSM computation of \cite{Erkinger:2022sqs}, now with the FJRW invariants.\footnote{In this case the invariants $\langle \cdot \rangle_k$ given in equation 4.20 of \cite{Erkinger:2022sqs} are related to ours by $\langle \cdot \rangle_k = 3^{2-3k} \hat{B}_k/k $. }
This function $B(q)$ encodes the Yukawa coupling of the underlying supergravity theory through just a derivative as \eqref{eq:KYukawa}. We can similarly write down its series expansion as
\begin{align}
    Y(q)=-\frac{4 \pi^2}{  L(f_{27},1)^2} \left(q+8q^4+\frac{501119}{4000}q^{7}+\frac{33777}{16}q^{10}+\dots \right)\, .
\end{align}

\paragraph{$\mathbf{X_{4,4}}$ geometry.}\label{parX44} For our next example we consider the Calabi-Yau threefold $X_{4,4}$, which has a K-point with local exponents $\tfrac{1}{4}, \tfrac{1}{4}, \tfrac{3}{4}, \tfrac{3}{4}$. The topological data required to rotate to the integral basis is given by
\begin{equation}
    \kappa= 4\, ,\qquad c_2=40\, , \qquad \chi=-144\, .
\end{equation}
The transition matrix from the complex Frobenius basis to the integral monodromy weight basis reads
\begin{align}
 \hat{m}_{\rm K} = &\left(
\begin{array}{cccc}
 \frac{16 \Gamma \left(\frac{5}{4}\right)^4}{\pi } & 0 & -\frac{\Gamma \left(\frac{3}{4}\right)^4}{256 \pi } & 0 \\
 \frac{16 i \Gamma \left(\frac{5}{4}\right)^4}{\pi } & 0 & \frac{i \Gamma \left(\frac{3}{4}\right)^4}{256 \pi } & 0 \\
 -\frac{8 (\pi -10 i \log (2)) \Gamma \left(\frac{5}{4}\right)^4}{\pi ^2} & \frac{8 \Gamma \left(\frac{5}{4}\right)^4}{\pi } & \frac{(\pi
   -i (2+\log (1024))) \Gamma \left(\frac{3}{4}\right)^4}{512 \pi ^2} & -\frac{\Gamma \left(\frac{3}{4}\right)^4}{512 \pi } \\
 -\frac{80 \log (2) \Gamma \left(\frac{5}{4}\right)^4}{\pi ^2} & \frac{8 i \Gamma \left(\frac{5}{4}\right)^4}{\pi } & -\frac{(1+\log
   (32)) \Gamma \left(\frac{3}{4}\right)^4}{256 \pi ^2} & \frac{i \Gamma \left(\frac{3}{4}\right)^4}{512 \pi } \\
\end{array}
\right) \, ,
\end{align}
Following the procedure outlined below \eqref{mfactorized} we can conveniently decompose this transition matrix into a LMHS matrix $\hat{m}_{\rm K}^{\rm LMHS}$ and a matrix $\hat{m}_{\rm K}^{\rm Frob}$ minimally rescaling and rotating the Frobenius solution. The LMHS factor \eqref{mC} is fixed in terms of the boundary data
\begin{equation}
    \begin{pmatrix}
        a & b \\
        b & c 
    \end{pmatrix} = \begin{pmatrix}
        1 & 0 \\
        0 & 1
    \end{pmatrix}\,, \qquad \tau=i\, , \qquad (\delta,\gamma) = \big(0,\, \tfrac{1}{2}\big)\, ,
\end{equation}
while the other factor is given by
\begin{equation}
    \hat{m}_{\rm K}^{\rm Frob} = \begin{pmatrix}
        8\pi L(f_{16},1)  & 0 & 0 & 0 \\
        8\pi h L(f_{16},1) & 4\pi L(f_{16},1) & 0 & 0 \\
        0 & 0 & -\frac{\pi}{512L(f_{16},1)} & 0 \\
        0 & 0 & -(h-\frac{1}{\pi i})\frac{\pi}{512L(f_{16},1)} & -\frac{\pi}{1024L(f_{27},1)}
    \end{pmatrix}\, ,
\end{equation}
where we defined the coordinate rescaling parameter
\begin{equation}
    h = \frac{1-5\log[2]}{2\pi i}\, .
\end{equation}
The modular form $f_{16}$ is $16.3.c.a$ which has CM by $d=-4$. Its L-values are
\begin{equation}
    L(f_{16},1)=\frac{\Gamma \left(\frac{1}{4}\right)^4}{32 \pi ^2}\, , \qquad L(f_{16},2)=\frac{\Gamma \left(\frac{1}{4}\right)^4}{64 \pi }\;.
\end{equation}
For the monodromy we find as semisimple factor in the monodromy weight basis 
\begin{equation}\label{eq:Tss44}
    T_{ss} = \left(
\begin{array}{cccc}
 0 & -1 & 0 & 0 \\
 1 & 0 & 0 & 0 \\
 0 & -\frac{1}{2} & 0 & -1 \\
 -\frac{1}{2} & 0 & 1 & 0 \\
\end{array}
\right)
\end{equation}
The $\mathbb{R}$-split boundary data is extracted by the standard procedure from the period vector in the monodromy weight frame
\begin{align}
N=\left(
\begin{array}{cccc}
 0 & 0 & 0 & 0 \\
 0 & 0 & 0 & 0 \\
 1 & 0 & 0 & 0 \\
 0 & 1 & 0 & 0 \\
\end{array}
\right) \,, \quad 
\omega_{3,1}=\left(
\begin{array}{c}
 1 \\
 i \\
 \frac{1}{2}\\ 0 \\
\end{array}
\right) \, , \quad \omega_{2,0}=\left(
\begin{array}{c}
 0 \\
 0 \\
 1 \\
 i \\
\end{array}
\right)\,.
\end{align}
Note that the semisimple monodromy factor \eqref{eq:Tss44} is an automorphism of this LMHS, as $\omega_{3,1}$ and $\omega_{2,0}$ are its eigenvectors with eigenvalues $i$. The coordinate change to the natural Hodge-theoretic coordinate $q$ is given by
\begin{align}
    z=1024q \left(1-20q^2-\frac{11590}{27}q^4-\frac{315400}{9}q^6-\frac{297965703095}{83349}q^8-\dots \right)\, .
\end{align}
The relevant holomorphic function appearing in the instanton map $\Gamma(q)$ starts as 
\begin{align}
    B(q)=-\frac{1}{4 L(f_{16},1)^2} \left( q+12q^3+\frac{474122}{675}q^5+\frac{187880}{3}q^7+\dots\right)\, .
\end{align}
in terms of which the Yukawa coupling is given by
\begin{align}
    Y=-\frac{2 \pi^2}{ L(f_{16},1)^2}\left(q+108q^3+\frac{474122}{27}q^5+\frac{9206120}{3}q^7+ \dots \right)\, .
\end{align}

\paragraph{$\mathbf{X_{6,6}}$ geometry.}\label{parX66} For our final hypergeometric example we consider the $X_{6,6}$, which has a K-point with Riemann indices $\tfrac{1}{6},\tfrac{1}{6}, \tfrac{5}{6}, \tfrac{5}{6}$. The topological data required to fix the integral basis is given by
\begin{equation}
    \kappa= 1\qquad c_2=22\qquad \chi=-120\, ,
\end{equation}
The transition matrix from the local Frobenius basis at the K-point to the monodromy weight basis reads
\begin{align}
\hat{m}_{\rm K} = &\scalemath{0.66}{\left(
\begin{array}{cccc}
 \frac{\sqrt{3} \pi  \Gamma \left(\frac{7}{6}\right)^2}{\Gamma \left(\frac{2}{3}\right)^2} & 0 & -\frac{\pi  \Gamma
   \left(\frac{5}{6}\right)^2}{248832 \sqrt{3} \Gamma \left(\frac{4}{3}\right)^2} & 0 \\
 -\frac{\left(\sqrt{3}+3 i\right) \pi  \Gamma \left(\frac{7}{6}\right)^2}{2 \Gamma \left(\frac{2}{3}\right)^2} & 0 &
   \frac{\left(\sqrt{3}-3 i\right) \pi  \Gamma \left(\frac{5}{6}\right)^2}{1492992 \Gamma \left(\frac{4}{3}\right)^2} & 0 \\
 \frac{3 \pi ^2 \left(\left(43 \sqrt{3}-5 i\right) \pi +\log \left(2^{48+48 i \sqrt{3}} 3^{27+27 i \sqrt{3}}\right)\right)}{\Gamma
   \left(-\frac{1}{3}\right)^2 \Gamma \left(-\frac{1}{6}\right)^2} & -\frac{(-1)^{5/6} \pi  \Gamma \left(\frac{7}{6}\right)^2}{\Gamma
   \left(\frac{2}{3}\right)^2} & \frac{\left(9-9 i \sqrt{3}-\left(43 \sqrt{3}+73 i\right) \pi +\log \left(2^{48-48 i \sqrt{3}} 3^{27-27 i
   \sqrt{3}}\right)\right) \Gamma \left(\frac{5}{6}\right)^2}{8957952 \Gamma \left(\frac{4}{3}\right)^2} & -\frac{\left(\sqrt{3}+i\right)
   \pi  \Gamma \left(\frac{5}{6}\right)^2}{1492992 \Gamma \left(\frac{4}{3}\right)^2} \\
 \frac{6 \pi ^2 \left(\left(\sqrt{3}-36 i\right) \pi +48 \log (2)+27 \log (3)\right)}{\Gamma \left(-\frac{1}{3}\right)^2 \Gamma
   \left(-\frac{1}{6}\right)^2} & -\frac{i \pi  \Gamma \left(\frac{7}{6}\right)^2}{\Gamma \left(\frac{2}{3}\right)^2} &
   \frac{\left(9-\left(\sqrt{3}+36 i\right) \pi +48 \log (2)+27 \log (3)\right) \Gamma \left(\frac{5}{6}\right)^2}{4478976 \Gamma
   \left(\frac{4}{3}\right)^2} & -\frac{i \pi  \Gamma \left(\frac{5}{6}\right)^2}{746496 \Gamma \left(\frac{4}{3}\right)^2} \\
\end{array}
\right)}\, .
\end{align}
Following our discussion below \eqref{mfactorized}, this transition matrix may be decomposed into an LMHS factor $\hat{m}_{\rm K}^{\rm LMHS}$ and a factor $\hat{m}_{\rm K}^{\rm Frob}$ acting directly on the Frobenius basis. The LMHS transition matrix \eqref{mK} is fixed in terms of the boundary data
\begin{equation}
    \begin{pmatrix}
        a & b \\
        b & c
    \end{pmatrix} = \begin{pmatrix}
        2 & -1 \\
        -1 & 2
    \end{pmatrix}\, , \qquad \tau=\frac{1}{2}+\frac{i\sqrt{3}}{2}\, , \qquad \gamma=\frac{1}{3}\, , \quad \delta=\frac{1}{6}\, ,
\end{equation}
while the other factor is found to be
\begin{equation}
    \hat{m}_{\rm K}^{\rm Frob} = \begin{pmatrix}
        8\pi L(f_{12},1) & 0 & 0 & 0 \\
        8\pi h L(f_{12},1) & -\frac{8}{3}\pi L(f_{12},1) & 0 & 0 \\
        0 & 0 & -\frac{\pi}{165888 L(f_{12},1)} & 0 \\
        0 & 0 & -(h+\frac{1}{2\pi i})\frac{\pi}{165888 L(f_{12},1)} & \frac{\pi}{497664 L(f_{12},1)}
    \end{pmatrix},
\end{equation}
where we defined a rescaling parameter
\begin{equation}
    h= \frac{16\log(2)+9\log(3)}{6\pi i}\, .
\end{equation}
The modular form appearing in the L-function values is $f_{12}=\eta(2\tau)^3\eta(6\tau)^3$, which corresponds to $12.3.c.a$ in \cite{lmfdb} with CM by $d=-3$. The numerical values of these L-function values are
\begin{equation}
    L(f_{12},1)=\frac{\sqrt{3} \Gamma \left(\frac{1}{3}\right)^6}{32\ 2^{2/3} \pi ^3}\; , \qquad L(f_{12},2)=\frac{\Gamma \left(\frac{1}{3}\right)^6}{32\ 2^{2/3} \pi ^2}\;.
\end{equation}
Let us next summarize the boundary data associated to this K-point. The semisimple factor of the monodromy is given by
\begin{equation}\label{eq:Tss66}
    T_{ss} = \left(
\begin{array}{cccc}
 1 & 0 & 0 & 0 \\
 1 & 0 & -1 & 0 \\
 0 & 1 & -1 & 0 \\
 \frac{1}{3} & -1 & 1 & 1 \\
\end{array}
\right)\, ,
\end{equation}
while the log-monodromy matrix and LMHS are
\begin{align}
 N=\left(
\begin{array}{cccc}
 0 & 0 & 0 & 0 \\
 0 & 0 & 0 & 0 \\
 2 & 1 & 0 & 0 \\
 1 & 2 & 0 & 0 \\
\end{array}
\right) \,, \quad 
\omega_{3,1}=\left(
\begin{array}{c}
 1 \\
 -\frac{1}{2}+\frac{i \sqrt{3}}{2} \\
\frac{1}{6}+\frac{i \sqrt{3}}{6}\\
 \frac{1}{3}  \\
\end{array}
\right) \, , \quad \omega_{2,0}=\left(
\begin{array}{c}
 0 \\
 0 \\
 1 \\
 \frac{1}{2}+\frac{i \sqrt{3}}{2} \\
\end{array}
\right)\, .
\end{align}
One straightforwardly checks that the semisimple monodromy factor \eqref{eq:Tss66} is an automorphism of the LMHS, as $\omega_{3,1}$ and $\omega_{2,0}$ are eigenvectors with eigenvalue $e^{2\pi i/6}$. The natural Hodge-theoretic coordinate $q$ is obtained in terms of the local normal coordinate $z$ of the Picard-Fuchs equation from the periods as
\begin{align}
    z=864 \cdot 2^{1/3} q\left(1+256q^3-246082q^6+562616064q^9+4222132996250 q^{12}+\dots\right)\, .
\end{align}
The relevant holomorphic function appearing in the instanton map $\Gamma(q)$ starts as 
\begin{align}
    B(q)=-\frac{9 \; 2^{2/3}}{16 L(f_{12},1)^2} \left( q^2-640 q^5+1251305q^8-3489651968q^{11}+\dots\right)\, ,
\end{align}
in terms of which the Yukawa coupling is given by
\begin{align}
    Y(q)=-\frac{27 \; 2^{2/3} \pi^2}{   L(f_{12},1)^2}\left(q^2-4000q^5+20020880 q^8-105561972032 q^{10}- \dots \right)\, .
\end{align}
We want to remark here that, after extracting the common factor, only integer coefficients remain.
\subsection{Other examples}
In this part we study models that are not hypergeometric in nature. This should serve as a nice illustration that the observations made in the previous part are not artefacts of the special hypergeometric property, but are rather tied to the nature of the singularities themselves. The different examples are labeled according to their operator number as found in the CY operator database. For completeness we also include the conifold point of the quintic, which -- in contrast to the conifold points at infinity from the hypergeometric models -- features more generic boundary data.

\subsubsection{Conifold cases}
\label{262}
We next study three conifold examples, considering both singularities with a semisimple monodromy and without. This includes the conifold point in the moduli space of the quintic, allowing the reader to compare our results with a familiar example.

\paragraph{\bf{Operator} $\mathbf{2.62}$   \, :} 
\label{par262}
In this section we study the operator 2.62 of the CY operator database. This operator is an interesting example as it contains one of the 15 quarter conifolds in the database with local exponents $\tfrac{3}{4}, 1, 1, \tfrac{5}{4}$. It is given by the PF system
\begin{equation}
   L=  \theta^4-x(65\theta^4+130\theta^3+105\theta^2+40\theta+6)+4x^2(4\theta+3)(4\theta+3)(\theta+1)^2(4\theta+5)\;.
\end{equation}
This operator has a geometric interpretation in terms of a complete intersection in the Grassmannian $G(3,6)$, for which the topological numbers are
\begin{equation*}
    \kappa=42\, ,\qquad c_2=84\, ,\qquad\chi=-96\;.
\end{equation*}
It has 4 special points, which are summarized in the Riemann symbol
\begin{equation}
{\cal  P}\left\{\begin{array}{cccc}
0& 1/64&1& \infty\\ \hline
0& 0 &0& \frac{3}{4}\\
0& 1 &1& 1\\
0& 1 &1& 1\\
0& 2 &2& \frac{5}{4} 
\end{array}\right\}\ .
\end{equation}
The special points are thus an LCS point at $x=0$, two regular conifolds at $x=1/64$ and $x=1$ as well as a quarter conifold at $x=\infty$. In this section we will focus on the latter. As can be seen from the Riemann indices, one has to go to a 4-fold cover of moduli space to make the monodromy unipotent. The transition matrix to the LCS frame for this singularity was already obtained in \cite{chmiel2021coefficients}. However, as the integral basis chosen was not the monodromy weight frame, it is not immediately obvious that this example follows the general pattern. In what follows we formulate these results in our framework. As an independent check, we performed the calculation from scratch and give the required details on computing the transition matrix from the local complex Frobenius solutions to the integer symplectic basis adopted to the LCS can be computed as explained in appendix \ref{appendix:transition}. The result for this transition matrix is
\begin{equation}
    m_{\rm C}=\left(
\begin{array}{cccc}
 -\frac{\left(\frac{5}{32}-\frac{5 i}{32}\right) \Gamma \left(\frac{1}{4}\right)^6}{\pi ^{9/2}} & \frac{105}{4 \pi ^2}-\frac{21 i}{4 \pi } & -\frac{21 i}{2 \pi } & \frac{(105+105 i) \pi ^{3/2}}{\Gamma \left(\frac{1}{4}\right)^6} \\
 -\frac{\left(\frac{3}{32}-\frac{i}{16}\right) \Gamma \left(\frac{1}{4}\right)^6}{\pi ^{9/2}} & \frac{105}{8 \pi ^2}-\frac{2 i}{\pi } & -\frac{21 i}{4 \pi } & \frac{(63+42 i) \pi ^{3/2}}{\Gamma \left(\frac{1}{4}\right)^6} \\
 \frac{21 \Gamma \left(\frac{1}{4}\right)^6}{32 \pi ^{9/2}} & -\frac{525}{8 \pi ^2} & \frac{105 i}{4 \pi } & -\frac{441 \pi ^{3/2}}{\Gamma \left(\frac{1}{4}\right)^6} \\
 -\frac{\left(\frac{9}{16}-\frac{5 i}{16}\right) \Gamma \left(\frac{1}{4}\right)^6}{\pi ^{9/2}} & \frac{315}{4 \pi ^2}-\frac{21 i}{2 \pi } & -\frac{63 i}{2 \pi } & \frac{(378+210 i) \pi ^{3/2}}{\Gamma \left(\frac{1}{4}\right)^6} \\
\end{array}
\right)\, .
\end{equation}
This can be brought into the monodromy weight frame via the following Sp$(4,\mathbb{Z})$ matrix 
\begin{equation}
  \hat{m}_{\rm C}(m_{\rm C})^{-1} =  \left(
\begin{array}{cccc}
 6 & -5 & -1 & -2 \\
 -12 & 16 & 2 & 3 \\
 -1 & 8 & 0 & -1 \\
 16 & 0 & -3 & -8 \\
\end{array}
\right).
\end{equation}
The rotation from the complex Frobenius basis directly to the monodromy weight basis can then be written conveniently as $\hat{m}_{\rm C}=\hat{m}_{\rm C}^{\rm LMHS} \hat{m}_{\rm C}^{\rm Frob}$, as outlined below \eqref{mfactorized}. The first factor $\hat{m}_{\rm C}^{\rm LMHS}$ is given by \eqref{mC} and determined by the boundary data
\begin{equation}
    \tau = \frac{1}{2}+\frac{i}{2}\, , \qquad k=42\, , \quad n=4\, , \qquad \gamma = 1\, , \quad \delta = \frac{1}{2}\, .
\end{equation}
The other factor acting directly on the Frobenius periods reads
\begin{equation}
    (2 \pi i)^3 \sqrt{42} \,  \hat{m}_{\rm C}^{\rm Frob} = \begin{pmatrix}
        16\sqrt{21} \pi^2 L(f_{32},1) & 0 & 0 & 0 \\
        0 & (2\pi i)^2 \sqrt{\kappa} & 0 & 0 \\
        0 & (2\pi i)^2 \sqrt{\kappa}\frac{k}{n}5 & -(2\pi i)^2 \sqrt{\kappa}\frac{k}{n} & 0 \\
        0 & 0 & 0 & \frac{21\sqrt{21} \pi}{L(f_{32},1)}  
    \end{pmatrix}\, .
\end{equation}
The modular form $f$ is the same level 32 CM modular form that has occurred in the $X_{4,2}$ geometry, with its L-function values given in \eqref{L42}. Before going to the covering space, we extract the finite order monodromy matrix 
\begin{equation}
    T_{ss} =  \left(
\begin{array}{cccc}
 1 & 0 & 0 & 0 \\
 3 & 1 & -2 & 0 \\
 2 & 1 & -1 & 0 \\
 \frac{3}{2} & 0 & -1 & 1 \\
\end{array}
\right) \, ,
\end{equation}
Going to the $n=4$ cover we extract the boundary data and normalize it as in section \ref{sec:construction}, yielding
\begin{align}
  N= \begin{pmatrix}
   0 & 0 & 0 & 0 \\
   0 & 0 & 0 & 0 \\
   0 & 0 & 0 & 0 \\
   -42 & 0 & 0 & 0 \\
   \end{pmatrix}\,, \quad \omega_{3,0} = \begin{pmatrix}
   0 \\ 1 \\ \frac{1}{2}+  \frac{i}{2} \\ -\frac{i}{2}
   \end{pmatrix} \,, \quad \omega_{2,2}= \begin{pmatrix}
   1 \\ 1 \\ \frac{1}{2} \\0
   \end{pmatrix} \,.
\end{align}
Using this boundary data and the procedure outlined in section \ref{sec:construction}, we can make the canonical coordinate change 
\begin{align}
    z = q \left(1-\frac{35}{384}q^4+\frac{1693}{1032192}q^{8}+\frac{98752219}{287758614528}q^{12}+ \dots \right)\, .
\end{align}
The first few terms in the holomorphic series $A(q)$ specifying the $\Gamma$-map are given by 
\begin{align}
    A(q)=  -\frac{1}{2\sqrt{2}L(f_{32},1)} \cdot \left(q- \frac{89}{2560}q^5-\frac{19407}{3670016}q^{9}-\frac{130005865679}{239415167287296}q^{13}+\dots \right)\,.
\end{align}
and the beginning of the Yukawa coupling has the form 
\begin{align}
   Y(q)= \frac{21 \pi^2}{L(f_{32},1)^2} \left(q^{2} - \frac{89}{256}q^6 - \frac{7451}{114688}q^{10} + \dots \right)\, .
\end{align}

\paragraph{\bf{Operator} $\mathbf{2.17}$.}\label{par217} This model features a conifold point with a semisimple monodromy that does not have the maximal number of distinct eigenvalues. The Riemann Symbol for this operator is given by 
\begin{align}
  \begin{Bmatrix}
   0 & 1/256 & \infty \\
  \hline 
   0 & 0 & 1/2 \\
   0 & 1/2 & 1/2 \\ 
   0 & 1/2 & 3/2 \\
   0 & 1 & 3/2
  \end{Bmatrix}  \,.
\end{align}
The operator has the usual LCS point a the origin $z=0$, a conifold singularity at $z=1/256$ while the monodromy at $z=\infty$ is a K-point. This operator can be realized via several double octic constructions, leading to the following topological data
\begin{align}
    \kappa=4 m\,, \qquad c_2=4m \,, \qquad \chi=24m \,,
\end{align}
where $m=3,4,5$, corresponding to the double octics (D.O.249, D.O.258, D.O.265), (D.O.36, D.O.73) and (D.O.8) respectively. The numbering and the Euler characteristics originate from the PhD thesis \cite{MeyerThesis}. The intersection number $\kappa$ and second Chern class $c_2$ are fixed by demanding the D6-brane period to correspond to the conifold state at $z=1/256$. We stress that we did not compute these topological numbers from the geometry itself, and it would be interesting to see if such a computation would reproduce our results.

We compute the transition matrix between the complex Frobenius basis and the monodromy weight frame at the conifold point numerically and identify the numerical quantities. We furthermore multiply an overall numerical constant and perform a polarization reversing integer rotation to adhere to our previous conventions. The result is given by
\begin{align}
  (2 \pi )^3 \frac{1}{16 \sqrt{m} } \, \hat{m}_{\rm C} = \scalebox{0.80}{$\left(
\begin{array}{cccc}
 0 &  (2 \pi i) ^2 2 \sqrt{m } & 0 & 0 \\
 8 \pi ^2 \sqrt{m} L(f_8,1) &  (2\pi i) ^2 4 \sqrt{m } & 0 &  \sqrt{m}\, \eta_+  \\
  8 i  \pi  \frac{L(f_8,2)}{\sqrt{m}} & (2\pi i) ^2\frac{2}{\sqrt{m}}   & 0 & i \frac{\eta_-}{\sqrt{m}} \\
 \frac{8}{\sqrt{m}}\left( \pi ^2 L(f_8,1)-i 2 \pi L(f_8,2) \right) &  - 2 \pi i \frac{16}{\sqrt{m}} (2\log(2)-1)  & -(2 \pi i)^2 \frac{8}{\sqrt{m}} & \frac{\eta_+ - 2 \eta_-}{\sqrt{m}}
\end{array}
\right)$} \,.
\end{align}
Following the approach outlined below \eqref{mfactorized} we may decompose this transition matrix into two factors, one based on the LMHS and minimally rescaling and rotating the Frobenius solution. The first is given by \eqref{mC} and is determined by the boundary data
\begin{equation}
    \tau = i \frac{L(f_8,2)}{\pi m L(f_8,1)}\, , \qquad k=\frac{8}{m}\, , \quad n =4\, ,\qquad \gamma = 2\, , \quad \delta = \frac{1}{m}\, ,
\end{equation}
while the other part of the transition matrix reads
\begin{equation}
      (2 \pi )^3 \frac{1}{16 \sqrt{m} } \, \hat{m}_{\rm C}^{\rm Frob} =    \scalebox{0.9}{$\begin{pmatrix} 8\pi^2 \sqrt{m} L(f_8,1) & 0 & 0 & 8 \pi^2 \sqrt{m} \theta L(f_8,1) \\
    0 & (2\pi i)^2 2\sqrt{m} & 0 & 0 \\
    0 & (2\pi i)^2 2\sqrt{m}\frac{k}{n} h & -(2\pi i)^2 2\sqrt{m} \frac{k}{n} & 0 \\
    0 & 0 & 0 & \frac{8\pi}{\sqrt{m}\tau_2 L(f_8,1)}
    \end{pmatrix}$},
\end{equation}
where we additionally defined the coordinate rescaling
\begin{equation}
    h= \frac{1-2\log(2)}{\pi i}\, .
\end{equation}
The modular form $f_8$ corresponds to the modular form $8.4.a.a$, whose L-values are given numerically by 
\begin{equation}
    L(f_8,1) \simeq 0.3545006837309647187655\, , \qquad L(f_8,2) \simeq 0.6900311631233975251191\, .
\end{equation}
The ratio of L-values does not yield an algebraic number in this case. This could have been expected as the modular form in question does not enjoy CM. In the cases when $m=3,5$ there are twists of this modular form by the character $\left( \frac{m}{\cdot} \right)$, that absorb the quadratic factors in the transition matrix. We suggestively write 
\begin{equation}
    \begin{aligned}
    \eta_+&= \frac {8 \pi^2  }{L(f_8,2)} +8 \pi^2   r_2    L(f_8,1)\, ,  \\
    \eta_-&=-\frac{8 \pi}{L(f_8,1)} +  8 \pi r_2  L(f_8,2) \,,  
    \end{aligned}
\end{equation}
so that the quasi-periods are expressed as a (transcendental) shift of the inverted L-values by
\begin{align}
    r_2 = \frac{\eta_+}{4 \pi^2 L(f,1)} + \frac{\eta_-}{2 \pi L(f,2)} \simeq -78.7923006226158 \,,
\end{align}
which is the K\"ahler transformation parameter defined in \eqref{eq:GFC}. The semisimple monodromy is given by
\begin{equation}\label{eq:Tss217}
    T_{ss} = \left(
\begin{array}{cccc}
 -1 & 0 & 0 & 0 \\
 -4 & 1 & 0 & 0 \\
 -\frac{2}{m} & 0 & 1 & 0 \\
 0 & \frac{2}{m} & -4 & -1 \\
\end{array}
\right)\, .
\end{equation}
Upon going to the covering space we extract the boundary data and normalize it as in section \ref{sec:construction}, which gives
\begin{align}
  N= \begin{pmatrix}
   0 & 0 & 0 & 0 \\
   0 & 0 & 0 & 0 \\
   0 & 0 & 0 & 0 \\
   -\frac{8}{m} & 0 & 0 & 0 \\
   \end{pmatrix}\,, \quad \omega_{3,0} = \begin{pmatrix}
   0 \\ 1 \\ i\frac{L(f_8,2)}{\pi m L(f_8,1)}\\ \frac{1}{m}-i \frac{2 L(f_8,2)}{\pi m L(f_8,1)}
   \end{pmatrix} \,, \quad \omega_{2,2}= \begin{pmatrix}
   1 \\ 2 \\ \frac{1}{m} \\ i \frac{ 8\log(2)}{m\pi}
   \end{pmatrix} \,.
\end{align}
Note that this LMHS indeed has the semisimple monodromy in \eqref{eq:Tss217} as an automorphism, with $\omega_{3,0}$ as eigenvector with eigenvalue $1$ and $\omega_{2,2}$ and $\omega_{1,1}=N\omega_{2,2}$ with $-1$.  Using this boundary data and the procedure outlined in section \ref{ssec:conifoldconstruction}, we can make the canonical coordinate change 
\begin{align}
    z = \frac{1}{4} q \left(1- \left(4 + \frac{1}{8} r_2 \right) q^2+\left(\frac{170}{3} + \frac{47}{18} r_2 + \frac{1}{32} (r_2)^2 \right)q^{4}-  \dots \right)\,.
\end{align}
As a result, the LMHS will be $\mathbb{R}$-split and the instanton map will be in canonical form. The first few terms in the holomorphic series $A(q)$ specifying the instanton map are given by 
\begin{align}
    A(q)= - \frac{ 1}{4 L(f_8,1)} \cdot \left(q- \left( \frac{40}{3} + \frac{3}{16} r_2 \right) q^3+ \left( \frac{1498}{5} + \frac{25}{3} r_2 +\frac{15}{256} (r_2)^2 \right)q^{5}+\dots \right)\, ,
\end{align}
while the Yukawa coupling reads
\begin{align}
    Y(q)= \frac{2 \pi^2}{ m L(f_8,1)^2}\left( q^2- \left(80 + \frac{9}{8} r_2 \right)q^4+ \left( 4596 + \frac{385}{3} r_2 + \frac{231}{256} (r_2)^2 \right)q^{6}+\dots \right)\, .
\end{align}
Compared to cases with a semisimple monodromy, note that these series are complicated by the presence of the numerical parameter $r_1$.

\paragraph{$\mathbf{X_5}$.}\label{parX5}
We close our conifold point discussions with one of the most familiar examples: the quintic $X_5$. We want to emphasize that this is mostly a rewriting of the structure already discussed in \cite{Bonisch:2022mgw}. We hope that this example makes it easier to compare the conventions used in both works. As the ordinary conifold point of the quintic, as many other conifold points, has no non-trivial finite order monodromy, it belongs to the most complicated class of singularities discussed in this paper and a certain degree of transcendentality is expected (already observed in \cite{Bonisch:2022mgw}). We hope that covering this class illustrates that our methods do not rely on finite order monodromy and can in general furnish a practical perspective. The topological numbers are
\begin{align}
    \kappa = 5\,  \qquad c_2 = 50\, , \qquad \chi= -200 \,.
\end{align}
The Picard-Fuchs operator is given by \eqref{eq:PF} with indices $\tfrac{1}{5},\tfrac{2}{5}, \tfrac{3}{5}, \tfrac{4}{5}$. We are interested in the periods near its conifold point, for which the transition matrix from the complex Frobenius basis to the integral basis is computed to be 
\begin{align}
(2\pi i)^3 5^{-5} \, \hat{m}_{\rm C}= &\scalebox{0.65}{$\begin{pmatrix}
    0 & (2\pi i)^2 \sqrt{\kappa} & 0 & 0 \\
    \frac{4}{625}\pi^2 L(1) & (2 \pi i)^2 \sqrt{\kappa} \gamma + r_{1}\frac{4}{625} \pi^2 L(1)  & 0 & \eta_+ \\
    \frac{2}{625} \pi^2 L(1) + i \frac{1}{10} \pi L(2) & \frac{1}{2}(2 \pi i)^2 \sqrt{\kappa} \gamma - r_{1}\left( \frac{2}{625} \pi^2 L(1) + i \frac{1}{10} \pi L(2) \right) & 0 & \frac{1}{2}\eta_+ + i \eta_- \\ - i \frac{1}{10} \pi L(2) \gamma & - i h- r_{1} i \frac{1}{10} \pi L(2) \gamma & -(2\pi i)^2 \sqrt{\kappa} & i \eta_- -\frac{\pi}{2}\sqrt{\kappa}\left( 5625-2 r_{1}\right)
\end{pmatrix}$} .
\end{align}
We may conveniently decompose this transition matrix according to the discussion below \eqref{mfactorized} by taking out a factor based on the LMHS data. This factor is given by \eqref{mC} and determined by the data 
\begin{equation}  
    \tau = \frac{1}{2}+ i \frac{125 L(f_{25},2)}{8\pi L(f_{25},1)}\, , \qquad \gamma = 2 \delta \, , \qquad \delta \simeq 0.08641832567733\, ,
\end{equation}
with $k=n=1$. The remaining factor acting directly on the Frobenius solution reads
\begin{equation}
    (2\pi i)^3 5^{-5} \hat{m}_{\rm C}^{\rm LMHS} = \begin{pmatrix}
    \frac{4 \pi^2L(f_{25},1)}{625} & -\frac{4 \pi^2 r_{1} L(f_{25},1)}{625} & 0 & \frac{4\pi^2 r_{2} L(f_{25},1)}{625} \\
    0 & -4\sqrt{5}\pi^2& 0 & 0 \\
    0 & -4\sqrt{5}\pi^2 h & -4\sqrt{5}\pi^2 & \frac{1}{2}i \sqrt{5} \pi(5625-2 r_{1}) \\
    0 & 0 & 0 & \frac{3125 \pi }{2 L(f_{25},1) \tau _2}
    \end{pmatrix}\, ,
\end{equation}
where we defined the parameters 
\begin{align}\label{quintic:etah}
    \eta_+ &= \frac{100 \pi^2}{L(f_{25},2)}+r_{2} \frac{4}{625} \pi^2 L(f_{25},1)\, , \\
    \eta_- &= -\frac{3125 \pi}{2L(f_{25},1)} + r_{2} \frac{1}{10} \pi L(f_{25},2) \, , \\
    h &= \frac{1}{2 \pi i}\left(1-\frac{10 \log(5)}{3}- \frac{2 L'(g_{25},2) }{3 \sqrt{5}L(f_{25},2)} \right)=\frac{1}{2 \pi i}\left(1-\frac{10 \log(5)}{3}- \log(\xi) \right)\, ,
\end{align}
where the K\"ahler transformation parameters have the values 
\begin{align}\label{quintic:r}
    r_1&= 512.8137841079743331\, , \\ \nonumber 
    r_{2}&=\frac{625 \eta_+}{8 \pi^2 L(f_{25},1)}+\frac{5 \eta_{-}}{\pi L(f_{25},2)} \simeq -215953.47912090389520\, .
\end{align}
The modular form $f_{25}$ refers to $25.4.a.b$, whose numerical L-values are 
\begin{equation}\label{quintic:Lf}
    L(f_{25},1) =1.6255530106 \, ,\qquad L(f_{25},2) =1.5652444137\, .
\end{equation}
Here $g_{25}$ refers to the modular form $25.4.a.a$ which is a twist of $25.4.b.a$ by the Kronecker character $\left( \frac{5}{\cdot}\right)$. The corresponding L-function is evaluated at its first derivative and takes the value 
\begin{equation}\label{quintic:Lg}
    L'(g_{25},2) = 0.75614145926\, .
\end{equation}
This identification was possible due to a progress in the upcoming work \cite{Bonischtoappear}. Its appearance in the LMHS as obstruction to being $\mathbb{R}$-split, suggests that this object is related to the Beilinson-Bloch height pairing as recently discussed in \cite{bloch2023computing}. The boundary data is given by  
\begin{align*}
  N= \begin{pmatrix}
   0 & 0 & 0 & 0 \\
   0 & 0 & 0 & 0 \\
   0 & 0 & 0 & 0 \\
   -1 & 0 & 0 & 0 \\
   \end{pmatrix}, \  \omega_{3,0} = \begin{pmatrix}
   0 \\ 1 \\ \frac{1}{2}+  i \frac{125 L(f_{25},2)}{8 \pi L(f_{25},1)} \\ -i\frac{125 L(f_{25},2)}{8 \pi L(f_{25},1) }\gamma
   \end{pmatrix} , \  \omega_{2,2}= \begin{pmatrix}
   1 \\ \gamma \\ \frac{\gamma}{2} \\  \frac{i}{3 \pi} \left( \frac{L'(g_{25},2)}{\sqrt{5} L(f_{25},2)}+5 \log(5) \right)
   \end{pmatrix} .
\end{align*}
Using this boundary data and the procedure outlined in section \ref{ssec:conifoldconstruction}, we can make the canonical coordinate change 
\begin{align}
    z =  \frac{\xi}{125 \; 5^{1/3}} q \left( 1- \frac{\xi}{125 \; 5^{1/3}} (625- 2 r_{1})q - \frac{\xi^2}{31250 \, 5^{2/3}} \left( 781250 + 1875 r_{1} - 11 r_{1}^2 +4 r_{2} \right)q^2 - \dots \right) \, . \nonumber
\end{align}
where $\xi$, which is known exactly by \eqref{quintic:etah}, is given numerically as
\begin{equation}\label{quintic:xi}
    \xi = 1.1549155637844015529502873148871509024 \, ,
\end{equation}
by using \eqref{quintic:Lf} and \eqref{quintic:Lg}. The function $A(q)$ specifying the instanton map reads
\begin{align}
    A(q)=- \frac{5 \; 5^{1/6} \xi}{L(f_{25},1)} \left(q+ \frac{\xi}{250 \; 5^{1/3}}(3125+6 r_{1}) q^2 + \dots \right)\, ,
\end{align}
while the Yukawa coupling with respect to the canonical coordinate is
\begin{align}
    Y(q) =\frac{100 \, 5^{1/3} \xi \; \pi^2}{L(f_{25},1)^2} \left(q^2 + \frac{\xi}{125 \; 5^{1/3}} (6250 + 6 r_{1}) q^3 + \dots \right)\, .
\end{align}
Note that in the above expansions the parameters $\xi,r_1,r_2$ complicate the series compared to cases with a semisimple monodromy.

\subsubsection{K-point cases}
We next consider two K-points in examples without hypergeometric periods. We picked one model with a semisimple monodromy factor and one without. This allows us to investigate what aspects of the boundary data are specific to the models. We can see what features carry over from the hypergeometric models to other examples that also have semisimple monodromy factors, and what is lost when we do not have such a semisimple part.

\paragraph{Operator 4.47 :}\label{par447} 
As the final example we take the operator 4.47 of the database. This operator is interesting as it is a non-hypergeometric example of a K-point with a finite part in the monodromy, thus we expect it to be fully solvable. The topological numbers are 
\begin{equation*}
    \kappa=4\, , \qquad c_2=28\, , \qquad\chi=-18\;.
\end{equation*}
The geometry corresponds to the PF operator\footnote{This is a rescaled version of the operator listed in the database. The rescaling is needed to ensure the usual form of the mirror map and it positions the closest conifold to the LCS point at x=1.}
\begin{equation*}
\begin{aligned}
        \theta^4+2^{-3} 3^{-2} x\left(9\theta^4-198\theta^3-131\theta^2-32\theta-3\right)-2^{-3} 3^{-4} x^{2}\left(486\theta^4+1215\theta^3+81\theta^2-27\theta-5\right)-\\2^{-5} 3^{-4} x^{3}\left(891\theta^4+972\theta^3+675\theta^2+216\theta+25\right)-2^{-5} 3^{-4} x^{4}(3\theta+1)^2(3\theta+2)^2
\end{aligned}
\end{equation*}
with the Riemann symbol
\begin{equation}
{\cal  P}\left\{\begin{array}{ccccc}
-8&-2&0& 1& \infty\\ \hline
0&0&0& 0 & \frac{1}{3}\\
1&1&0& 1 & \frac{1}{3}\\
1&3&0& 1 & \frac{2}{3}\\
2&4&0& 2 & \frac{2}{3} 
\end{array}\right\}\ 
\end{equation}
Thus this geometry has a MUM point, 2 conifold points an apparent singularity as well as a generalized K-point. We will be mostly interested in the K-point, although the methods developed in this paper are applicable to the other points as well. The region of convergence between the bases around $\infty,-8,-2 $ and $0$ all overlap, allowing to compute the transition matrix numerically. The transition matrix from the integral LCS to the monodromy weight basis at the K-point reads
\begin{align}
    \hat{m}_{\rm K} (m_{\rm K})^{-1} = \left(
\begin{array}{cccc}
 0 & 2 & 0 & -1 \\
 2 & 0 & -1 & -1 \\
 -1 & 3 & 0 & -1 \\
 4 & -1 & -2 & -2 \\
\end{array}
\right)\, .
\end{align}
The transition matrix from the complex Frobenius basis directly to the monodromy weight basis at the K-point is given by
\begin{align*}
    \hat{m}_{\rm K}=\scalemath{0.75}{\left(
\begin{array}{cccc}
 \frac{9 (-1)^{5/6} \Gamma \left(\frac{1}{3}\right)^6}{2 \sqrt[3]{2} \pi ^2} & 0 & -\frac{96 \sqrt[6]{-1} \sqrt[3]{2} \pi ^4}{\Gamma \left(\frac{1}{3}\right)^6} & 0 \\
 -\frac{3 \sqrt[3]{-\frac{1}{2}} \sqrt{3} \Gamma \left(\frac{1}{3}\right)^6}{2 \pi ^2} & 0 & \frac{32 (-1)^{2/3} \sqrt[3]{2} \sqrt{3} \pi ^4}{\Gamma \left(\frac{1}{3}\right)^6}
   & 0 \\
 \frac{\left(\left(-35 \sqrt{3}+3 i\right) \pi +3(1+ i \sqrt{3})\log \left(\frac{27}{65536}\right)\right) \Gamma \left(\frac{1}{3}\right)^6}{16 \sqrt[3]{2} \pi
   ^3} & \frac{3 (-1)^{5/6} \Gamma \left(\frac{1}{3}\right)^6}{\sqrt[3]{2} \pi ^2} & \frac{4 \sqrt[3]{2} \pi ^3 \left(\left(\sqrt{3}-3 i\right) \left(24+\log
   \left(\frac{65536}{27}\right)\right)-5 i \left(\sqrt{3}-7 i\right) \pi \right)}{\sqrt{3} \Gamma \left(\frac{1}{3}\right)^6} & -\frac{64 \sqrt[6]{-1} \sqrt[3]{2} \pi
   ^4}{\Gamma \left(\frac{1}{3}\right)^6} \\
 -\frac{3 \left(\left(5 \sqrt{3}-i\right) \pi +(3-i \sqrt{3})\log \left(\frac{27}{65536}\right)\right) \Gamma \left(\frac{1}{3}\right)^6}{16 \sqrt[3]{2} \pi ^3} &
   -\frac{3 \sqrt[3]{-\frac{1}{2}} \sqrt{3} \Gamma \left(\frac{1}{3}\right)^6}{\pi ^2} & -\frac{4 \sqrt[3]{2} \pi ^3 \left(5 \left(3+i \sqrt{3}\right) \pi +3
   \left(\sqrt{3}+i\right) \left(24+\log \left(\frac{65536}{27}\right)\right)\right)}{\sqrt{3} \Gamma \left(\frac{1}{3}\right)^6} & \frac{64 (-1)^{2/3} \sqrt[3]{2} \sqrt{3} \pi
   ^4}{\Gamma \left(\frac{1}{3}\right)^6} \\
\end{array}
\right)}
\end{align*}
This transition matrix is conveniently decomposed into the form \eqref{mfactorized}, where the LMHS matrix \eqref{mK} is determined by the boundary data
\begin{equation}
    \begin{pmatrix}
        a & b \\
        b & c
    \end{pmatrix} = \begin{pmatrix}
        2 & 0 \\
        0 & 6
    \end{pmatrix}\, , \qquad \tau= \frac{i}{\sqrt{3}}\, , \qquad \delta =\frac{5}{6}  \, , \quad  \gamma= \frac{5}{12}\, .
\end{equation}
The other factor acting directly on the Frobenius basis is given by
\begin{equation}
   \hat{m}_{\rm K}^{\rm Frob} = \scalebox{1}{$4 \pi i \sqrt{3} \begin{pmatrix}
       3(-1)^{\frac{1}{3}}  L(f_{108},1) & 0 & 0 & 0 \\
       3(-1)^{\frac{1}{3}} h L(f_{108},1)  & (-1)^{\frac{1}{3}} L(f_{108},1) & 0 & 0 \\
       0 & 0 & \frac{3(-1)^{\frac{2}{3}}}{ L(f_{108},1)} & 0 \\
       0 & 0 & \frac{3(-1)^{\frac{2}{3}}}{ L(f_{108},1)}(h-\frac{1}{\pi i})  & \frac{(-1)^{\frac{2}{3}}}{L(f_{108},1)}
   \end{pmatrix}$} \, ,
\end{equation}
where we defined the coordinate rescaling parameter
\begin{equation}
    h = \frac{3\log(3)-16\log(2)}{24\pi i}\, .
\end{equation}
The modular form $f_{108}$ refers to $108.3.c.a$, whose L-values are 
\begin{equation}
    L(f_{108},1) =\frac{\sqrt{3} \Gamma \left(\frac{1}{3}\right)^6}{8 \sqrt[3]{2} \pi ^3} \, ,\qquad L(f_{108},2)=\frac{\Gamma \left(\frac{1}{3}\right)^6}{24 \sqrt[3]{2} \pi ^2}\, .
\end{equation}
Let us now write down the extracted boundary data. The semisimple factor of the monodromy is computed to be
\begin{equation}
    T_{ss} = \left(
\begin{array}{cccc}
 -\frac{1}{2} & \frac{3}{2} & 0 & 0 \\
 -\frac{1}{2} & -\frac{1}{2} & 0 & 0 \\
 -\frac{5}{12} & \frac{5}{4} & -\frac{1}{2} & \frac{1}{2} \\
 \frac{5}{4} & \frac{5}{4} & -\frac{3}{2} & -\frac{1}{2} \\
\end{array}
\right)\, ,
\end{equation}
while the log-monodromy and $\mathbb{R}$-split LMHS is specified by
\begin{align}
N=\left(
\begin{array}{cccc}
 0 & 0 & 0 & 0 \\
 0 & 0 & 0 & 0 \\
 2 & 0 & 0 & 0 \\
 0 & 6 & 0 & 0 \\
\end{array}
\right) \,, \quad 
\omega_{3,1}=\left(
\begin{array}{c}
 1 \\
\frac{i }{\sqrt{3}} \\
 \frac{5}{6}+\frac{i 5}{12\sqrt{3}}\\ \frac{5}{12} \\
\end{array}
\right) \,.
\end{align}
One straightforwardly verifies that the semisimple monodromy commutes with $N$, and has $\omega_{3,1}$ as one of its eigenvectors with eigenvalue $e^{2\pi i/3}$. We do want to stress that this example has a rational monodromy $T = T_{ss} e^{N/3}$, as can be seen directly from the upper-left and lower-right subblocks of the semisimple factor.

We then turn to the expansion of the period vector around the K-point. The canonical coordinate change for this example is given by
\begin{align}
    z= \frac{2^{5/6}}{3^{1/4}} q \left(1+ \frac{4}{3^{3/4}} q^3 -  \frac{16088}{375 \sqrt{3}} q^6 + \frac{1737377}{3375 \, 3^{1/4} } q^9 + \frac{15760767614557}{1684546875}q^{12}+ \dots \right)\, .
\end{align}
The holomorphic function encoding the instanton map $\Gamma(q)$ starts as 
\begin{align}
    B(q)= \frac{2^{4/3}}{3^{1/4} L(f,1)^2} \left(q-\frac{7
   q^4}{3\ 3^{3/4}}+\frac{5217112 q^7}{165375 \sqrt{3}}-\frac{9088541 q^{10}}{18225 \sqrt[4]{3}}
   +\ldots\right).
\end{align}
From this data we determine the Yukawa coupling to be
\begin{align}
    Y= \frac{8\; 2^{1/3}\;3^{3/4}\pi^2}{ L(f,1)^2 }\left( 1-\frac{112 q^3}{3\
   3^{3/4}}+\frac{5217112 q^6}{3375 \sqrt{3}}-\frac{36354164 q^9}{729 \sqrt[4]{3}}+\ldots \right)\, .
\end{align}
In this case the coefficients take algebraic values in $\mathbb{Q}(3^{1/4})$ instead of rational values. Notice, however, that in principle one could absorb these quartic roots into the coordinate $q$ by a rescaling $q \to 3^{1/4} q$ such that these factors drop out of the series. The catch is that this transformation would destroy the $\mathbb{R}$-splitness of the LMHS, which is why we have chosen not to perform this redefinition.

\paragraph{Operator 3.7.}\label{par37} This example features a K-point without any semisimple monodromy. The Riemann Symbol for this operator is given by 
\begin{align}
  \begin{Bmatrix}
  -1/108 & 0 & 1/324 & \infty \\
  \hline 
  0 & 0 & 0 & 1/2 \\
  0 & 0 & 1 & 1 \\ 
  1 & 0 & 1 & 2 \\
  1 & 0 & 2 & 5/2
  \end{Bmatrix}  \,.
\end{align}
The operator has the usual LCS point a the origin $z=0$, a K-point at $z=-1/108$ and a conifold singularity at $z=1/324$ while the monodromy at $z=\infty$ is simply of finite order type and thus not interesting for our purposes. The underlying geometry has the following topological data
\begin{align}
    \kappa=9 \,, \qquad c_2=30 \,, \qquad \chi=12 \,.
\end{align}
We will be interested in the transition matrix and LMHS of the K-point. The numerical computation of the transition matrix from the Frobenius basis to the monodromy weight frame yields
\begin{align}
\hat{m}_K=  \scalebox{0.8}{$
\left(
\begin{array}{cccc}
 22.1636 & 0 & 945.376 & 0 \\
 -11.0818+11.0818 i & 0 & -472.688+499.74 i & 0 \\
 6.57905\, +12.392 i & -11.0818-11.0818 i & 234.593\, +603.583 i & -472.688-499.74 i \\
 2.0763\, +11.0818 i &  -22.1636 i & -3.502+499.74 i & -999.481 \\
\end{array}
\right) $} .
\end{align}
Next we identify these numerical values in terms of the appropriate geometrical counterparts. To this end we use the decomposition \eqref{mfactorized} where we take out a factor based on LMHS data. This factor \eqref{mK} is fixed by the data
\begin{equation}
    \begin{pmatrix}
        a & b \\
        b & c
    \end{pmatrix} = \begin{pmatrix}
        1 & 1 \\
        1 & 2
    \end{pmatrix}\, , \quad \tau =-\frac{1}{2}+\frac{i}{2}\, , \qquad \gamma = \delta = 0.355955\, ,
\end{equation}
The other factor acts directly on the Frobenius periods and reads 
\begin{equation}
    \hat{m}_{\rm K}^{\rm Frob} = \begin{pmatrix}
        9\pi L(f_{36},1) & 0 & \frac{3159 L(f_{36},1)}{8} & 0  \\
        9\pi h L(f_{36},1) & 9\pi L(f_{36},1) & \frac{9L(f_{36},1)}{16}(351 i+702 h \pi -8i\rho) & \frac{3159 L(f_{36},1) \pi}{8}  \\
        0 & 0 & -\frac{27 \pi}{4 L(f_{36},1)} & 0 \\
        0 & 0 & \frac{27(3i+2h\pi)}{8L(f_{36},1)} & -\frac{27}{4L(f_{36},1)}
    \end{pmatrix}\, ,
\end{equation}
where we defined the coordinate rescaling
\begin{equation}
    h = \frac{3 \log[3]}{4\pi i}\, .
\end{equation}
The modular form $f_{36}$ refers to $36.3.d.a$ which has CM by $d=-4$, whose L-values are given by
\begin{equation}
    L(f_{36},1) = \frac{\Gamma \left(\frac{1}{4}\right)^4\Gamma \left(\frac{1}{2}\right)^2}{12 \sqrt{2} \sqrt[4]{3} \pi ^3} \,, \quad L(f_{36},2) =\frac{\Gamma \left(\frac{1}{4}\right)^4\Gamma \left(\frac{1}{2}\right)^2}{36 \sqrt{2} \sqrt[4]{3} \pi^2 }
\end{equation}
We furthermore defined the numerical parameter
\begin{equation}
    \rho \simeq 69.61364705994566\,.
\end{equation}
for which it would be interesting to give a more precise interpretation in the future.
The $\mathbb{R}$-split boundary data is extracted by the standard procedure from the period vector in the monodromy weight frame
\begin{align}
N=\left(
\begin{array}{cccc}
 0 & 0 & 0 & 0 \\
 0 & 0 & 0 & 0 \\
 1 & 1 & 0 & 0 \\
 1 & 2 & 0 & 0 \\
\end{array}
\right) \,, \quad 
\omega_{3,1}=\left(
\begin{array}{c}
 1 \\
 -\frac{1}{2}+\frac{i}{2} \\
  \gamma(\frac{1}{2}+\frac{i}{2}) \\ \gamma \\
\end{array}
\right) \, .
\end{align}
The canonical coordinate change for this example is given by 
\begin{align}
    z=-\frac{1}{3 \sqrt{3}}q\left(1-\sqrt{3}\frac{\rho}{9}q + \frac{1}{1152} \left( 100359-3384 \rho +64 \rho^2 \right)q^2 + \dots \right)\, .
\end{align}
The holomorphic series $B(q)$ specifying the instanton map reads
\begin{align}
    B(q)= -\frac{1}{4 \sqrt{3} L(f_{36},1)^2} \left(q + \frac{1}{48 \sqrt{3}} \left(  1269-16 \rho \right)q^2 + \frac{1}{1152} \left( 394317 -10152 \rho + 64 \rho^2 \right) q^3 + \dots \right)\, , \nonumber
\end{align}
which may be used to compute the Yukawa coupling as
\begin{align}
    Y(q)= \frac{\pi^2}{ \sqrt{3} L(f_{36},1)^2} \left(q + \frac{1}{12 \sqrt{3}} \left(  1269-16 \rho \right)q^2 + \frac{1}{128} \left( 394317 -10152 \rho + 64 \rho^2 \right) q^3 + \dots \right)\, . \nonumber
\end{align} 
The series here are complicated by the presence of the numerical parameter $\rho$. It would be interesting to figure out what this number, as well as $\gamma$ in the boundary data, corresponds to in the degenerate geometry that arises at the K-point.

\section{Conclusions}\label{sec:conclusions}
The wide array of degenerations possible for Calabi-Yau manifolds give rise to a rich variety in phases for effective field theories arising from string theory. We have explored such regions in one-dimensional complex structure moduli spaces, focusing on those away from the familiar LCS point: conifold points and K-points.

Our study started from the framework of Hodge theory, which we used to describe the periods in each of these phases. The main take-away from this perspective is that it provides a universal frame suitable for any example. It does so through providing a convenient integral basis and a natural coordinate that are both specialized to the phase under consideration. This choice elucidates the geometrical properties underlying the Calabi-Yau manifold in this regime, which would not have been directly accessible in the usual LCS basis and coordinate. We do note however that this methodology, when applied to the LCS phase, does give in fact both the mirror D-brane basis and the mirror map familiar from the physics literature.

From this Hodge-theoretic viewpoint we classified the boundary data for every phase, sorting it into monodromy data, rigid periods and extension data. We found that the information in each data class can be understood from the degenerate geometry that arises at the singularity: 
\begin{itemize}
    \item The monodromy matrix contains intersection data about the geometry. In the LCS phase it gives the triple intersection numbers; for conifold points it specifies the quotient order of the conifold cycle; for K-points it encodes the intersection pairing of a rigid K3 surface. 
    \item The rigid periods belong to the degenerate variety at the singularity: for conifold points it is the two-component period vector of the respective conifold; for K-points it is the period vector of the aforementioned K3 surface.
    \item The extension data parametrizes mixing among the periods. This information is crucial in order to quantize the integral basis correctly. At LCS points this is described by the integrated second Chern class of the mirror manifold; for other phases the extension data gives us similar prescriptions.
\end{itemize} 
The numerical values of the data in the latter two classes have a natural interpretation through modularity: they are specified by L-function values of modular forms associated to the variety. We identified these modular forms explicitly by matching the L-values in the studied examples using the database \cite{lmfdb} and PariGP. We collected our findings for all boundary data in tables \ref{table:C} and \ref{table:K}.

In order to bridge the gap to physics, we computed the couplings in the 4d $\mathcal{N}=2$ and $\mathcal{N}=1$ supergravity theories. Here the gauge-kinetic functions of the two U(1) vectors in particular illuminated the underlying geometry: at conifold points one coupling is constant and given by the rigid conifold period, while the other is moduli-dependent and proportional to the covering coordinate; at K-points the coupling matrix was given by the intersection form of the rigid K3 surface, multiplied by the covering coordinate. 

Our results also provide a good future starting point for model building of phenomenological scenarios, both from an abstract perspective as well as by using the explicit data we have given. The examples with algebraic data provide simple models where it is straightforward to tune flux quanta. On the other hand, these geometries are complemented by other examples where more transcendental data appears, enabling one to push such constructions to more general settings. It would be interesting to carefully track the period data such as L-values in these computations and see how they determine physical observables such as moduli masses and vacuum superpotentials. Instead of resorting to numerical approximations, this approach could also lead to exact algebraic relations in for instance solving the vacuum equations.

Finally, it would be interesting to continue the research initiated in this paper towards settings with multiple moduli. With the methods laid out in this work it should be straightforward to generalize the monodromy weight basis and period expressions to multi-moduli boundaries, adapting for instance the two-moduli models of \cite{Bastian:2021eom} to a convenient integral basis. In this framework one can then extract the boundary data of examples, again matching numerical parameters with topological and arithmetic numbers associated to the geometry. 

\subsubsection*{Acknowledgements}
We would like to thank Thomas Grimm for early collaboration on this work. Moreover, it is a pleasure to thank Kilian B\"onisch, Chuck Doran, Matt Kerr, Erik Plauschinn, John Stout, and Max Wiesner for discussions and correspondence. The research of LS is supported, in part, by the Dutch Research Council (NWO) via a Start-Up grant and
a Vici grant.

\appendix 
\section{Review of 4d supergravity theories}
In this section we briefly review the physical couplings in 4d supergravity theories arising from Type IIB Calabi-Yau compactifications. These expressions are used in section \ref{sec:summary} in order to clarify the physical role of the geometrical data appearing in the periods for the different asymptotic regimes in complex structure moduli space. We consider both $\mathcal{N}=2$ supergravity theories where we compute the K\"ahler potential, metric and gauge kinetic functions, as well as the flux superpotential and scalar potential in $\mathcal{N}=1$ orientifold compactifications.

\paragraph{4d action.} We start off with $\mathcal{N}=2$ supergravity theories, focusing on the gravity and vector multiplet sector. The scalars $z^i$ in the vector multiplets $(i=1,\ldots,h^{2,1}$) arise from the complex structure moduli of the Calabi-Yau threefold. The vectors $A^I$ ($I=0,\ldots, h^{2,1}$) arise from expanding the R-R three-form potential of Type IIB along a basis of three-forms in its middle cohomology $H^3(Y_3,\mathbb{Z})$, with the additional vector coming from the gravity multiplet. For completeness, let us write down the 4d action for this sector as
\begin{equation}
    S^{(4)} = \int_{M^{3,1}} \bigg( \frac{1}{2} R \, \hat{\ast} \, 1 - \frac{1}{2} K_{i\bar{j}} \, \mathrm{d} z^i \wedge \hat{\ast} \, \mathrm{d}\bar{z}^{\bar j} +\frac{1}{4} \mathcal{I}_{IJ} \,  F^I \wedge \hat{\ast} \, F^J +\frac{1}{4} \mathcal{R}_{IJ} \, F^I \wedge F^J \bigg) \, ,
\end{equation}
where we denoted the field strengths by $F^I = \mathrm{d}A^I$, and $\hat{\ast}$ denotes the 4d Hodge star. The kinetic terms of the scalars are determined by the K\"ahler metric $K_{i\bar{j}}$ on the moduli space. The gauge kinetic functions for the field strengths are computed by $\mathcal{N}_{IJ} = \mathcal{R}_{IJ} + i \mathcal{I}_{IJ}$, which from a geometrical perspective correspond to the Hodge star operator of the three-form cohomology, cf.~\eqref{eq:cM}. In the following we describe precisely how these moduli-dependent physical couplings are computed from the period data of the Calabi-Yau threefold.

\paragraph{Periods and prepotential.} Let us begin by briefly recalling the relation between the periods and the prepotential, since the prepotential is used as the starting point for the computation of most physical couplings. The periods can be computed from the prepotential as
\begin{equation}
    \Pi = \begin{pmatrix}
        X^I \\
        F_I
    \end{pmatrix},
\end{equation}
where the index runs over $I=0,\ldots, h^{2,1}$, and the magnetic periods are computed as derivatives $F_I = \partial_I F$ from the prepotential $F(X^I)$. In this formulation the prepotential is a homogeneous function of degree two in the coordinates $X^I$. In practice it is often convenient to switch to the coordinate $z = X^1/X^0 $ and fix $X^0=1$ by a rescaling of the period vector. The period vector then reads
\begin{equation}
    \Pi = \begin{pmatrix}
        1 \\
        z \\
        \partial_z F \\
        2F - z \partial_z F
    \end{pmatrix}
\end{equation}
where we used that $F_0 = \partial_0 \big( (X^0)^2 F(z)\big) = 2F - \partial_z F$ setting $X^0=1$ at the very end. In the following we phrase the physical couplings both in terms of the periods $(X^I ,F_I)$ as well as using the prepotential $F$.

\paragraph{K\"ahler potential and metric.} Having set up the periods and the prepotential, we are now in the position to write down the K\"ahler potential. By writing out the volume as computed by the periods we can express it in terms of the prepotential as
\begin{equation}
    K = -\log [ i\langle \Pi , \bar{\Pi} \rangle] = -\log[ 2i (F-\bar{F}) - i(z-\bar{z})(\partial_z F +\partial_{\bar z}\bar{F}) ]\, .
\end{equation}
By taking two derivatives we find as K\"ahler metric
\begin{equation}
    K_{z \bar{z}} = \partial_z \partial_{\bar z} K\, ,
\end{equation}
We note that any holomorphic rescaling of the period vector $\Pi \to e^{f(z)} \Pi$ transforms the K\"ahler potential as $K \to K+f(z)+\bar{f}(\bar{z})$, but this K\"ahler transformation leaves the K\"ahler metric invariant.

\paragraph{Yukawa coupling.} From the periods we can also obtain the Yukawa coupling of the 4d $\mathcal{N}=2$ supergravity theory, which characterizes certain terms in the action coupling two fermions and the scalar in the vector multiplet.  We can compute it by taking three derivatives of the period vector as
\begin{equation}
    Y=(2 \pi i z)^3 \langle \Pi, \, \partial_z^3 \Pi \rangle\, ,
\end{equation}
where $z$ denotes a normal coordinate for the boundary. The overall $(2\pi i z)^3$ has been included for normalization of the coefficients.

\paragraph{Gauge kinetic functions.} Having characterized the kinetic terms for the scalars, we next move on to the gauge-kinetic functions for the vectors. In terms of the electric and magnetic periods $X^I$ and $F_I$ we can compute these periods as
\begin{equation}\label{eq:NIJ}
    \mathcal{N}_{IJ} = \begin{pmatrix}
        F_I & D_{\bar{z}} \bar{F}_I
    \end{pmatrix}\begin{pmatrix}
        X^J & D_{\bar z} \bar{X}^J
    \end{pmatrix}^{-1}\, .
\end{equation}
where we introduced the K\"ahler covariant derivative $D_z X^I = \partial_z X^I + (\partial_z K) X^I$ (and equivalently $D_z F_I = \partial_z F_I + (\partial_z K) F_I$). Alternatively, the gauge kinetic functions can be formulated in terms of the prepotential
\begin{equation}
    \mathcal{N}_{IJ} = \bar{F}_{IJ} + 2i\frac{\text{Im}(\mathcal{F})_{IK}X^K \text{Im}(\mathcal{F})_{JL}X^L}{\text{Im}(\mathcal{F})_{MN}X^M X^N}\, .
\end{equation}
where we used the shorthand $F_{IJ} = \partial_I F_J$. These gauge-kinetic functions correspond geometrically to the Hodge star operator, as is made more precise below in \eqref{eq:cM} in the context of the $\mathcal{N}=1$ flux potential.

\paragraph{Flux superpotential and scalar potential.} Having discussed the couplings for the $\mathcal{N}=2$ supergravity theory, we now move on to $\mathcal{N}=1$ Type IIB Calabi-Yau orientifolds. We consider the Gukov-Vafa-Witten flux superpotential and the resulting scalar potential for the complex structure moduli and axio-dilaton. From the 10d Type IIB perspective this scalar potential is induced by turning R-R and NS-NS three-form fluxes $F_3$ and $H_3$ along three-cycles of the Calabi-Yau manifold. The flux superpotential is given by
\begin{equation}
    W = \langle G_3, \Pi \rangle = g_I X^I - g^I F_I = g_0 +g_1 z -g^1 \partial_z F -g^0 (2F -z \partial_z F)\, .
\end{equation}
where we expanded the three-form flux in terms of an integral basis $G_3 = (g^I, g_I)$. It may be decomposed in terms of the R-R and NS-NS fluxes by $G_3 = F_3 - \tau H_3$ with $\tau$ the axio-dilaton, but we choose to leave this implicit in our expressions. In turn, the scalar potential follows 
\begin{equation}\label{eq:potential}
    V = e^K K^{\mathcal{I}\mathcal{J}} D_{\mathcal{I}} W D_{\bar{\mathcal{J}}} \bar{W} = \frac{1}{\tau_2}\big(\langle \bar{G}_3 , \star G_3 \rangle -i \langle \bar{G}_3 , \, G_3 \rangle \big) \, ,
\end{equation}
where the sum over $\mathcal{I}$ runs over the complex structure moduli and the axio-dilaton. The expression on the right-hand side depends on the complex structure moduli through the Hodge star operator $\star$ of the Calabi-Yau manifold. We can make this more explicit by writing the scalar potential into a bilinear form as
\begin{equation}
    V = \frac{1}{\tau_2}\bar{G}_3^T \big( \mathcal{M} + i\Sigma \big) G_3\, ,
\end{equation}
with the Hodge star matrix given by
\begin{equation}\label{eq:cM}
    \mathcal{M} = \begin{pmatrix}
        \mathcal{I} + \mathcal{R}\mathcal{I}^{-1}\mathcal{R} & \mathcal{R}\mathcal{I}^{-1} \\
        \mathcal{R}\mathcal{I}^{-1} & \mathcal{I}^{-1}
    \end{pmatrix}\, .
\end{equation}
We expressed this Hodge star operator in terms of the $\mathcal{N}=2$ gauge kinetic functions defined before in \eqref{eq:NIJ}. Now \eqref{eq:cM} clarifies how these gauge kinetic functions are precisely related to the Hodge star operator of the Calabi-Yau manifold.

\section{L-function values and modular forms}
\label{app:Lvalues}
In this section we give a brief definition of critical L-function values. For more details we refer to the literature.
Given the $q$-series expansion of a weight $k$ modular function $f$

\begin{equation}
    f(\tau)=\sum_{n\ge 0} a_n \, q^n\;,
\end{equation}
where $q=e^{2\pi i \tau}$, its corresponding L-function is defined as

\begin{equation}
    L(f,x)=\sum_{n\ge 0} \frac{a_n}{n^x}\;.
\end{equation}
A value $L(f,j)$ is called a critical L-value if $j\in \{1,2,\dots,k-1\}$. They follow a simple functional equation. Defining the completed L-values as
\begin{equation}
    L^{\star}(f,s)=(2\pi)^{-s} L(f,s), \label{eq:LFuncRel}
\end{equation}
they fulfill the equation
\begin{equation}
    L^{\star}(f,s)=\pm L^{\star}(f,k-s)
\end{equation}
L-values also have a simple integral representations in terms of the underlying modular form
\begin{equation}
    L(f,j)=\frac{(2\pi)^s}{ \Gamma(s)}\int_0^\infty f(i \tau) \tau^{s-1}{\rm d}\tau.
\end{equation}
\begin{equation}
    L^{\star}(f,j)=\int_0^\infty f(i \tau) \tau^{s-1}{\rm d}\tau.
\end{equation}
The Hecke operators $T_m$ are defined by their action on a modular form as
\begin{equation}
    T_m f(\tau)= m^{k-1}\sum_{d|m}d^{-k}\sum\limits_{b=0}^{d-1}f\left(\frac{m\tau+bd}{d^2}\right)\;.
\end{equation}
A modular form which is an eigenfunction of all Hecke operators is called a Hecke eigenform, i.e.
\begin{equation}
    T_m f(\tau)=\lambda_mf(\tau)\;.
\end{equation}
The eigenvalues are the expansion coefficients of the form, $\lambda_n=c_n$ and are multiplicative $\lambda_{n+m}=\lambda_n\cdot \lambda_m$. Thus the knowledge of the prime coefficients is sufficient. The examples of modular forms appearing at the points at infinity all have CM, which allows to express them in terms of $\Gamma$-functions. This representation makes it easier to identify the relations between the different L values as well as the relation to the L-values of elliptic curves. 

\begin{table}[h!]
{{ 
\begin{center}
	\begin{tabular}{|c|c|c|c|c|}
		\hline
	     Geometry&modular form&d&  $L(f,1)$	&$L(f,2)$ 					   \\\hline
      $X_{3,2,2}$&9.4.a.a&-3&$\frac{\Gamma(1/3)^9}{32\sqrt{3}\pi^5}$&$\frac{\Gamma(1/3)^9}{96\pi^4}$\\
      \hline
     $X_{6,2}$&108.4.a.c&-3&$\frac{3\sqrt{3}\Gamma(1/3)^9}{16\, 2^{1/3}\pi^5}$&$\frac{\Gamma(1/3)^9}{32\, 2^{1/3}\pi^4}$\\
     \hline
     $X_{4,2}$&32.4.a.b&-4&$\frac{\Gamma(1/4)^6\Gamma(1/2)^3}{16\sqrt{2}\pi^5}$&$\frac{\Gamma(1/4)^6\Gamma(1/2)^3}{64\sqrt{2}\pi^4}$\\
     \hline
	 \end{tabular}	
\end{center}}}
\caption{\label{table:Lvalues} L-values of weight 4 CM forms.}
\end{table}
Note that the L-values for the conifold points all have the same form, including 9 $\Gamma$-factors. This allows to rewrite the weight 4 L-values in terms of qubes of weight 2 L-values. By the Chowla-Selberg formula any period of an elliptic curve can be expressed as an algebraic multiple of a special transcendental number $b_{\mathbb{Q}(\sqrt{d})}$ depending on the imaginary quadratic field.
For CM with $d=-3$ and $d=-4$ these are given by
\begin{equation}
    b_{\mathbb{Q}(\sqrt{-4})}=\frac{1}{\sqrt{2}\pi}\Gamma(\frac{1}{2})\Gamma(\frac{1}{4})^2 \qquad b_{\mathbb{Q}(\sqrt{-3})}=\left(\frac{3}{4}\right)^{3/4}\frac{1}{\pi}\Gamma(\frac{1}{3})^3\;.
\end{equation}
From table \ref{table:Lvalues} one can see that the transcendantal part of the L-values of CM conifolds are exactly qubes of these numbers, i.e. $L(f,1)=\frac{a}{\pi^2}\cdot b_{\mathbb{Q}(\sqrt{d})}^3$ with $a$ an algebraic number.
This was already observed in \cite{li2018computing} for the 9.4.a.a modular form. The explicit expression for this modular form is
\begin{equation}
    f_9=\eta(3\tau)^8\;.
\end{equation}
\cite{li2018computing} shows that the following identity holds
\begin{equation}
   \frac{8}{3} L(\eta(6\tau)^4,1)^3=L(\eta(3\tau)^8,3)\;.
\end{equation}
This together with $L(\eta(3\tau)^8,3)=\frac{4}{\pi^2}L(\eta(3\tau)^8,1)$ gives the relation between the elliptic curve L-value and the L-value at the CM conifold. The origin of these relations are the Shioda-Inose structures relating the underlying geometries.
Moreover, \cite{li2018computing} shows that
\begin{equation}
   \frac{3}{2} L(\eta(6\tau)^4,1)^2=L(\eta(2\tau)^3\eta(6\tau)^3,2)\;.
\end{equation}
Note that the modular form $\eta(2\tau)^3\eta(6\tau)^3$ on the right hand side is the modular form 12.3.c.a appearing in the $X_{6,6}$ K-point. We observe this property at all K-points, i.e. we find a relation of the form

\begin{equation}
    L(f_3,2)=a\cdot L(f_2,1)^2\qquad a\in \mathbb{Q}\;.
\end{equation}
For example the modular form associated to $X_{3,3}$ is related to the  modular curve $X_0(27)$ and the associated weight 2 modular form 27.2.a.a . The square L value $L(E,1)$ then gives the required L value of the weight 3 form:

\begin{align}
 L(f,2)=3 L(E,1)^2&=\frac{\Gamma \left(\frac{1}{3}\right)^4}{27 \Gamma \left(\frac{2}{3}\right)^2}=\frac{\Gamma(\frac{1}{3})^6}{36\pi^2}
\end{align}
Table \ref{table:relations} lists the related modular forms. It is interesting to observe that the level of the weight 2 form is given by $N_2=\sqrt{\mu}$.
\begin{table}[h!]
{{ 
\begin{center}
	\begin{tabular}{|c|c|c|c|c|}
		\hline
	     Geometry&modular form&  $N_2$	&$d$ & $j$					   \\\hline
      $X_{3,3}$&27.3.b.a&27&-3&0\\
		\hline
        $X_{4,4}$&16.3.c.a&64&-4&1728\\
		\hline
$X_{6,6}$&12.3.c.a&432&-3&0\\
		\hline
	 \end{tabular}	
\end{center}}}
\caption{\label{table:relations} Relations between weight 2 and weight 3 L-values.}
\end{table}

\section{Instanton map formulation of periods}\label{app:instantonmap}
In this appendix we provide details on the instanton map formulation of the periods discussed in section \ref{ssec:Hodge1}. In the first subsection we lay out the complete set of conditions that this instanton map $\Gamma(z)$ needs to obey. In the other two subsections we turn to the three conifold and K-points we studied in the hypergeometric examples, where we explicitly identify the combinations of hypergeometric functions that specify the components of $\Gamma(z)$.

\subsection{Constraints on the instanton map}
The instanton expansion in \eqref{eq:periodexpansion} can be conveniently encoded in terms of a holomorphic, Lie algebra-valued operator $\Gamma(z)$ as described in \cite{CattaniFernandez2000,CattaniFernandez2008, Bastian:2021eom}. In \cite{Bastian:2021eom} this map was therefore referred to as the instanton map. In this formulation the period vector takes the following form 
\begin{equation}
    \Pi(t) = e^{t N} e^{\Gamma(e^{2\pi i t})} a_0\, .
\end{equation}
The instanton map $\Gamma(z)$ here satisfies the following properties:
\begin{itemize}
    \item It is holomorphic in $z$ and vanishes at the boundary $\Gamma(0)=0$. This condition ensures that the instanton terms go to zero when we approach the boundary.
    \item It is valued in the algebra of the isometry group of the symplectic pairing. For our purposes this means that
    \begin{equation}
        \Gamma(z) \in \mathfrak{sp}(4)\, .
    \end{equation}
    \item It acts in a decreasing manner on the first index of the Deligne splitting \eqref{Ipq}. To be precise, it is valued in
    \begin{equation}
        \Gamma(z) \in \Lambda_-\, , \qquad \Lambda_- = \oplus_{p<0} \oplus_q\Lambda^{p,q}\, ,
    \end{equation}
    where we defined the operator subspaces
    \begin{equation}
        \mathcal{O}_{p,q} \in \Lambda_{p,q}\, , \qquad \mathcal{O}_{p,q} I^{r,s} \subseteq I^{r+p, q+s}\, . 
    \end{equation}
    \item Its higher-charged components $\Gamma_{-p}$ can be related through  differential equations to the lower-charged operators $\Gamma_{-1}, \ldots , \Gamma_{-p+1}$. To be precise, these conditions read
    \begin{equation}\label{eq:recursiongamma}
        \partial_{z} \exp[\Gamma(z)]_{-p} = [\exp[\Gamma(z)]_{-p+1}, N_{i}]+  \exp[\Gamma(z)]_{-p+1}  \partial_{i} \Gamma_{-1}(z)\, .
    \end{equation}
    We thus find that the first component $\Gamma_{-1}$ fixes all other components recursively through this set of differential equations. 
    We note that in the multi-moduli case this recursion needs to be supplemented by a constraint on the $\Gamma_{-1}$ component, but for this paper focusing on the one-modulus case this will not be necessary.
    \item Finally, the charge minus one component $\Gamma_{-1}$ needs to satisfy a rank condition, which in the one-modulus case reads
    \begin{equation}
    \text{rk}(N+\partial_t \Gamma_{-1}(e^{2\pi i t})) = 3\, .
    \end{equation}
    For the large complex structure point the log-monodromy matrix satisfies $\text{rk}(N_{\rm LCS})=3$, and hence instanton corrections can be consistently dropped in a leading order approximation. However, for boundaries such as a conifold point (type $\mathrm{I}_1$) or K-point (type $\mathrm{II}_0$) these ranks are given by $\text{rk}(N_c)=1$ and $\text{rk}(N_{K})=2$ respectively. Hence $\Gamma$ cannot be set to zero, and thus exponential instantons have to be present; in \cite{Bastian:2021eom} these were therefore termed \textit{essential instantons}.
\end{itemize}

\subsection{Conifold points in hypergeometric models}
In this section we consider the three conifold points at infinity arising in the hypergeometric models. We determine explicitly the two functions $A(z),n(z)$ that specify the instanton map as discussed in section \ref{ssec:conifoldconstruction}. This is done by rotating the periods into the LMHS basis, allowing us to identify these functions from \eqref{eq:Cperiod}.

\paragraph{$\mathbf{X_{4,2}}$ geometry.} The instanton map functions are given by
\begin{equation}
\begin{aligned}
A(z) &= -\frac{32 i \pi ^{7/2}}{\Gamma \left(\frac{1}{4}\right)^6} \frac{z \, _4F_3\left(\frac{1}{2},\frac{1}{2},\frac{1}{2},\frac{1}{2};\frac{3}{4},1,\frac{5}{4};4 z^4\right)}{\,_4F_3\left(\frac{1}{4},\frac{1}{4},\frac{1}{4},\frac{1}{4};\frac{1}{2},\frac{3}{4},\frac{3}{4};4 z^4\right)} \, , \\
n(z) &=i  \frac{\log[z]}{\pi} + \frac{2\pi^{3/2}}{\Gamma(\frac{1}{4})^6} \frac{1}{\partial_z A(z)} \partial_z \Big( \frac{G_{4,4}^{2,4}\left(4 z^4|\begin{array}{c}
 \frac{3}{4},\frac{3}{4},\frac{3}{4},\frac{3}{4} \\
 \frac{1}{4},\frac{1}{4},0,\frac{1}{2} \\
\end{array}\right)}{\, _4F_3\left(\frac{1}{4},\frac{1}{4},\frac{1}{4},\frac{1}{4};\frac{1}{2},\frac{3}{4},\frac{3}{4};4 z^4\right)} \Big)\, . \\
\end{aligned}
\end{equation}
Their series expansions may be found in paragraph \hyperref[parX42]{$X_{4,2}$} in section \ref{sec:examples}: $n(z)$ encodes the change to the canonical coordinate $q(z)$, while the expansion of $A(q)$ has been given there directly in terms of this coordinate.

\paragraph{$\mathbf{X_{3,2,2}}$.} The instanton map functions are given by
\begin{equation}
\begin{aligned}
A(z) &= -\frac{32 \,i \, 2^{2/3} \pi ^4}{\sqrt{3} \Gamma \left(\frac{1}{6}\right)^2 \Gamma \left(\frac{1}{3}\right)^5} \frac{z \, _4F_3\left(\frac{1}{2},\frac{1}{2},\frac{1}{2},\frac{1}{2};\frac{5}{6},1,\frac{7}{6};\frac{16 z^6}{27}\right)}{\, _4F_3\left(\frac{1}{3},\frac{1}{3},\frac{1}{3},\frac{1}{3};\frac{2}{3},\frac{5}{6},\frac{5}{6};\frac{16 z^6}{27}\right)}\, , \\
   n(z) &= \frac{3i \log[z]}{\pi} + \frac{16 \pi ^2}{3 \Gamma \left(\frac{1}{6}\right)^2 \Gamma \left(\frac{1}{3}\right)^5} \frac{1}{\partial_z A(z)} \partial_z \Big(\frac{G_{4,4}^{2,4}\left(\frac{16 z^6}{27}|\begin{array}{c}
 \frac{2}{3},\frac{2}{3},\frac{2}{3},\frac{2}{3} \\
 \frac{1}{6},\frac{1}{6},0,\frac{1}{3} \\
 \end{array}\right)}{\,_4F_3\left(\frac{1}{3},\frac{1}{3},\frac{1}{3},\frac{1}{3};\frac{2}{3},\frac{5}{6},\frac{5}{6};\frac{16 z^6}{27}\right)}\Big)\, . \\
\end{aligned}
\end{equation}
Their series expansions may be found in paragraph \hyperref[parX322]{$X_{3,2,2}$} in section \ref{sec:examples}: $n(z)$ encodes the coordinate change to $q(z)$, while the expansion of $A(q)$ has been given there directly in $q$.

\paragraph{$\mathbf{X_{6,2}}$ geometry.} The instanton map functions are given by
\begin{equation}
\begin{aligned}
A(z) &= \frac{32 i 2^{2/3} \pi ^3}{\Gamma \left(\frac{1}{6}\right)^4 \Gamma \left(\frac{1}{3}\right)}   \frac{z \, _4F_3\left(\frac{1}{2},\frac{1}{2},\frac{1}{2},\frac{1}{2};\frac{2}{3},1,\frac{4}{3};\frac{256z^3}{27}\right)}{\,_4F_3\left(\frac{1}{6},\frac{1}{6},\frac{1}{6},\frac{1}{6};\frac{1}{3},\frac{2}{3},\frac{2}{3};\frac{256 z^3}{27}\right)}  \, , \\
   n(z) &= \frac{i \log[z]}{2\pi} - \frac{8\pi}{3\sqrt{3} \Gamma(\frac{1}{3}) \Gamma(\frac{1}{6})^4} \frac{1}{\partial_z A(z)} \partial_z \Big( \frac{G_{4,4}^{2,4}\left(\frac{256 z^3}{27}|
\begin{array}{c}
 \frac{5}{6},\frac{5}{6},\frac{5}{6},\frac{5}{6} \\
 \frac{1}{3},\frac{1}{3},0,\frac{2}{3} \\
\end{array}\right)}{\,_4F_3\left(\frac{1}{6},\frac{1}{6},\frac{1}{6},\frac{1}{6};\frac{1}{3},\frac{2}{3},\frac{2}{3};\frac{256 z^3}{27}\right)} \Big)\, . \\
\end{aligned}
\end{equation}
Their series expansions may be found in paragraph \hyperref[parX62]{$X_{6,2}$} in section \ref{sec:examples}: $n(z)$ encodes the coordinate change to $q(z)$, while the expansion of $A(q)$ has been given there directly in $q$.

\subsection{K-points in hypergeometric models}
In this section we consider the three K-points arising in the hypergeometric models. We identify the two functions $B(z), n(z)$ that specify the instanton map as we discussed in section \ref{ssec:Kconstruction}. Similar to the conifold points, this is done by rotating the periods into the LMHS basis, allowing us to identify these functions from \eqref{eq:Kperiods}.

\paragraph{$\mathbf{X_{3,3}}$ geometry.} The instanton map functions, in terms of the $3$-fold covering coordinate, are
\begin{equation}
    \begin{aligned}
        n(z) &= \frac{5}{6}(1-\frac{i}{\sqrt{3}})-\frac{\log[z]}{2\pi i}+ \frac{i \sqrt[3]{2} \pi  G_{4,4}^{2,4}\left(6912 z^3|
\begin{array}{c}
 \frac{5}{6},\frac{5}{6},\frac{5}{6},\frac{5}{6} \\
 0,0,\frac{2}{3},\frac{2}{3} \\
\end{array}
\right)}{27 \Gamma \left(\frac{1}{3}\right)^6 \, _4F_3\left(\frac{1}{6},\frac{1}{6},\frac{1}{6},\frac{1}{6};\frac{1}{3},\frac{1}{3},1;6912 z^3\right)}\, , \\
        B(z) &= \frac{384 \left(1+i \sqrt{3}\right) \pi ^6 z^2 \, _4F_3\left(\frac{5}{6},\frac{5}{6},\frac{5}{6},\frac{5}{6};1,\frac{5}{3},\frac{5}{3};6912 z^3\right)}{\Gamma
   \left(\frac{1}{3}\right)^{12} \, _4F_3\left(\frac{1}{6},\frac{1}{6},\frac{1}{6},\frac{1}{6};\frac{1}{3},\frac{1}{3},1;6912 z^3\right)}\, .
    \end{aligned}
\end{equation}
Their series expansions have been given in paragraph \hyperref[parX33]{$X_{3,3}$} of section \ref{sec:examples}: $n(z)$ gives the coordinate change to $q(z)$, while $B(q)$ is directly given in the canonical coordinate $q$ instead of $z$. The constant and logarithmic term in $n(z)$ are present to cancel off such terms in the expansion of the hypergeometric functions; similar terms will be present in the other two examples.

\paragraph{$\mathbf{X_{4,4}}$ geometry.} The instanton map functions, in terms of the $2$-fold covering coordinate, are
\begin{equation}
    \begin{aligned}
        n(z) &= -\frac{i}{2} -\frac{\log[z]}{2\pi i} +\frac{i G_{4,4}^{2,4}\left(256 z^2|
\begin{array}{c}
 \frac{3}{4},\frac{3}{4},\frac{3}{4},\frac{3}{4} \\
 0,0,\frac{1}{2},\frac{1}{2} \\
\end{array}
\right)}{1024 \Gamma \left(\frac{5}{4}\right)^4 \, _4F_3\left(\frac{1}{4},\frac{1}{4},\frac{1}{4},\frac{1}{4};\frac{1}{2},\frac{1}{2},1;256 z^2\right)}\, , \\
        B(z) &= \frac{256 \pi ^4 z \, _4F_3\left(\frac{3}{4},\frac{3}{4},\frac{3}{4},\frac{3}{4};1,\frac{3}{2},\frac{3}{2};256 z^2\right)}{\Gamma \left(\frac{1}{4}\right)^8 \,
   _4F_3\left(\frac{1}{4},\frac{1}{4},\frac{1}{4},\frac{1}{4};\frac{1}{2},\frac{1}{2},1;256 z^2\right)}\, . 
    \end{aligned}
\end{equation}
Their series expansions have been given in paragraph \hyperref[parX44]{$X_{4,4}$} of section \ref{sec:examples}: $n(z)$ gives the coordinate change to $q(z)$, while $B(q)$ is directly given in the canonical coordinate $q$ instead of $z$.

\paragraph{$\mathbf{X_{6,6}}$ geometry.} The instanton map functions, in terms of the $3$-fold covering coordinate, are
\begin{equation}
    \begin{aligned}
        n(z) &= \frac{13-i\sqrt{3}}{6}-\frac{\log[z]}{2\pi i}+\frac{2 i \pi  G_{4,4}^{2,4}\left(\frac{z^3}{27}|
\begin{array}{c}
 \frac{2}{3},\frac{2}{3},\frac{2}{3},\frac{2}{3} \\
 0,0,\frac{1}{3},\frac{1}{3} \\
\end{array}
\right)}{9 \Gamma \left(\frac{1}{3}\right)^6 \, _4F_3\left(\frac{1}{3},\frac{1}{3},\frac{1}{3},\frac{1}{3};\frac{2}{3},\frac{2}{3},1;\frac{z^3}{27}\right)} \, , \\
        B(z) &= \frac{32 \left(1+i \sqrt{3}\right) \pi ^6 z \, _4F_3\left(\frac{2}{3},\frac{2}{3},\frac{2}{3},\frac{2}{3};1,\frac{4}{3},\frac{4}{3};\frac{z^3}{27}\right)}{9 \Gamma
   \left(\frac{1}{3}\right)^{12} \, _4F_3\left(\frac{1}{3},\frac{1}{3},\frac{1}{3},\frac{1}{3};\frac{2}{3},\frac{2}{3},1;\frac{z^3}{27}\right)}\, .
    \end{aligned}
\end{equation}
Their series expansions have been given in paragraph \hyperref[parX66]{$X_{6,6}$} of section \ref{sec:examples}: $n(z)$ gives the coordinate change to $q(z)$, while $B(q)$ is directly given in the canonical coordinate $q$.

\section{Transition matrices between moduli space patches}\label{appendix:transition}

In this appendix we discuss some details on the numerical evaluation of the transition matrices between the different local bases. The general idea is to expand both bases around a point at which both power series converge. In a one-parameter model with at most 3 singularities one can always find a chain of boundaries for which such a point exists. In the case of at least four boundaries it can happen that there is no overlap between the regions of convergence. Moreover, even if the regions overlap the convergence can be very slow and numerically unfeasible. An example of such a situation is the operator 2.62 of section \ref{262}. This operator has the Riemann symbol
\begin{equation}
{\cal  P}\left\{\begin{array}{cccc}
0& 1/64&1& \infty\\ \hline
0& 0 &0& \frac{3}{4}\\
0& 1 &1& 1\\
0& 1 &1& 1\\
0& 2 &2& \frac{5}{4} 
\end{array}\right\}\ 
\end{equation}
As an example we want to compute the transition matrix between the two conifolds located at $x=1/64$ and $x=1$. The radius of convergence of a basis is determined by the distance to the next boundary. Thus the basis around $x=1/64$ has a radius of convergence of $1/64$ and the conifold at $x=1$ of $63/64$. Both bases converge in the interval $(1/64,2/64)$ and in principle any point inside of this interval could be used to determine the transition matrix. In practice the numerical stability at any point in this interval is poor. To determine the optimal point of expansion $x_0$ between two bases located at $x=x_a$ and $x=a_b$\footnote{For the basis at $x=\infty$ one changes the coordinate to $z=1/x$.} we solve the equation
\begin{equation}
\label{optimumequation}
    \frac{|x_a-x_0|}{r_a}=\frac{|x_b-x_0|}{r_b},
\end{equation}
where $r_a$ and $r_b$ are the radii of convergence of the respective bases.  For the two conifolds in our example this results in 
\begin{equation}
    \frac{|\frac{1}{64}-x_0|}{\frac{1}{64}}=\frac{|1-x_0|}{\frac{63}{64}}\quad\rightarrow\quad x_0=\frac{127}{4096}.
\end{equation}
The numerical issue with this is that this point is in the local conifold coordinates located at $|x_c|=\frac{63}{64}\approx 0.98$. Thus even at at order 500 in the power series expansion the suppression is only $3.8\cdot 10^{-4}$. Reaching a suppression of machine precision requires 2339 terms, explaining the slow convergence.

In these cases it is necessary to introduce an expansion around additional points in the bulk of the moduli space. As there is no monodromy around these points, the basis is formed purely out of holomorphic power series:

\begin{equation}
    \omega_i=x^i\sum_{n=0}^\infty c_n x^n\;.
\end{equation}
Using this, one can determine a chain of transition matrices
\begin{equation}
    \Pi_1=\mathcal{T}_{\rm bulk}^1\Pi_{\rm bulk}=\mathcal{T}_{\rm bulk}^1\mathcal{T}_{1/64}^{\rm bulk}\Pi_{1/64},
\end{equation}
thus the transition matrix one is interested in is given by 
\begin{equation}
  \mathcal{T}_{1/64}^1=\mathcal{T}_{\rm bulk}^1\mathcal{T}_{1/64}^{\rm bulk}.
\end{equation}
The numerical errors in determining the transition matrices to the bulk basis can be much smaller than computing the transition matrix directly. To find the optimal point of expansion one solves the analogue of \eqref{optimumequation} for both transitions. Without loss of generality we assume that the radii of convergence fulfill $r_a<r_b$, other wise one just switches the labels. This ensures that the location of bulk basis will be closer to $x_a$ than to $x_b$. Thus the radius of convergence of the bulk basis is $x_{\text{bulk}}-x_a$, denoting the position of the bulkbasis as $x_{\rm bulk}$ and the two points around which we expand as $x_1$ and $x_2$. The optimal expansion point is then determined by the three equations
\begin{equation}
\frac{x_1-x_a}{r_a}=\frac{x_{\text{bulk}}-x_1}{x_{\text{bulk}}-x_a}\qquad\frac{x_b-x_2}{r_b}=\frac{x_2-x_{\text{bulk}}}{x_{\text{bulk}}-x_a}\qquad\frac{x_2-x_{\text{bulk}}}{x_{\text{bulk}}-x_a}=\frac{x_{\text{bulk}}-x_1}{x_{\text{bulk}}-x_a},
\end{equation}
for the three unknowns $x_1,x_2$ and $x_\text{bulk}$. While these have a bit lengthy algebraic solutions, solving them numerically suffices. In the example 2.62 this results in the choices 
\begin{align*}
x_1&= \frac{251-\sqrt{505}}{7936}\approx 0.0287963\\x_{\rm bulk}&= \frac{3+\sqrt{505}}{256} \approx 0.0995008\\x_2&=\frac{63 \sqrt{505}-65}{7936}\approx 0.170205
\end{align*}
With this choice, the expansion points are positioned in local coordinates at $x_c\approx 0.843$ which is much further into the region of convergence compared to the $x_c\approx 0.984375$ without the additional basis. The suppression at 500 terms is now of order $10^{-38}$, which suffices for most computations. If even higher precision is required one can add additional bulk bases and repeat the above steps. 
\clearpage
\bibliography{literature}
\bibliographystyle{utphys}

\end{document}